%% file: main.tex
\documentclass[a4paper,11pt]{article}
\usepackage{jheppub}
\usepackage[utf8]{inputenc}

\input{preamble}

\usepackage{multirow}
\usepackage{array}

\usepackage{hhline}

\usepackage[edge-length=1cm,root-radius=0.1cm]{dynkin-diagrams}

\usepackage{tabu}

\usepackage{subfiles}

\usepackage{graphicx}
\usepackage{subcaption}

\title{Domain walls in super Yang-Mills: worldvolume TQFTs and deconfinement from semiclassics on $\mathbf{\RS}$}
\author{Andrew A. Cox}
\affiliation{Department of Physics, University of Toronto, Toronto, ON M5S 1A7, Canada}
\emailAdd{aacox@physics.utoronto.ca}

\abstract{This work studies domain walls between chirally-separated vacua in supersymmetric Yang-Mills theory (SYM) on $\RS_L$ in the semiclassical limit. For all gauge groups we explicitly find the electric fluxes of all BPS domain walls and fully characterize the representation that they form under the global symmetry of SYM. We  compute the characters of these representations formed by the semiclassical domain walls. We also compute these characters for the worldvolume TQFTs appearing in the literature for $\SU{N}$ and $\Sp{N}$ gauge groups. We find complete agreement between the two computations, providing thus a dynamical test of the proposed worldvolume TQFTs. We also propose a new worldvolume TQFT for $\E{6}$ domain walls, subjecting it to the same tests. Finally, we study deconfinement of quarks on domain walls for all gauge groups. We show that for all gauge groups confining strings (stable in the abelianized regime) can end on domain walls, regardless of whether or not the group has a center.}

\begin{document}

\maketitle

\section{Introduction}
    \subfile{intro.tex}

\section[Review of SYM on $\RS_L$]{Review of SYM on $\boldsymbol{\RS_L}$\label{sec:SYM}}
    \subfile{sym.tex}

\section{Domain walls in SYM\label{sec:DW}}
    \subfile{DomainWalls.tex}

\section{Domain wall worldvolume theories\label{sec:TQFT}}
    \subfile{worldvolume.tex}

\section{Deconfinement on domain walls\label{sec:deconfinement}}
    \subfile{deconfinement.tex}

\acknowledgments
I am grateful to Erich Poppitz for his help and guidance throughout this project and for comments on this manuscript. I acknowledge the support of the Natural Sciences and Engineering Research Council of Canada (NSERC), funding reference number 559071 / 2021.

\appendix
\section{Group theory background\label{appendix:group}}
    \subfile{group.tex}

\section{Center and charge conjugation symmetries\label{appendix:symmetries}}
    \subfile{symmetries.tex}

\section{Charge conjugation proof for \texorpdfstring{$\boldsymbol{U(u)}$ Chern-Simons theories}{Charge conjugation proof for $U(u)$ Chern-Simons theories}\label{appendix:U(N)_CS}}
    \subfile{unitaryCS.tex}

\section{Proofs of claims about deconfinement on domain walls\label{appendix:deconProofs}}
    \subfile{deconfinementProofs.tex}

\clearpage

\bibliographystyle{JHEP}
\bibliography{bibfile.bib}

\end{document}

%% file: preamble.tex
\usepackage{amsmath}
\usepackage{amssymb}
\usepackage{amsfonts}
\usepackage{dsfont}
\usepackage{mathrsfs}
\usepackage{mathtools}
\usepackage{bm}

\usepackage{enumerate}

\usepackage[arrowdel]{physics}

\newcommand{\operator}[1]{\hat{#1}}
\newcommand{\identity}{\mathds{1}}
\newcommand{\R}{\mathbb{R}}
\newcommand{\C}{\mathbb{C}}
\newcommand{\Z}{\mathbb{Z}}
\newcommand{\s}{\mathbb{S}}
\newcommand{\RS}{\R^3\times\s^1}
\newcommand{\T}{\mathbb{T}}

\newcommand{\SU}[1]{SU(#1)}
\newcommand{\SO}[1]{SO(#1)}
\newcommand{\Sp}[1]{Sp(#1)}
\newcommand{\Spin}[1]{Spin(#1)}
\newcommand{\E}[1]{E_{#1}}
\newcommand{\G}{G_2}
\newcommand{\F}{F_4}

\newcommand{\su}[1]{\mathfrak{su}(#1)}

\newcommand{\lie}[1]{\mathfrak{#1}}
\newcommand{\weyl}[1]{\mathsf{#1}}

\let\svec\vec
\let\vec\boldsymbol

%% file: intro.tex
$\mathcal{N}=1$ supersymmetric Yang-Mills theory (SYM) in four dimensions has been a remarkably useful testing ground for studying QCD-like theories due to its similarity to pure Yang-Mills and its tractability owing to the supersymmetry nonrenormalization theorems. The $\Z_{2c_2}$ 0-form discrete chiral symmetry of SYM is spontaneously broken down to $\Z_2$ leading to $c_2$ vacua, where $c_2$ is the dual Coxeter number of the gauge group. The vacua may be labelled by integers $n=0,1,\dots,c_2-1$ and are physically distinguished by the phase of the gluino condensate, $\expval{\lambda\lambda}_n = e^{2\pi in/c_2} \Lambda^3$, with $\Lambda$ the strong coupling scale. Thus, there are domain walls between the $\Z_{2c_2}$-breaking vacua. Importantly, the precise physics of the domain walls depends only on the difference between the vacua at the two sides of the domain wall. As such, we call a domain wall interpolating between vacua $n$ and $n+u$ a $u$-wall. Domain walls in SYM, and in particular BPS (Bogomolnyi-Prasad-Sommerfeld, or lowest tension) domain walls, have been studied from a variety of perspectives, see \cite{acharya_domain_2001,Ritz2002,ritz_note_2003,anber_strings_2015,Cox2019,delmastro_domain_2021} for a highly incomplete list of references. 

\paragraph{}
The low energy physics of SYM in the presence of a BPS domain wall is thought to be described by a topological quantum field theory (TQFT) which lives on the domain wall worldvolume, separating the two vacua in the bulk. In this framework, for each value of $u$, the physics of $u$-walls is described by a TQFT which depends on both $u$ and the gauge group. In these TQFTs the states of the Hilbert spaces are associated with the domain walls. Furthermore, the TQFTs must have the same symmetries as the $u$-walls, namely center symmetry and charge conjugation, and the mapping between the set of $u$-walls and the corresponding TQFT Hilbert space ought to preserve those symmetries and their 't Hooft anomalies. 

\paragraph{}
In this paper, we study the domain walls in SYM and their worldvolume TQFTs using the fact that when one of the spatial dimensions is compactified on a small circle, SYM abelianizes and the non-perturbative physics is accessible to semiclassical studies \cite{davies_gluino_1999,Davies2000,unsal_abelian_2007,Unsal2009}. In particular confinement is understood to be due to the proliferation of magnetic bions comprised of monopole-instantons \cite{unsal_abelian_2007,Unsal2009}. More recently, using semiclassical anaylses studies have shown  that heavy (probe) quarks of any non-zero $N$-ality can be deconfined on domain walls in $\SU{N}$ SYM, showing that confining strings can end on domain walls \cite{anber_strings_2015,Cox2019,bub_confinement_2021}. The semiclassical deconfinement mechanism provides an explicit realization of the mixed 0-form/1-form 't Hooft anomaly between the discrete chiral symmetry and center symmetry \cite{Gaiotto2015,Cox2019}.

\paragraph{}
Here we continue that work for all gauge groups. In particular, we find the number of BPS domain walls and their fluxes for all $u$-walls with every gauge group $G$ using semiclassics. Our counting of domain walls matches results found using more formal tools \cite{acharya_domain_2001,Ritz2002}. Further, we study how the domain walls transform under the global symmetries of SYM. More specifically, the domain walls form a representation of the global symmetry group which is understood by computing its characters.

\paragraph{}
To verify that a given proposed $u$-wall TQFT \cite{bashmakov_living_2019,delmastro_domain_2021} is correct we proceed as follows. We first semiclassically identify all the BPS $u$-walls of SYM compactified on $\RS$ and note how they transform under the appropriate symmetries. These data comprise the semiclassical Hilbert space of the domain wall worldvolume theory. We then compare our semiclassical Hilbert space to the Hilbert space of the proposed TQFT, checking both the dimension and the symmetry transformations. Concretely, in refs.~\cite{bashmakov_living_2019,delmastro_domain_2021} it was proposed that the worlvolume TQFT of $u$-walls in SYM with gauge group $G$ is three dimensional $\mathcal{N}=1$ SYM, also with gauge group $G$, with a supersymmetric Chern-Simons term at level $c_2/2-u$. Here we explicitly verify these proposals for $\SU{N}$ and $\Sp{N}$ gauge groups, using semiclassical means. We also use our semiclassical results to propose a new worlvolume TQFT for the $6$-walls of $\E{6}$.

\subsection{Summary of results}
\begin{enumerate}
    \item We find all BPS domain wall fluxes for all simple gauge groups. This extends the work of \cite{Cox2019} to all gauge groups and the results agree with previous BPS-wall counting arguments \cite{acharya_domain_2001,Ritz2002}.
    \item We semiclassically verify the proposed domain wall TQFTs for $\SU{N}$ and $\Sp{N}$ gauge groups (put forth by \cite{bashmakov_living_2019,delmastro_domain_2021}), via a detailed comparison of the TQFT Hilbert space on a torus to the semiclassical $u$-wall properties. 
    \begin{itemize}
        \item Using our semiclassical analysis, we calculate the characters of the representations of the global symmetries formed by the $u$-walls of $\SU{N}$ and $\Sp{N}$ SYM, as well as all other gauge groups for completeness. The results are summarized in tables \ref{tab:SU_uWall_characters} and \ref{tab:Sp_uWall_characters} for $\SU{N}$ and $\Sp{N}$ respectively.
        \item Starting from the canonical quantization on $\T^2$ described in, for example, ref.~\cite{elitzur_remarks_1989}, we construct the Hilbert spaces of the proposed (in \cite{bashmakov_living_2019,delmastro_domain_2021}) $u$-wall worldvolume TQFTs for $\SU{N}$ and $\Sp{N}$ explicitly. For $\SU{N}$ this construction is quite technical because the worldvolume TQFT is $U(u)$ Chern-Simons theory, which involves first constructing the Hilbert space of $\SU{u}\times U(1)$ Chern-Simons theory, then taking the $\Z_u$ quotient.
        \item We compute the characters of the Hilbert spaces of the worldvolume TQFTs. The results are given in equations \eqref{eqn:U(u)_identity_character}, \eqref{eqn:U(u)_center_character}, \eqref{eqn:U(u)_CC_character}, and \eqref{eqn:U(u)_center_CC_character} for $\SU{N}$, and \eqref{eqn:Sp_CS_identity_character} and \eqref{eqn:Sp_CS_center_character} for $\Sp{N}$, showing that they agree with those calculated using semiclassical techniques in SYM, tables \ref{tab:SU_uWall_characters} and \ref{tab:Sp_uWall_characters} mentioned above. Our results agree with calculations of the twisted Witten index\footnote{For $\SU{N}$ and $\Sp{N}$ the twisted Witten index, twisted by an element $g$ if the global symmetry group, is the same as the character of $g$ up to a sign.} in ref.~\cite{delmastro_domain_2021}. 
        
        Demonstrating the agreement of the semiclassical and TQFT calculations outlined above constitutes the main result of this paper.
        \item Note that we calculate the character of the combined action of charge conjugation and center symmetry in $\SU{2N}$ SYM, and in the corresponding worldvolume TQFT, which is necessary to fully describe representations of the global symmetry and was previously missed \cite{delmastro_domain_2021}.
    \end{itemize}
    \item We also propose, to the best of our knowledge,  for the first time a domain wall TQFT for the $u=6$-walls of $\E{6}$ SYM, arguing that it should be $(\E{6})_3$ Chern-Simons theory. We check this proposal in the same way as for $\SU{N}$ and $\Sp{N}$, finding agreement between the characters of the global symmetry of the $u=6$-walls of SYM, summarized in table \ref{tab:E6_uWall_characters}, and the Hilbert space of $(\E{6})_3$, given in equations \eqref{eqn:E6_CS_identity_character}, \eqref{eqn:E6_CS_center_character}, and \eqref{eqn:E6_CS_CC_character}. This proposal, combined with the hypothesis that the $u$-wall TQFT of four dimensional SYM with gauge group $G$ is three dimensional SYM with gauge group $G$ and a Chern-Simons term at level $c_2/2-u$, could shed light on the low energy phase of $(\E{6})_3$ SYM.
    \item We find which representations of probe quarks are deconfined on the worldvolume of the various domain walls. For almost all groups, all representations are deconfined  on all $u$-walls. The exceptions are $\SU{N}$, $\Spin{2N}$, and $\E{8}$, where $N$-ality zero representations are not deconfined in the worldvolume of $1$- and $c_2-1$-walls. We note that our analysis is carried out in the abelianized regime, and in particular $N$-ality zero quarks will in general be deconfined in the bulk in the full theory.
\end{enumerate}

\subsection{Future work}
We note that the construction of the torus Hilbert space for the $u$-wall TQFT of $\SU{N}$ SYM, section \ref{sec:SU(N)_TQFT}, is already technically quite involved; hence, the extension of the work of this paper to SYM with gauge groups other than $\SU{N}$ and $\Sp{N}$ is left for future work. It would also be interesting to use the results here about domain wall fluxes and deconfinement to study the confining strings of SYM on $\RS$, extending the work done in \cite{bub_confinement_2021} from $\SU{N}$ to all gauge groups. Further, it might be interesting to study the Abelian large-$N$ limit of $\SU{N}$ domain walls, where the semiclassical vacua become dense in field space. 

\subsection{Organization of this paper}
In section \ref{sec:SYM} we review the basics of the dynamics of SYM on $\RS_{L}$, discussing the abelianization of the theory and the action of both the center and charge conjugation symmetries on the low energy Cartan degrees of freedom. In section \ref{sec:DW} we first review the basics of domain walls in SYM, and discuss the problem of determining which domain walls are BPS. We then find all of the BPS domain walls of SYM in the semiclassical regime, and study their transformations under center symmetry and charge conjugation. This comprises all semiclassical data that we use to verify the proposal of \cite{bashmakov_living_2019,delmastro_domain_2021}.

\paragraph{}
In section \ref{sec:TQFT}, in order to facilitate the comparison to the proposed worldvolume TQFTs, we first review the Hilbert spaces of both Abelian and non-Abelian Chern-Simons theory on spatial $\T^2$, studying how the states transform under center symmetry and charge conjugation. We then explicitly construct the states in the Hilbert space of the proposed $u$-wall TQFTs for $\SU{N}$ and $\Sp{N}$, and one theory for $\E{6}$. We show that these states furnish the same representations of the global symmetries as the semiclassical $u$-walls described in section \ref{sec:DW}, providing an  explicit semiclassical check of the proposal \cite{bashmakov_living_2019,delmastro_domain_2021}.

\paragraph{}
In section \ref{sec:deconfinement} we study the deconfinement of quarks on domain walls, characterizing the deconfinement of quarks by $N$-ality for all groups. Supplementary proofs are given in appendix \ref{appendix:deconProofs}. 

\paragraph{}
All of the necessary group theory, and the accompanying notation, used throughout this paper is reviewed in appendix \ref{appendix:group}. Appendix \ref{appendix:symmetries} reviews how center symmetry and charge conjugation act on physical degrees of freedom, and gives the specific data of these symmetries needed for computations.

%% file: sym.tex
\subsection{EFT and vacua}
For small $\s^1_L$ size $L$, such that $c_2\Lambda L\ll 2\pi$ where $c_2$ is the dual Coxeter number of the gauge group\footnote{The dual Coxeter number can be understood in many ways, but perhaps the most familiar in the context of physics is as the Dynkin index of the adjoint representation.} and $\Lambda$ is the strong coupling scale, SYM abelianizes, breaking $G$ down to its maximal torus, $U(1)^r$, where $r$ is the rank of the gauge group. In the abelianized regime, $U(1)^r$ is generated by a choice of Cartan subalgebra, spanned by $r$ mutually commuting generators $H^1,H^2,\dots,H^r$, which we arrange into $\vec{H}=(H^1,\dots,H^r)$ (see appendix \ref{appendix:group} for more details). Upon integrating out the massive Kaluza-Klein modes, the remaining bosonic degrees of freedom are the holonomy scalar, $\varphi=\vec{\varphi}\cdot\vec{H}\sim \oint_{\s^1}A$, and the dual photon, $\sigma=\vec{\sigma}\cdot\vec{H}$. The full low energy description is given by a generalized Wess-Zumino model \cite{Shifman_2022}, where the K\"{a}hler metric is the identity up to subleading corrections which are negligible in the semiclassical limit, see \cite{Anber2014} for 1-loop corrections. The dual photon is related to the $\R^3$ field strength via the duality transformation $F^{\mu\nu}=\frac{g^2}{4\pi L}\varepsilon^{\mu\nu\rho}\partial_\rho\sigma$, where $g$ is the SYM coupling at the scale of order $1/L$. After shifting the fields to be centered around the center-symmetric vev, the low energy bosonic action is compactly written in terms of the complex scalar $\vec{z}=i(\vec{\sigma}+\tau\vec{\varphi})$ where $\tau=\frac{\theta}{2\pi}+\frac{4\pi i}{g^2}$ is the usual complexified coupling, as
\begin{equation}
    S=\int\dd[3]{x}M\left(\partial_\mu\vec{z}\cdot\partial^\mu\vec{z}^\dagger - \frac{m^2}{4}\pdv{W}{\vec{z}}\cdot\pdv{W^*}{\vec{z}^\dagger}\right),
\end{equation}
where $W(\vec{z})$ is the superpotential given by
\begin{equation}
    W(\vec{z})=\sum_{a=0}^rk_a^*e^{\vec{\alpha}_a^*\cdot\vec{z}},\label{eqn:superpotential}
\end{equation}
where $\vec{\alpha}_a^*$ are the simple ($1\leq a\leq r$) and affine ($a=0$) co-roots, and $k_a^*$ are the dual Kac labels, defined by $\sum_{a=0}^rk_a^*\vec{\alpha}_a^*=0$ with $k_0^*=1$. The scales are set by $M\sim\frac{g^2}{L}$ and $m\sim Me^{-8\pi^2/(g^2c_2)}$. There are $c_2$ classical vacua given by $\vec{z}_n=\frac{2\pi i n}{c_2}\vec{\rho}$, up to $2\pi i$ additions of weights as explained below, where $\vec{\rho}$ is the Weyl vector, given by the sum of the fundamental weights. For more background on Cartan subalgebras and the notation used throughout this paper, see appendix \ref{appendix:group}.

\paragraph{}
After fixing $\varphi$ to be in the Cartan, we may still perform large gauge transformations which wind around the $\s^1$, allowing us to shift $\vec{\varphi}$ by $2\pi$ times a co-root. Further, we may perform constant gauge transformations which preserve the Cartan, ie the Weyl group $W$\footnote{See appendix \ref{appendix:gauge_Weyl} for explicit construction of the gauge transformations which correspond to Weyl group elements.}, noting that $\sigma$ also transforms under constant gauge transformations since $\sigma \sim F$. Thus, the moduli space for $\vec{\varphi}$ is
\begin{equation*}
    \mathcal{M}_{\vec{\varphi}}=\frac{\R^r}{W\ltimes \Lambda_r^*},
\end{equation*}
where $\Lambda_r^*$ is the co-root lattice, representing co-root translations. Note that we have the semi-direct product $W\ltimes \Lambda_r^*$ because the Weyl group and co-root shifts do not commute\footnote{This is fairly easy to see with an example: first shift by $\vec{\alpha}^*$ then do a simple Weyl reflection $s_{\vec{\beta}}$ to get $s_{\vec{\beta}}(\vec{\varphi}+\vec{\alpha}^*)=\vec{\varphi}+\vec{\alpha}^*-(\vec{\beta}\cdot(\vec{\varphi}+\vec{\alpha}^*))\vec{\beta}^*$. Doing the transformation in the opposite order gives something else: $s_{\vec{\beta}}(\vec{\varphi})+\vec{\alpha}^*=\vec{\varphi}+\vec{\alpha}^*-(\vec{\beta}\cdot\vec{\varphi})\vec{\beta}^*$. Notice that the two differ by $-(\vec{\beta}\cdot\vec{\alpha}^*)\vec{\beta}^*$, and hence in general are not the same transformation. Note that for convenience factors of $2\pi$ have been omitted here.}. The fundamental domain for $\vec{\varphi}$, the region where no two points on the interior are identified via $W\ltimes \Lambda_r^*$, is given by
\begin{equation*}
    \hat{T}_{\vec{\varphi}}=\left\{\vec{v}\in\R^r\mid \vec{\alpha}_a^*\cdot\vec{v}\geq0,\ a=1,\dots,r,\ -\vec{\alpha}_0^*\cdot\vec{v}<2\pi\right\}.
\end{equation*}
The moduli space of the dual photon is similar, except that $\vec{\sigma}$ is identified with $2\pi$ times any weight\footnote{This identification comes from the introduction of $\sigma$ as a Lagrange multiplier for the Bianchi identity through the action $S_\sigma=\frac{i}{4\pi}\int\dd[3]{x}\vec{\sigma}\cdot\partial_\mu\vec{B}^\mu$, where $\vec{B}^\mu=\epsilon^{\mu\nu\rho}\vec{F}_{\nu\rho}$. Since the magnetic charge over the surface integral at spatial infinity is quantized in the co-root lattice, $-\frac{1}{2\pi}\int_{\s^2_\infty}\dd[2]S_\mu\vec{B}^\mu\in\Lambda_r^*$, shifting $\vec{\sigma}$ by anything in $2\pi\Lambda_w$ does not change the path integral.}, so the appropriate moduli space is
\begin{equation*}
    \mathcal{M}_{\vec{\sigma}}=\frac{\R^r}{W\ltimes \Lambda_w},
\end{equation*}
where $\Lambda_w$ is the weight-lattice. The fundamental domain for $\vec{\sigma}$ is given by
\begin{equation*}
    \hat{T}_{\vec{\sigma}}=\left\{\vec{v}\in\R^r\mid 0\leq\vec{\alpha}_a^*\cdot\vec{v}\leq2\pi,\ a=1,\dots,r\right\}.
\end{equation*}

\subsection{Symmetries}
\subsubsection*{Two important Weyl group elements}
We will show below that the action of the relevant global symmetries, center symmetry in the $\s^1$ direction and charge conjugation, on the Cartan degrees of freedom will ultimately come from two special Weyl group elements. These elements are described in more detail in appendix \ref{appendix:Special_Weyl}, and their actions for each group are summarized in table \ref{tab:Weyl_longest}. The first, which is relevant for both center symmetry and charge conjugation, we call $\weyl{w}_\Pi$ because it is the unique Weyl group element which maps the set of simple roots, $\Pi=\left\{\vec{\alpha}_a\mid a=1,\dots,r\right\}$, to itself with a sign flip. In other words, $\weyl{w}_\Pi$ maps $\Pi$ to $-\Pi$ setwise, meaning that there is a (not necessarily non-trivial) permutation $\varpi\in S_{r+1}$\footnote{We call the permutation, or symmetric, group of $n$ elements $S_n$.} such that $\weyl{w}_\Pi(\vec{\alpha}_a)=-\vec{\alpha}_{\varpi(a)}$. Note that we chose $\varpi$ to be a permutation of $r+1$ elements and not $r$ so that we could capture the action of $\weyl{w}_\Pi$ on the affine root. It is not too hard to show that $\weyl{w}_\Pi$ must flip the sign of the affine root, so that $\varpi(0)=0$. Since the only Weyl group element which preserves $\Pi$ is the identity, we see that $\weyl{w}_\Pi$, and hence $\varpi$, must have order 2. Finally, it will be important to note that $\varpi$ preserves the dual Kac labels, in other words $k_a^*=k_{\varpi(a)}^*$.

\paragraph{}
The second special Weyl group element is relevant only for center symmetry. We call it $\weyl{w}_{\Pi_c}$ because, similar to $\weyl{w}_\Pi$, it is the unique Weyl group element which maps $\Pi\setminus\{\vec{\alpha}_c\}$ to $-\left(\Pi\setminus\{\vec{\alpha}_c\}\right)$. Here $1\leq c\leq r$ is an index such that the Kac label $k_c$ is unity and $\vec{\alpha}_c$ is a long root\footnote{Note that such a $c$ is not always possible to find; each $c$ corresponds to an element of the center of the corresponding Lie group.}. Like $\weyl{w}_\Pi$, we capture the action of $\weyl{w}_{\Pi_c}$ with a permutation $\gamma_c\in S_{r+1}$ such that $\weyl{w}_{\Pi_c}(\vec{\alpha}_a)=-\vec{\alpha}_{\gamma_c(a)}$. In addition to mapping $\Pi\setminus\{\vec{\alpha}_c\}$ to $-\left(\Pi\setminus\{\vec{\alpha}_c\}\right)$, $\weyl{w}_{\Pi_c}$ maps $\vec{\alpha}_c$ and $-\vec{\alpha}_0$ to each other. Thus, the permutation $\gamma_c$ satisfies $\gamma_c(0)=c$ and $\gamma_c(c)=0$. In the same way that $\weyl{w}_\Pi$ has order 2, $\weyl{w}_{\Pi_c}$, and hence $\gamma_c$, also have order 2. Like $\varpi$, $\gamma_c$ also preserves dual Kac labels.

\subsubsection*{1-form center symmetry}
The $\s^1$ part of the center symmetry, when it exists, acts on Wilson loops that wind around $\s^1$ by a phase. In the dimensional reduction of $\s^1_L$, this action appears to be a 0-form symmetry, and hence we will sometimes refer to it as the ``0-form" center symmetry. Acting on the $A_4$ gauge field, we can take a center element to act by an improper ``gauge transformation" $g_c(x)=e^{2\pi i x_4/L \vec{w}_c^*\cdot\vec{H}}$, shifting $\vec{A}_4$ by $2\pi\frac{x_4}{L}\vec{w}_c^*$, where $\vec{w}_c^*$ is a fundamental co-weight and $c$ is an index chosen such that $g_c(L)$ is the desired element of the center (refer to appendix \ref{appendix:center} for more details). Recall that we shifted the holonomy scalar to be centered around the center-symmetric vev $\vec{\varphi}_0$, so that $\vec{\varphi}+\vec{\varphi}_0\sim\oint_{\s^1}\vec{A}$. Accordingly, $\vec{\varphi}+\vec{\varphi}_0$ transforms by a shift of $2\pi\vec{w}_c^*$ which does not preserve the fundamental domain for $\vec{\varphi}$. We remedy this by supplementing the improper gauge transformation with a Weyl transformation, $\mathcal{T}=\weyl{w}_{\Pi_c}\circ\weyl{w}_\Pi$, which will also act on $\vec{\sigma}$. The Weyl transformation is the unique Weyl group element which, combined with the improper gauge transformation, preserves $\hat{T}_{\vec{\varphi}}$. Further, since $\vec{\varphi}_0$ is the center symmetric vev it absorbs the shift by $2\pi\vec{w}_c^*$, and $\vec{\varphi}$ and $\vec{\sigma}$ simply transform by the Weyl transformation,
\begin{equation}
    Z^{\s^1}(G): \vec{z}\rightarrow \mathcal{T}(\vec{z})=\weyl{w}_{\Pi_c}\circ\weyl{w}_\Pi(\vec{z}).\label{eqn:center_fields}
\end{equation}
While the specifics depend on the gauge group, the action of $\mathcal{T}$ is to permute the simple and affine roots,
\begin{equation}
    Z^{\s^1}(G): \vec{\alpha}_a^*\cdot\vec{z}\rightarrow \vec{\alpha}^*_{\varpi\circ\gamma_c(a)}\cdot\vec{z},
\end{equation}
where $\varpi$ and $\gamma_c$ are discussed above. See appendix \ref{sec:all_groups_symmetries} and table \ref{tab:Weyl_longest} for details for each group. Though it will not be necessary for this work, it is worth noting that the $\R^3$ part of the 1-form center symmetry acts on Wilson lines in $\R^3$ with a phase.

\subsubsection*{0-form charge conjugation symmetry}
Under charge conjugation the Cartan part of the gauge field flips sign, and hence both $\vec{\sigma}$ and $\vec{\varphi}$ flip sign as well. The na\"{i}ve action of charge conjugation does not preserve the superpotential, nor the fundamental domains for $\vec{\sigma}$ and $\vec{\varphi}$. We can supplement charge conjugation with the Weyl transformation $\weyl{w}_\Pi$, so that our new charge conjugation symmetry acts as
\begin{equation}
    \Z_2^{(0)}: \vec{z}\rightarrow \mathcal{C}(\vec{z})=-\weyl{w}_\Pi(\vec{z}),\label{eqn:cc_fields}
\end{equation}
where we note that $\vec{A}$ transforms in the same way. Thus, the action of $\mathcal{C}$ is to permute the simple roots and keep the affine root constant, and hence permute the terms in the superpotential \eqref{eqn:superpotential} according to $\varpi$ while keeping the fundamental domains for $\vec{\varphi}$ and $\vec{\sigma}$ constant,
\begin{equation}
    \Z_2^{(0)}: \vec{\alpha}_a^*\cdot\vec{z} \rightarrow \vec{\alpha}_{\varpi(a)}^*\cdot\vec{z}.
\end{equation}
See appendix \ref{appendix:charge_conjugation} for a more in-depth discussion including a note on $\Spin{4n}$ where $\mathcal{C}$ as defined above is trivial, but a non-trivial notion of charge conjugation may still be defined. See appendix \ref{sec:all_groups_symmetries} and table \ref{tab:Weyl_longest} for the explicit action of $\mathcal{C}$ for each group.

%% file: DomainWalls.tex
One can show that the lowest energy, or BPS, domain walls satisfy the so-called BPS equations \cite{cecotti_classification_1992},
\begin{equation}
    \dv{z^i}{x}=m\frac{\alpha}{2}g^{ij}\pdv{\bar{W}}{\bar{z}^j},\label{eqn:BPS}
\end{equation}
where $x$ is the coordinate along the domain wall, $g^{ij}\sim\delta^{ij}+\dots$ is the inverse K\"{a}hler metric, and $\alpha$ is the complex phase of $W(x=\infty)-W(x=-\infty)$. We call a domain wall interpolating from $\vec{z}_n$ to $\vec{z}_{n+u}$ a $u$-wall, and without loss of generality we consider only $n=0$, in which case $\alpha=ie^{i\pi u/c_2}$. The energy of a BPS $u$-wall is given by
\begin{equation}
    E_\text{BPS}(u)=2mMc_2\sin\frac{\pi u}{c_2}.
\end{equation}

\paragraph{}
Each domain wall is labelled by its (electric\footnote{The term electric flux comes from the duality relation; in the abelianized regime the $\R^2$ electric field components are dual to the spatial derivatives of the dual photon, $\vec{E}_i=\vec{F}_{0i}=\frac{g^2}{4\pi L}\varepsilon_{ij}\partial^j\vec{\sigma}$.}) flux $\vec{\Phi}=\frac{i}{2\pi}(\vec{z}(\infty)-\vec{z}(-\infty))$. Working within $\hat{T}_{\vec{\sigma}}$, we assume that $\vec{z}(-\infty)=\vec{z}_0$ is given by $2\pi i\sum_{a=1}^rq_a\vec{w}_a$ for $q_a\in\{0,1\}$ and that $\vec{z}(\infty)=\vec{z}_u=\frac{2\pi iu}{c_2}\vec{\rho}$, so that each flux uniquely defines a domain wall. It will be most convenient to represent domain walls by either their flux directly, or as a $(r+1)$-tuple $(q_0,q_1,\dots,q_r)$ where $q_0$ is defined by $\sum_{a=0}^rk_a^*q_a=u$. The latter representation will be particularly useful when discussing various symmetries.

\subsection{Which domain walls are BPS?}
It has been argued \cite{Ritz2002,ritz_note_2003} that the domain walls for which $X_a=e^{\vec{\alpha}_a^*\cdot\vec{z}}$ does not wind more than once around the origin in the complex plane, for every $a=0,1,\dots,r$, are the BPS domain walls. The winding number of $X_a$ is readily computed as
\begin{equation*}
    \operatorname{winding}(X_a)=\frac{1}{2\pi i}\int_\text{DW}\frac{\dd{X_a}}{X_a}=-\vec{\alpha}_a^*\cdot\vec{\Phi},
\end{equation*}
for any domain wall with flux $\vec{\Phi}$. With boundary conditions $\vec{z}(-\infty)=2\pi i\sum_{a=1}^rq_a\vec{w}_a$ and $\vec{z}(\infty)=\frac{2\pi iu}{c_2}\vec{\rho}$, the winding of $X_a$ is given by $\frac{u}{c_2}-q_a$ for all $a$. Thus, the argument says that BPS domain walls must have $q_a=0,1$ for each $a$, which introduces a non-trivial constraint on $q_0$ since $\sum_{a=0}^rk_a^*q_a=u$. The problem of finding the BPS domain walls comes down to solving
\begin{equation}
    \sum_{a=0}^rk_a^*q_a=u,\quad q_a\in\{0,1\}\label{eqn:BPS_condition}.
\end{equation}
The solutions are given in section \ref{sec:DW_Counting} for each gauge group. Using numerics, with the same approach as in \cite{Cox2019}, we have verified that domain walls saturate the BPS bound if and only if equation \eqref{eqn:BPS_condition} is satisfied, checking all domain walls with $G=\Sp{N}$ for $3\leq N\leq 7$, $G=\Spin{N}$ for $7\leq N\leq 14$\footnote{Note that $\Spin{N}$ for $N<7$ is isomorphic to one of the other gauge groups studied.}, as well as all the exceptional groups, noting that $\SU{N}$ was studied in \cite{Cox2019}.

\begin{figure}[h]
    \centering
    \includegraphics[width=0.75\linewidth]{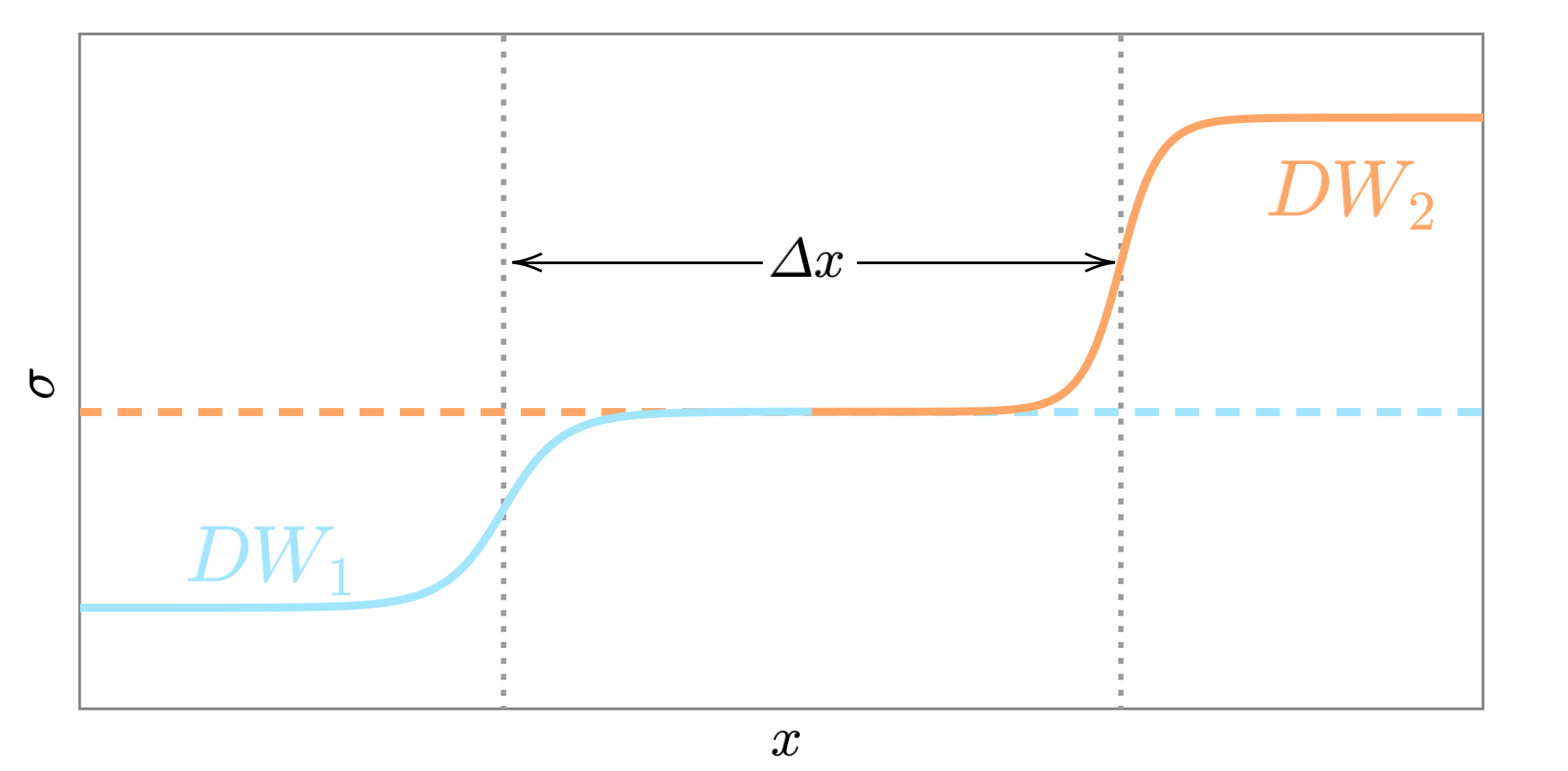}
    \caption{Schematic of combining two domain walls, whose centers are separated by a distance $\Delta x$. All scales are arbitrary.}
    \label{fig:dw_combining}
\end{figure}

\paragraph{}
Working towards a more dynamical explanation of equation \eqref{eqn:BPS_condition}, we have numerically solved for domain wall configurations by minimizing energy while keeping the boundaries fixed, as done in \cite{Cox2019}, for several groups and all values of $u$. Then, we were able to build ``composite" domain walls by adding together two BPS domain walls, which are the domain walls that both solve the equations of motion and have energy equal to the BPS energy, possibly reversing the orientation of one of the walls. For example, let $\vec{z}^{(1)}(x)$ and $\vec{z}^{(2)}(x)$ be BPS $u_1$- and $u_2$-walls both centered at $x=0$. We can add the two walls together, separated by a distance $\Delta x$, to get a $(u_1+u_2)$-wall
\begin{equation*}
    \vec{z}^{(1)+(2)}(x)=\vec{z}^{(1)}\left(x+\frac{\Delta x}{2}\right)+\vec{z}^{(2)}\left(x-\frac{\Delta x}{2}\right),
\end{equation*}
so that for $\Delta x$ much larger than the domain wall widths, $\vec{z}^{(1)+(2)}(x<0)=\vec{z}^{(1)}\left(x+\frac{\Delta x}{2}\right)+const$ and $\vec{z}^{(1)+(2)}(x>0)=\vec{z}^{(2)}\left(x+\frac{\Delta x}{2}\right)+const$, which is depicted schematically in figure \ref{fig:dw_combining}. In the limit of $\Delta x\rightarrow \infty$, $\vec{z}^{(1)+(2)}(x)$ exactly solves the equations of motion, but will cost more energy than a BPS $(u_1+u_2)$-wall. In general, the flux of the resulting domain wall is the sum of the fluxes of the two walls being merged, so a 1-wall and a 1-wall would combine to form a 2-wall, a 2-wall and an anti 1-wall would combine to form a 1-wall, etc. 

\begin{figure}[h]
    \centering
    \begin{subfigure}{.48\textwidth}
        \centering
        \includegraphics[width=\linewidth]{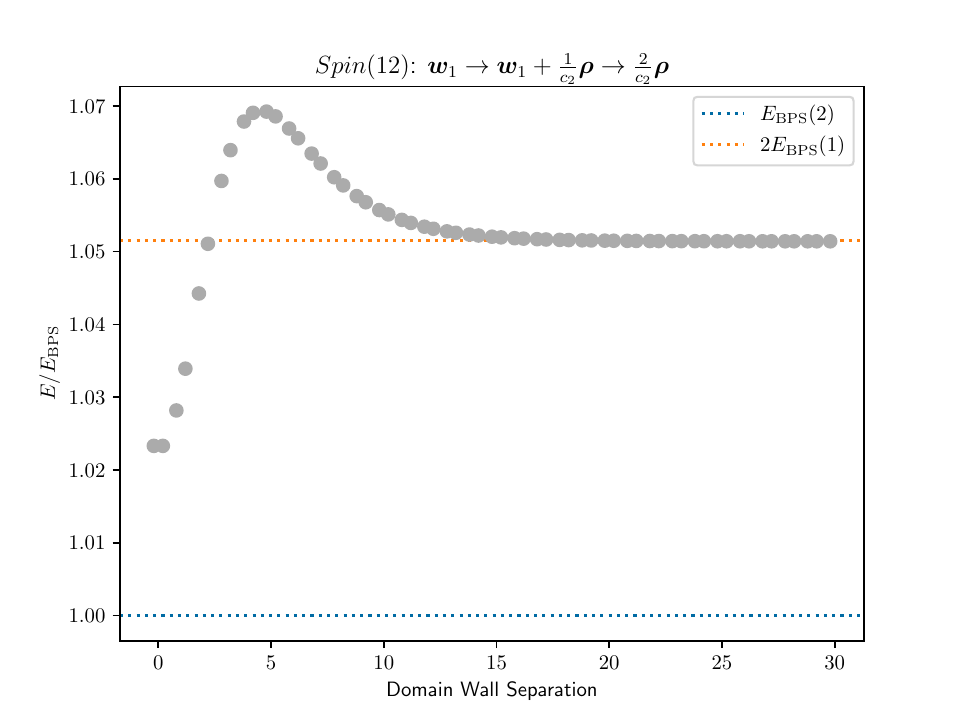}
        \caption{}
        \label{fig:mergeDW_EvenSpin6}
    \end{subfigure}
    \begin{subfigure}{.48\textwidth}
        \centering
        \includegraphics[width=\linewidth]{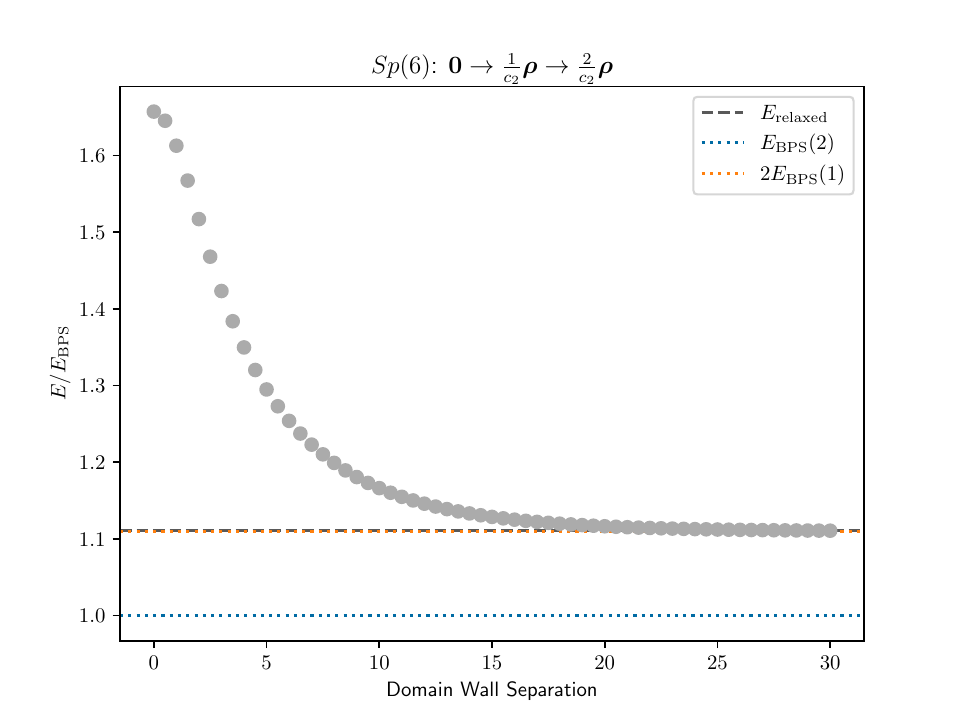}
        \caption{}
        \label{fig:mergeDW_Sp6}
    \end{subfigure}
    \begin{subfigure}{.48\textwidth}
        \centering
        \includegraphics[width=\linewidth]{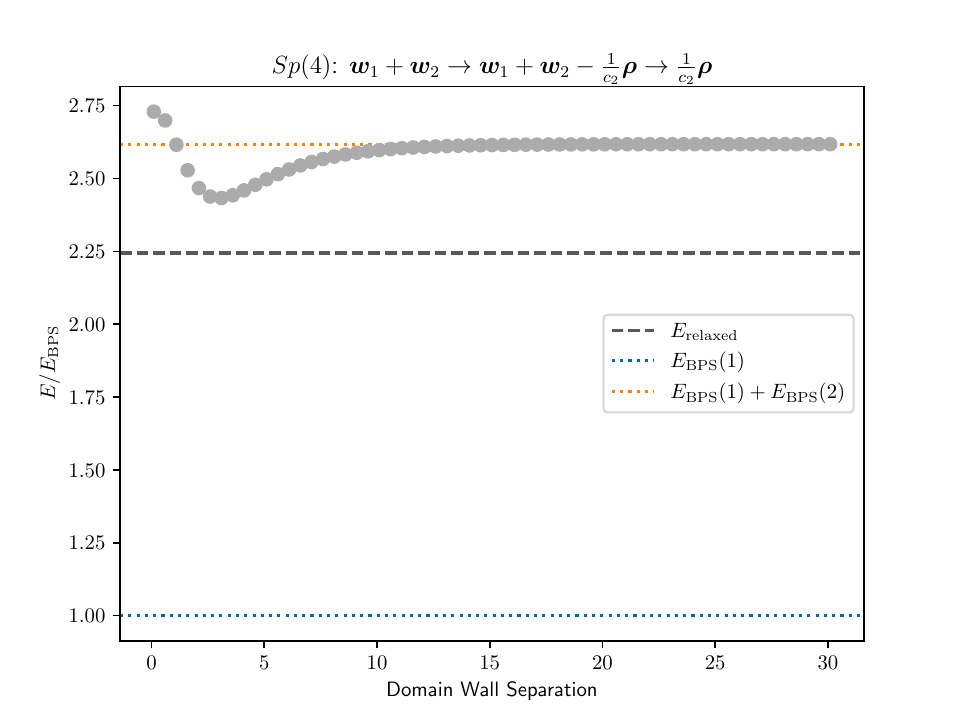}
        \caption{}
        \label{fig:mergeDW_Sp4}
    \end{subfigure}
    \caption{Energy of merging domain wall configurations, where each figure indicates the corresponding gauge group and fluxes of the domain walls being merged. The two walls merged were both numerically found to be BPS in each case. In \ref{fig:mergeDW_EvenSpin6} the overall configuration, a domain wall going from $\vec{w}_1$ to $\frac{2}{c_2}\vec{\rho}$ is BPS, while in figures \ref{fig:mergeDW_Sp6} and \ref{fig:mergeDW_Sp4} the resulting configurations are not BPS. Each figure shows the BPS energy for the corresponding domain wall, as well as the sum of the BPS energies for the two domain walls being merged. For the non-BPS domain walls, the energy corresponding to minimizing the energy with the appropriate boundary conditions is also shown.}
    \label{fig:mergeDW}
\end{figure}

\paragraph{}
We then studied the energy of the resulting configuration as a function of the separation of the two domain wall centers\footnote{The domain wall center was numerically calculated to be the position corresponding to the midpoint of the superpotential. We note that the image of a BPS domain wall in the complex $W$-plane is a straight line, which follows directly from the BPS equation \eqref{eqn:BPS}.}. Figure \ref{fig:mergeDW_EvenSpin6} shows an example where there is a clear attraction between the two domain walls being merged, and in fact corresponds to a BPS domain wall. In contrast, figure \ref{fig:mergeDW_Sp6} clearly shows that the two domain walls repel, and suggests that the lowest energy configuration will be of two well separated BPS 1-walls. Indeed, when minimizing the energy of a configuration with the corresponding boundary conditions, the minimum energy is in fact the same as the sum of the two BPS 1-wall energies, as shown in the figure. Finally, figure \ref{fig:mergeDW_Sp4} shows a case where the corresponding domain wall is not BPS, but there is a region where the two domain walls are attracted. A general model of domain wall interactions would help elucidate this phenomenon, and it seems would show that whenever two BPS fluxes add to another BPS flux, the interaction at short distances is attractive, and repulsive when they do not.

\subsection{Action of symmetries on domain walls}
\subsubsection*{``0-form" center symmetry}
Domain walls, and hence their fluxes, transform under the $\s^1$ part of the center symmetry according to equation \eqref{eqn:center_fields}, which acts on the labels $q_a$ as
\begin{equation}
    Z^{\s^1}(G): q_a\rightarrow q_{\varpi\circ\gamma_c(a)}.
\end{equation}
See appendix \ref{sec:all_groups_symmetries} and table \ref{tab:Weyl_longest} for more details for each group.

\subsubsection*{0-form charge conjugation}
Domain walls transform under charge conjugation according to equation \eqref{eqn:cc_fields}, and hence the labels $q_a$ transform as
\begin{equation}
    \Z_2^{(0)}: q_a\rightarrow q_{\varpi(a)}.
\end{equation}
See appendix \ref{sec:all_groups_symmetries} and table \ref{tab:Weyl_longest} for more details for each group.

\subsection{Counting and characterizing BPS domain walls\label{sec:DW_Counting}}
The domain walls form a representation of the global symmetry of SYM, which will be a combination of charge conjugation and center symmetry. To specify the representation that they form, we compute the characters of that representation. The character of a representation $R$ of a finite group $G$, $\chi_R$, is a function from $G$ to $\C$ defined as
\begin{equation*}
    \chi_R(g)=\Tr\left(R(g)\right).
\end{equation*}
The characters of $R$, computed for each $g\in G$, fully determine the representation, and if two representations have the same characters then they are equivalent up to isomorphism. It is important to notice that $\chi_R$ is a class function, meaning that it takes the same value across a conjugacy class,
\begin{equation*}
    \chi_R(g)=\chi_R(hgh^{-1}),\ \forall g,h\in G.
\end{equation*}
Thus, to compute the characters of a representation of $G$, we simply have to compute the character of a representative from each conjugacy class of $G$. We will label the conjugacy class of $g\in G$ by $[g]=\left\{ hgh^{-1}\mid h\in G\right\}$.

\paragraph{}
To construct the representations that the domain walls of SYM form, we first determine the number of domain walls, say $n$. We then associate to each domain wall a unit vector $\svec{e}_i\in \R^n$. Since the global symmetries permute the domain walls, they will simply permute the unit vectors so that the character of $g\in G$ is the number of domain walls which are fixed by $g$. We will denote the character of the representation formed by $u$-walls of SYM with gauge group $G$ by $\chi_u^G$.

\paragraph{}
In the most basic cases, charge conjugation and center symmetry are trivial, so the only character to compute is that of the identity. In other words, the best we can do is count how many domain walls there are. For some gauge groups charge conjugation is trivial and the global symmetry of SYM is simply $\Z_m$, so the characters are simply the number of domain walls fixed by each power of the $\Z_m$ generator. Finally, in cases, except for $\Spin{8}$, where both charge conjugation and center symmetry are non-trivial the global symmetry will be a dihedral group, $D_{2N}$, which is the group of symmetries of an $N$-sided regular polygon, and is generated by a $\Z_N$ rotation, $\mathcal{T}$, and a $\Z_2$ reflection, $\mathcal{C}$, which obey a dihedral algebra
\begin{equation*}
    D_{2N}=\left\langle \mathcal{C},\mathcal{T}\mid \mathcal{C}^2=\mathcal{T}^N=\left(\mathcal{C}\circ\mathcal{T}\right)^2=\identity\right\rangle.
\end{equation*}
When $N=2k+1$ is odd, representatives of the conjugacy classes of $D_{2N}$ are $\identity$, $\mathcal{T}^m$ for $1\leq m\leq k$, and $\mathcal{C}$. When $N=2k$ is even, representatives of the conjugacy classes of $D_{2N}$ are $\identity$, $\mathcal{T}^m$ for $1\leq m\leq k$, $\mathcal{C}$, and $\mathcal{C}\circ\mathcal{T}$. Thus, for both $N$ even and $N$ odd, we have to count the number of domain walls fixed by $\mathcal{T}^m$ and $\mathcal{C}$, and when $N$ is even we must also count the number of domain walls fixed by $\mathcal{C}\circ\mathcal{T}$.

\paragraph{}
Let us first illustrate with an example: 2-walls in $\SU{4}$. Here the rank is three, the dual Kac labels are all unity, the center symmetry acts as $\mathcal{T}: q_a\rightarrow q_{a+1\bmod{4}}$, and the charge conjugation symmetry acts as $\mathcal{C}: q_a\rightarrow q_{4-a\bmod{4}}$. The global symmetry is $D_8=\Z_2\ltimes \Z_4$, the symmetry group of the square, whose conjugacy classes are $[\identity]$, $[\mathcal{T}]$, $[\mathcal{T}^2]$, $[\mathcal{C}]$, and $[\mathcal{C}\circ\mathcal{T}]$. To count the number of BPS 2-walls, we need to solve
\begin{equation*}
    q_0+q_1+q_2+q_3=2,\ q_a\in\{0,1\},
\end{equation*}
which has $\chi_R(\identity)=\binom{4}{2}=6$ solutions: $q_{a_1}=q_{a_2}=1$ and $q_{a_3}=q_{a_4}=0$ where all $a_i$ are different. Under the $\Z_4^{(1),\s^1}$ center symmetry $(q_0,q_1,q_2,q_3)$ is rotated to the left to become $(q_1,q_2,q_3,q_0)$. A domain wall fixed by $\mathcal{T}$ must satisfy $q_0=q_1=q_2=q_3$ which is clearly not BPS by the above condition. $\mathcal{T}^2$ maps $(q_0,q_1,q_2,q_3)$ to $(q_2,q_3,q_0,q_1)$, and thus fixes two BPS domain walls: $(1,0,1,0)$ and $(0,1,0,1)$. Then, $\chi_R(\mathcal{T})=0$ and $\chi_R(\mathcal{T}^2)=2$. Under charge conjugation only $q_1$ and $q_3$ are swapped, so there are two BPS domain walls fixed by $\mathcal{C}$: $(1,0,1,0)$ and $(0,1,0,1)$, giving us $\chi_R(\mathcal{C})=2$. Finally, $\mathcal{C}\circ\mathcal{T}$ acts as $q_a\rightarrow q_{3-a\bmod{4}}$, so there are two BPS domain walls fixed by $\mathcal{C}\circ\mathcal{T}$: $(1,0,0,1)$ and $(0,1,1,0)$, giving $\chi_R(\mathcal{C}\circ\mathcal{T})=2$. Thus, the full character table of the domain walls is
\begin{table}[h!]
    \centering
    \begin{tabular}{l|lllll}
         & $[\identity]$ & $[\mathcal{T}]$ & $[\mathcal{T}^2]$ & $[\mathcal{C}]$ & $[\mathcal{C}\circ\mathcal{T}]$ \\ \hline
         $u=2$ & 6 & 0 & 2 & 2 & 2
    \end{tabular}
\end{table}

\paragraph{}
In the following, we compute the characters of the global symmetries formed by the domain walls of SYM for all gauge groups. For the specific group theoretic data used for each group, see appendix \ref{sec:all_groups_symmetries}.

\subsubsection[$\SU{N}$]{$\boldsymbol{\SU{N}}$}
For $\SU{N}$, the $\Z_N^{(1),\s^1}$ center symmetry generator acts as $\mathcal{T}:q_a\rightarrow q_{a+1\bmod{N}}$, while the $\Z_2^{(0)}$ charge conjugation generator acts as $\mathcal{C}: q_a\rightarrow q_{N-a\bmod{N}}$, so that the global symmetry group is the dihedral group with $2N$ elements, $D_{2N}$. The conjugacy classes for $D_{2N}$ are $[\identity]$, $[\mathcal{C}]$, $[\mathcal{T}^n]$ for $n=1,\dots,\left\lfloor \frac{N}{2}\right\rfloor$, and when $N$ is even $[\mathcal{C}\circ\mathcal{T}]$. The dual Kac labels are all unity, so equation \eqref{eqn:BPS_condition} becomes
\begin{equation}
    u=\sum_{a=0}^{N-1}q_a.\label{eqn:SU_uWall_BPS}
\end{equation}
Computing the character of the identity amounts to counting solutions to equation \eqref{eqn:SU_uWall_BPS} with $q_a\in\{0,1\}$. The number of solutions is the number of ways to choose $u$ of $\left\{q_0,q_1,\dots,q_{N-1}\right\}$ to be one, $\binom{N}{u}$.

\paragraph{}
Next let us compute the character of $\mathcal{T}^n$. Suppose that a domain wall is invariant under $\mathcal{T}^n$, so that $q_a=q_{a+n\bmod{N}}$. We can then ask when does the sequence $\{q_a,q_{a+n},q_{a+2n},\dots\}$ start repeating, or in other words when is $kn\equiv 0\bmod{N}$ for a strictly positive integer $k$? To see the answer, it is first helpful to write $n$ as $\frac{n}{\gcd(N,n)}\gcd(N,n)$, where $\frac{n}{\gcd(N,n)}$ is co-prime\footnote{Two integers are called co-prime when they share no common prime factors, or equivalently when their $\gcd$ is one.} with $N$. It is well known from number theory that $ab\equiv0\bmod{c}$ is the same as $a\equiv 0\bmod{c}$ whenever $b$ and $c$ are co-prime. Thus, we have that $\left(k\gcd(N,n)\right)\frac{n}{\gcd(N,n)}\equiv0\bmod{N}$ is the same as $k\gcd(N,n)\equiv 0\bmod{N}$. The smallest postive solution is then $k\gcd(N,n)=N$. In other words, the sequence $\{q_a,q_{a+n\bmod{N}},q_{a+2n\bmod{N}},\dots\}$ has $\frac{N}{\gcd(N,n)}$ unique elements. Furthermore, there are $\gcd(N,n)$ such sequences. Starting from B\'{e}zout's identity it is not too difficult to show that each $q_a$ for $a=0,\dots,\gcd(N,n)-1$ are all in different sequences. Thus, whenever a domain wall is invariant under $\mathcal{T}^n$ we can write equation \eqref{eqn:SU_uWall_BPS} as
\begin{equation*}
    u=\frac{N}{\gcd(N,n)}\sum_{a=0}^{\gcd(N,n)-1}q_a.
\end{equation*}
From basic combinatorics, there are $\binom{\gcd(N,n)}{u/(N/\gcd(N,n))}$ solutions whenever $\frac{N}{\gcd(N,n)}$ divides $u$, and no solutions otherwise.

\paragraph{}
To compute the character of $\mathcal{C}$, we simply have to solve equation \eqref{eqn:SU_uWall_BPS} subject to $q_a=q_{N-a\bmod{N}}$
\begin{equation*}
    u=\begin{dcases}q_0+2\sum_{a=1}^{N/2-1}q_a+q_{N/2} & N\text{ even} \\ q_0+2\sum_{a=1}^{(N-1)/2}q_a & N\text{ odd}\end{dcases}.
\end{equation*}
The counting of invariant domain walls depends on whether $N$ is even or odd. For $N$ even and $u$ odd, we need $q_0+q_{N/2}=1$ which can be done two ways, giving us $u-1=2\sum_{a=1}^{N/2-1}q_a$ which has $\binom{N/2-1}{(u-1)/2}$ different solutions, giving $2\binom{N/2-1}{(u-1)/2}$ possible domain walls. For $N$ and $u$ even, we know that $q_0+q_{N/2}$ must be even implying that $q_0=q_{N/2}$, giving us $u=2\sum_{a=0}^{N/2-1}q_a$, which has $\binom{N/2}{u/2}$ solutions. For $N$ odd, $q_0$ is exactly determined by $u$ so that $u-q_0$ is even. In other words, $q_0$ must be such that $u-q_0=2\left\lfloor u/2\right\rfloor$. We then have $\left\lfloor u/2\right\rfloor=\sum_{a=1}^{(N-1)/2}q_a$, giving us $\binom{(N-1)/2}{\left\lfloor u/2\right\rfloor}$ charge conjugation invariant domain walls.

\paragraph{}
When $N$ is even, we also have to compute the character of $\mathcal{C}\circ\mathcal{T}$, which acts as $q_a\rightarrow q_{N-a-1\bmod{N}}$. Requiring that a domain wall is fixed by $\mathcal{C}\circ\mathcal{T}$, equation \eqref{eqn:SU_uWall_BPS} becomes
\begin{equation*}
    u=2\sum_{a=0}^{N/2-1}q_a,
\end{equation*}
so we see that $u$ must be even, and when it is there are $\binom{N/2}{u/2}$ solutions.

\paragraph{}
Table \ref{tab:SU_uWall_characters} summarizes the results in a character table for $\SU{N}$ $u$-walls.
\begin{table}[h!]
    \centering
    {\tabulinesep=1.0mm
    \begin{tabu}{l|ccccc}

        & $[\identity]$ & $[\mathcal{T}^n]$ for $\frac{N}{\gcd(N,n)}\mid u$ & $[\mathcal{T}^n]$ for $\frac{N}{\gcd(N,n)}\nmid u$ & $[\mathcal{C}]$ & $[\mathcal{C}\circ\mathcal{T}]$ \\ \hline
        $N$ odd & \multirow{3}{*}{$\binom{N}{u}$} & \multirow{3}{*}{$\binom{\gcd(N,n)}{u\gcd(N,n)/N}$} & \multirow{3}{*}{$0$} & $\binom{\left\lfloor N/2\right\rfloor}{\left\lfloor u/2\right\rfloor}$ & - \\
        $N$ even, $u$ odd & & & & $2\binom{N/2-1}{(u-1)/2}$ & 0 \\
        $N$ even, $u$ even & & & & $\binom{N/2}{u/2}$ & $\binom{N/2}{u/2}$
    \end{tabu}}
    \caption{Character table for $\SU{N}$ $u$-walls. When $N$ is odd, $[\mathcal{C}\circ\mathcal{T}]$ is equal to $[\mathcal{C}]$, so the characters are not listed.}
    \label{tab:SU_uWall_characters}
\end{table}

\subsubsection[$\Sp{N}$]{$\boldsymbol{\Sp{N}}$}
For $\Sp{N}$, there is no charge conjugation symmetry, so we only have to consider the action of the $\Z_2^{(1),\s^1}$ center symmetry, $q_a\rightarrow q_{N-a}$. The dual Kac labels are all unity and the rank is $N$, so the counting of domain walls is the same as $\SU{N+1}$, $\chi_u^{\Sp{N}}(\identity)=\binom{N+1}{u}$. Requiring that domain walls are invariant under the center, equation \eqref{eqn:BPS_condition} becomes
\begin{equation*}
    u=\begin{dcases}2\sum_{a=0}^{N/2-1}q_a + q_{N/2} & N\text{ even} \\ 2\sum_{a=0}^{(N-1)/2}q_a & N\text{ odd}\end{dcases}.
\end{equation*}
For $N$ even, $q_{N/2}$ is determined so that $u-q_{N/2}$ is even, satisfying $u-q_{N/2}=2\left\lfloor u/2\right\rfloor$. We then have $2\left\lfloor u/2\right\rfloor=2\sum_{a=0}^{N/2-1}q_a$, which has $\binom{N/2}{\left\lfloor u/2\right\rfloor}$ solutions. For $N$ odd, we see that there will be no center-invariant domain walls if $u$ is odd, and for $u$ even there are $\binom{(N+1)/2}{u/2}$ solutions.

\paragraph{}
Table \ref{tab:Sp_uWall_characters} summarizes the results in a character table for $\Sp{N}$ $u$-walls.
\begin{table}[h!]
    \centering
    {\tabulinesep=1.0mm
    \begin{tabu}{l|cc}
        & $[\identity]$ & $[\mathcal{T}]$ \\ \hline
        $N$ and $u$ odd & \multirow{3}{*}{$\displaystyle\binom{N+1}{u}$} & 0 \\
        Otherwise & & $\displaystyle\binom{\left\lfloor (N+1)/2\right\rfloor}{\left\lfloor u/2\right\rfloor}$
    \end{tabu}}
    \caption{Character table for $\Sp{N}$ $u$-walls.}
    \label{tab:Sp_uWall_characters}
\end{table}

\subsubsection[$\Spin{2N}$]{$\boldsymbol{\Spin{2N}}$}
For both $N$ even and $N$ odd, except $N=4$ which is treated separately below, the global symmetry is the dihedral group of order eight. The specific structure of the group with respect to center and charge conjugation generators differs between the two cases however, since for $N$ even the center is $\Z_2\times\Z_2$ while for $N$ odd it is $\Z_4$. In both cases however, charge conjugation acts the same and the counting of domain walls is the same, so we can compute the characters of the identity and charge conjugation without having to specify $N$ even or odd. The BPS condition for $\Spin{2N}$ is given by
\begin{equation}
    u=q_0+q_1+q_-+q_++2\sum_{a=2}^{N-2}q_a.\label{eqn:EvenSpin_uWall_BPS}
\end{equation}
Counting the number of solutions is easier when we split into cases of $u$ even/odd. First for $u$ even, we could have $q_0+q_1+q_-+q_+=0,2,4$ which can be done one, six, and one way respectively. For each of these, we'll have $u-(q_0+q_1+q_-+q_+)=2\sum_{a=2}^{N-2}q_a$ which has $\binom{N-3}{(u-(q_0+q_1+q_-+q_+))/2}$ solutions. Putting the two pieces together, we find that the number of $u$-walls is
\begin{equation*}
    \binom{N-3}{u/2}+6\binom{N-3}{u/2-1}+\binom{N-3}{u/2-2}=\binom{N-1}{u/2}+4\binom{N-3}{u/2-1}.
\end{equation*}
For $u=2$, the third term on the left doesn't contribute, but the right hand side is still valid, similarly for $u=2(N-2)=c_2-2$ the left hand side isn't technically correct, but the right hand side is valid.

\paragraph{}
For $u$ odd, we could have $q_0+q_1+q_-+q_+=1,3$ both of which can be done four ways. Then, we have $u-(q_0+q_1+q_-+q_+)=2\sum_{a=2}^{N-2}q_a$ which has $\binom{N-3}{(u-(q_0+q_1+q_-+q_+))/2}$ solutions. Thus, the total number of $u$-walls is
\begin{equation*}
    4\binom{N-3}{(u-1)/2}+4\binom{N-3}{(u-3)/2}=4\binom{N-2}{(u-1)/2}.
\end{equation*}
For $u=1$ the second term on the left hand side is not valid, but the right hand side still holds, similarly when $u=2N-3=c_2-1$ the first term on the left hand side is not valid, but again the right hand side holds. Thus, we find for general $u$,
\begin{equation}
    \chi_u^{\Spin{2N}}(\identity)=\begin{dcases}
        \binom{N-1}{u/2}+4\binom{N-3}{u/2-1} & u\text{ even} \\
        4\binom{N-2}{(u-1)/2} & u\text{ odd}
    \end{dcases}.
\end{equation}

\paragraph{}
Charge conjugation acts by swapping $q_+$ and $q_-$, so a domain wall invariant under charge conjugation has $q_-=q_+$ which modifies equation \eqref{eqn:EvenSpin_uWall_BPS} to be
\begin{equation*}
    u=q_0+q_1+2\left(\sum_{a=2}^{N-2}q_a+q_-\right).
\end{equation*}
For $u$ even we necessarily have $q_0+q_1$ even, which implies $q_0=q_1$ giving us $u=2\left(\sum_{a=1}^{N-2}q_a+q_-\right)$ which has $\binom{N-1}{u/2}$ solutions. For $u$ odd, $q_0+q_1$ is odd so we have $q_0=1-q_1$ giving us $u-1=2\left(\sum_{a=2}^{N-2}q_a+q_-\right)$, which has $\binom{N-2}{(u-1)/2}$ solutions. Combining these we get
\begin{equation}
    \chi_u^{\Spin{2N}}(\mathcal{C})=\begin{dcases}
        \binom{N-1}{u/2} & u\text{ even}\\
        \binom{N-2}{(u-1)/2} & u\text{ odd}
    \end{dcases}.
\end{equation}

\paragraph{}
In general, the dihedral group of order $8$ is generated by a $\Z_2$ ``reflection", $s$, and a $\Z_4$ ``rotation", $r$, such that $(sr)^2=1$. The conjugacy classes are $[\identity]$, $[s]$, $[r]$, $[r^2]$, and $[sr]$. For both $N$ even and $N$ odd, we identify $\mathcal{C}$ with $s$. When $N$ is odd the center is $\Z_4^{(1),\s^1}$, so $\mathcal{T}$ naturally is identified with $r$, and acts by cycling $q_0\rightarrow q_-\rightarrow q_1\rightarrow q_+\rightarrow q_0$ and $q_a\rightarrow q_{N-a}$ for $2\leq a\leq N-2$. When $N$ is even the center is $\Z_2^+\times\Z_2^-$, where each $\Z_2^\pm$ is generated by $\mathcal{T}^\pm$ which acts by swapping $q_0\leftrightarrow q_\pm$, $q_1\leftrightarrow q_\mp$, and $q_a\leftrightarrow q_{N-a}$ for $a=2,\dots,N-2$. Clearly neither of $\mathcal{T}^\pm$ can be identified with $r$ since they have order two, but $\mathcal{C}\circ\mathcal{T}^\pm$ has order four and can be identified with $r$. Further, $\mathcal{C}\circ\mathcal{T}^+$ acts exactly the same way as $\mathcal{T}$ did when $N$ was odd. Thus the equivalence classes of the dihedral group are
\begin{align}
\begin{split}
    [\identity]&\\
    [s]&=[\mathcal{C}]\\
    [r]&=\begin{dcases}
        [\mathcal{T}] & N\text{ odd} \\ [\mathcal{C}\circ\mathcal{T}^+] & N\text{ even}
    \end{dcases}\\
    [r^2]&=\begin{dcases}
        [\mathcal{T}^2] & N\text{ odd} \\ [\mathcal{T}^+\circ\mathcal{T}^-] & N\text{ even}
    \end{dcases}\\
    [sr]&=\begin{dcases}
        [\mathcal{C}\circ\mathcal{T}] & N\text{ odd} \\ [T^+] & N\text{ even}
    \end{dcases},
\end{split}\label{eqn:Spin(2N)_conjugacy_classes}
\end{align}
where $r$ and $s$ act the same way on domain walls for $N$ even and odd. Now we just have to compute the characters of $r$, $r^2$, and $sr$. First, as described above, $r$ acts by cycling $q_0, q_-, q_1, q_+$ and swapping $q_a$ with $q_{N-a}$ for $a=2,\dots,N-2$, so that when a domain wall is invariant under $r$ equation \eqref{eqn:EvenSpin_uWall_BPS} becomes
\begin{equation*}
    u=\begin{dcases}
        4\sum_{a=1}^{(N-1)/2}q_a & N\text{ odd} \\
        4\sum_{a=1}^{N/2-1}q_a+2q_{N/2} & N\text{ even}
    \end{dcases}.
\end{equation*}
When $N$ is odd, the number of invariant domain walls is $\binom{(N-1)/2}{u/4}$ for $u$ a multiple of four and zero otherwise. When $N$ is even, domain walls can only be invariant when $u$ is even. Further, $q_{N/2}$ is determined by whether $u/2$ is even or odd, so that $u/2-q_{N/2}=2\left\lfloor u/4\right\rfloor$ is even. From the remaining $N/2-1$ degrees of freedom we can construct $\binom{N/2-1}{\left\lfloor u/4\right\rfloor}$ invariant domain walls. Thus we find
\begin{equation*}
    \chi_u^{\Spin{2N}}(r)=\begin{dcases}
        \binom{(N-1)/2}{u/4} & N\text{ odd, }u\equiv 0\bmod{4}\\
        \binom{N/2-1}{\left\lfloor u/4\right\rfloor} & N,u\text{ even}\\
        0 & \text{otherwise}
    \end{dcases}.
\end{equation*}

\paragraph{}
Next $r^2$ acts on domain walls by swapping $q_0\leftrightarrow q_1$ and $q_-\leftrightarrow q_+$, so invariant domain walls satisfy
\begin{equation*}
    u=2\sum_{a=1}^{N-2}q_a+2q_-.
\end{equation*}
We see that $u$ must be even, and that there are $\chi_u^{\Spin{2N}}(r^2)=\binom{N-1}{u/2}$ invariant domain walls. 

\paragraph{}
Finally, $sr$ acts by swapping $q_0\leftrightarrow q_+$, $q_1\leftrightarrow q_-$, and $q_a\leftrightarrow q_{N-a}$ for $a=2,\dots,N-2$. Domain walls invariant under $sr$ satisfy
\begin{equation*}
    u=\begin{dcases}
        2\left(q_0+q_1\right)+4\sum_{a=2}^{(N-1)/2}q_a & N\text{ odd} \\
        2\left(q_0+q_1+q_{N/2}\right)+4\sum_{a=2}^{N/2-1}q_a & N\text{ even}
    \end{dcases}.
\end{equation*}
Clearly $u$ must be even in order to have invariant domain walls. For $N$ and $u/2$ odd $q_0+q_1=1$ which can be done two ways, and $u/2-1=2\sum_{a=2}^{(N-1)/2}q_a$ which has $\binom{(N-3)/2}{(u/2-1)/2}$ solutions, giving us $2\binom{(N-3)/2}{(u/2-1)/2}$ invariant domain walls. For $N$ odd and $u/2$ even we must have $q_0=q_1$, giving us $\binom{(N-1)/2}{u/4}$ invariant domain walls. For $N$ even and $u/2$ odd, $q_0+q_1+q_{N/2}$ must be odd, with either $q_0+q_1+q_{N/2}=1$, which can be done three ways, or $q_0+q_1+q_{N/2}=3$, which can only be done one way. We then have $u/2-(q_0+q_1+q_{N/2})=2\sum_{a=2}^{N/2-1}q_a$, which has $\binom{N/2-2}{(u/2-(q_0+q_1+q_{N/2}))/2}$ solutions. For $N$ and $u/2$ even, $q_0+q_1+q_{N/2}$ must be even, with either $q_0+q_1+q_{N/2}=0$, which can be done one way, or $q_0+q_1+q_{N/2}=2$, which can be done three ways. We then have $u/2-(q_0+q_1+q_{N/2})=2\sum_{a=2}^{N/2-1}q_a$ which has $\binom{N/2-2}{(u/2-(q_0+q_1+q_{N/2}))/2}$ solutions. Then, the character of $sr$ is given by
\begin{equation*}
    \chi_u^{\Spin{2N}}(sr)=\begin{dcases}
        \binom{(N-1)/2}{u/4} & N\text{ odd, }u\equiv 0\bmod{4}\\
        2\binom{(N-3)/2}{(u-2)/4} & N\text{ odd, }u\equiv 2\bmod{4}\\
        \binom{N/2-2}{u/4}+3\binom{N/2-2}{(u-4)/4} & N\text{ even, }u\equiv0\bmod{4}\\
        3\binom{N/2-2}{(u-2)/4}+\binom{N/2-2}{(u-6)/2} & N\text{ even, }u\equiv2\bmod{4}\\
        0&\text{otherwise}
    \end{dcases}
\end{equation*}

\paragraph{}
Table \ref{tab:EvenSpin_uWall_characters} summarizes the results in a character table for $\Spin{2N}$ $u$-walls.
\begin{table}[h!]
    \centering
    
    {\tabulinesep=1.0mm
    \begin{tabu}{l|ccccc}

        & $[\identity]$ & $[s]$ & $[r]$ & $[r^2]$ & $[sr]$ \\ \hline
        $N$ odd, $u\equiv 0\bmod{4}$ & \multirow{4}{*}{$\binom{N-1}{u/2}+4\binom{N-3}{u/2-1}$} & \multirow{4}{*}{$\binom{N-1}{u/2}$} & $\binom{(N-1)/2}{u/4}$ & \multirow{4}{*}{$\binom{N-1}{u/2}$} & $\binom{(N-1)/2}{u/4}$ \\
        $N$ odd, $u\equiv 2\bmod{4}$ & & & 0 & & $2\binom{(N-3)/2}{\left\lfloor u/4\right\rfloor}$\\
        $N$ even, $u\equiv 0\bmod{4}$ & & & \multirow{2}{*}{$\binom{N/2-1}{\left\lfloor u/4\right\rfloor}$} & & $\binom{N/2-2}{u/4}+3\binom{N/2-2}{(u-4)/4}$ \\
        $N$ even, $u\equiv 2\bmod{4}$ & & & & & $3\binom{N/2-2}{\left\lfloor u/4\right\rfloor}+\binom{N/2-2}{\left\lfloor (u-4)/4\right\rfloor}$\\
        $u$ odd & $4\binom{N-2}{(u-1)/2}$ & $\binom{N-2}{(u-1)/2}$ & $0$ & $0$ & $0$
    \end{tabu}}
    \caption{Character table for $\Spin{2N}$ $u$-walls for $N>4$, where the relations between $r$ and $s$ which generate the global $D_8$ symmetry and $\mathcal{C}$ and $\mathcal{T}$ which generate charge conjugation and the ``0-form" center symmetry are given by \eqref{eqn:Spin(2N)_conjugacy_classes}.}
    \label{tab:EvenSpin_uWall_characters}
\end{table}

\paragraph{}
When $N=4$ the charge conjugation symmetry becomes $S_3$, the group of permutations of $\{q_1,q_-,q_+\}$. Along with the $\Z_2^+\times\Z_2^-$ center symmetry, the total symmetry group is $S_4$, the group of permutations of $\{q_0,q_1,q_-,q_+\}$. The conjugacy classes of $S_4$ are $[\identity]$, $[(q_0q_1)(q_-q_+)]$, $[(q_0q_1q_-)]$, $[(q_0q_1)]$, and $[(q_0q_1q_-q_+)]$\footnote{We use a standard notation for permutation group elements, where $(x_1x_2\dots x_n)$ is the element which maps $x_1\rightarrow x_2$, $x_2\rightarrow x_3$, and so on until $x_n\rightarrow x_1$. These cycles may be multiplied together and are read as acting right to left, so $(ab)(ac)$ maps $c\rightarrow a\rightarrow b$ and is equal to $(cba)$.}. The character of the identity is unchanged from the larger $N$ cases. The rest of the characters may be worked out by hand, to produce the character table \ref{tab:EvenSpin4_uWall_characters}.
\begin{table}[h!]
    \centering
    
    {\tabulinesep=1.0mm
    \begin{tabu}{l|ccccc}
        & $[\identity]$ & $[(q_0q_1)(q_-q_+)]$ & $[(q_0q_1q_-)]$ & $[(q_0q_1)]$ & $[(q_0q_1q_-q_+)]$ \\ \hline
        $u=1,5$ & 4 & 0 & 1 & 2 & 0 \\
        $u=2,4$ & 7 & 3 & 1 & 3 & 1 \\
        $u=3$ & 8 & 0 & 2 & 4 & 0
    \end{tabu}}
    \caption{Character table for $\Spin{8}$ $u$-walls.}
    \label{tab:EvenSpin4_uWall_characters}
\end{table}

\subsubsection[$\Spin{2N+1}$]{$\boldsymbol{\Spin{2N+1}}$}
The center symmetry is $\Z_2$ and there is no charge conjugation symmetry, so we only have to compute characters of the identity and the $\Z_2$ generator. The BPS condition for $\Spin{2N+1}$ is
\begin{equation*}
    u=q_0+q_1+q_N+2\sum_{a=2}^{N-1}q_a.
\end{equation*}
Counting the number of domain walls, when $u$ is even we can either have $q_0=q_1=q_N=0$ and $u=2\sum_{a=2}^{N-1}q_a$ which has $\binom{N-2}{u/2}$ solutions, or have $q_0+q_1+q_N=2$, which can be done three ways, and $u=2+2\sum_{a=2}^{N-1}q_a$ which has $\binom{N-2}{u/2-1}$ solutions. When $u$ is odd we can either have $q_0+q_1+q_N=1$, which can be done three ways, and $u=1+2\sum_{a=2}^{N-1}q_a$, which has $\binom{N-2}{(u-1)/2}$ solutions, or we could have $q_0=q_1=q_N=1$ and $u=3+2\sum_{a=2}^{N-1}q_a$ which has $\binom{N-2}{(u-3)/2}$ solutions. Combining both $u$ even and odd and simplifying we find
\begin{equation*}
    \chi_u^{\Spin{2N+1}}(\identity)=\binom{N-1}{\left\lfloor u/2\right\rfloor} + 2\binom{N-2}{\left\lfloor (u-1)/2\right\rfloor}.
\end{equation*}

\paragraph{}
The $\Z_2^{(1),\s^1}$ center symmetry generator of $\Spin{2N+1}$ acts by exchanging $q_0$ and $q_1$, thus domain walls invariant under $\Z_2^{(1),\s^1}$ have $q_0=q_1$ giving us
\begin{equation*}
    u=2\sum_{a=1}^{N-1}q_a+q_N.
\end{equation*}
Notice that $q_N$ is completely determined by $u$ with $q_N=2\left(\frac{u}{2}-\left\lfloor\frac{u}{2}\right\rfloor\right)$ so that we have $2\left\lfloor\frac{u}{2}\right\rfloor=2\sum_{a=1}^{N-1}q_a$ which has $\binom{N-1}{\left\lfloor u/2\right\rfloor}$ solutions,
\begin{equation*}
    \chi_u^{\Spin{2N+1}}(\mathcal{T})=\binom{N-1}{\left\lfloor u/2\right\rfloor}.
\end{equation*}

\paragraph{}
Table \ref{tab:OddSpin_uWall_characters} summarizes the results in a character table for $\Spin{2N+1}$ $u$-walls.

\begin{table}[h!]
    \centering
    {\tabulinesep=1.0mm
    \begin{tabu}{l|cc}
        & $[\identity]$ & $[\mathcal{T}]$ \\ \hline
        All $u$ & $\displaystyle\binom{N-1}{\left\lfloor u/2\right\rfloor} + 2\binom{N-2}{\left\lfloor (u-1)/2\right\rfloor}$ & $\displaystyle\binom{N-1}{\left\lfloor u/2\right\rfloor}$
    \end{tabu}}
    \caption{Character table for $\Spin{2N+1}$ $u$-walls.}
    \label{tab:OddSpin_uWall_characters}
\end{table}

\subsubsection[$\E{6}$]{$\boldsymbol{\E{6}}$}
The center symmetry for $\E{6}$ is $\Z_3^{(1)}$ and the charge conjugation symmetry is $\Z_2^{(0)}$, which together form a dihedral algebra making the global symmetry $D_6$ with conjugacy classes $[\identity]$, $[\mathcal{T}]$, and $[\mathcal{C}]$. The BPS condition for $\E{6}$ is
\begin{equation*}
    u=q_0+q_1+q_5+2(q_2+q_4+q_6)+3q_3.
\end{equation*}
We can directly compute the number of solutions by brute force, listing the results here
\begin{equation*}
    \chi_u^{\E{6}}(\identity)=\begin{dcases}
        3 & u=1,11 \\ 6 & u=2,10 \\ 11 & u=3,9 \\ 15 & u=4,8 \\ 18 & u=5,7 \\ 20 & u=6
    \end{dcases}.
\end{equation*}

\paragraph{}
The $\Z_3^{(1),\s^1}$ center symmetry generator of $\E{6}$ acts by cyclically permuting $(q_0,q_1,q_5)$ and $(q_2,q_4,q_6)$, thus a domain wall invariant under $\Z_3^{(1),\s^1}$ has $q_0=q_1=q_5$ and $q_2=q_4=q_6$ giving us
\begin{equation*}
    u=3\left(q_1+2q_2+q_3\right),
\end{equation*}
so $u$ must be a multiple of three, meaning that only $u=3,6,9$ can have domain walls fixed by $\Z_3^{(1),\s^1}$. In each case there are two walls fixed, for example in $u=3$ we could have $(q_1,q_2,q_3)=(1,0,0)$ or $(q_1,q_2,q_3)=(0,0,1)$,
\begin{equation*}
    \chi_u^{\E{6}}(\mathcal{T})=\begin{dcases}
        2 & u\equiv 0\bmod{3} \\ 0 & \text{otherwise}
    \end{dcases}.
\end{equation*}

\paragraph{}
The $\Z_2^{(0)}$ charge conjugation symmetry in $\E{6}$ acts by swapping $q_1\leftrightarrow q_5$ and $q_2\leftrightarrow q_4$, so domain walls invariant under charge conjugation have $q_1=q_5$ and $q_2=q_4$ giving us
\begin{equation*}
    u=q_0+2q_1+4q_2+3q_3+2q_6,
\end{equation*}
which we may solve by brute force to obtain
\begin{equation*}
    \chi_u^{\E{6}}(\mathcal{C})=\begin{dcases}
        1 & u=1,11 \\ 2 & u=2,10 \\ 3 & u=3,9 \\ 3 & u=4,8 \\ 4 & u=5,7 \\ 4 & u=6
    \end{dcases}.
\end{equation*}

\paragraph{}
Table \ref{tab:E6_uWall_characters} summarizes the results in a character table for $\E{6}$ $u$-walls.
\begin{table}[h!]
    \centering
    {\tabulinesep=1.0mm
    \begin{tabu}{l|ccc}
        & $[\identity]$ & $[\mathcal{T}]$ & $[\mathcal{C}]$ \\ \hline
        $u=1,11$ & $3$ & $0$ & $1$ \\
        $u=2,10$ & $6$ & $0$ & $2$ \\
        $u=3,9$ & $11$ & $2$ & $3$ \\
        $u=4,8$ & $15$ & $0$ & $3$ \\
        $u=5,7$ & $18$ & $0$ & $4$ \\
        $u=6$ & $20$ & $2$ & $4$
    \end{tabu}}
    \caption{Character table for $\E{6}$ $u$-walls.}
    \label{tab:E6_uWall_characters}
\end{table}

\subsubsection[$\E{7}$]{$\boldsymbol{\E{7}}$}
The center symmetry for $\E{7}$ is $\Z_2^{(1)}$ and there is no charge conjugation symmetry, so the global symmetry is $\Z_2$ and we just have to compute the characters of $\identity$ and $\mathcal{T}$. The BPS condition for $\E{7}$ $u$-walls is
\begin{equation*}
    u=q_0+q_6+2(q_1+q_5+q_7)+3(q_2+q_4)+4q_3.
\end{equation*}
Computing the number of solutions by brute force we obtain
\begin{equation}
    \chi_u^{\E{7}}(\identity)=\begin{dcases}
        2 & u=1,17 \\ 4 & u=2,16 \\ 8 & u=3,15 \\ 11 & u=4,14 \\ 16 & u=5,13 \\ 21 & u=6,12 \\ 24 & u=7,11 \\ 27 & u=8,10 \\ 28 & u=9
    \end{dcases}.
\end{equation}

\paragraph{}
The $\Z_2^{(1),\s^1}$ center symmetry of $\E{7}$ acts by swapping $q_0\leftrightarrow q_6$, $q_1\leftrightarrow q_5$, $q_2\leftrightarrow q_4$, so a domain wall fixed by $\Z_2^{(1),\s^1}$ has
\begin{equation*}
    u=2q_0+4q_1+6q_2+4q_3+2q_7=2(q_0+2q_1+3q_2+2q_3+q_7),
\end{equation*}
so $u$ must be even. Solving by brute force we obtain
\begin{equation}
    \chi_u^{\E{7}}(\mathcal{T})=\begin{dcases}
        2 & u=2,16 \\ 3 & u=4,14 \\ 5 & u=6,12 \\ 5 & u=8,10 \\ 0 & u\text{ odd}
    \end{dcases}.
\end{equation}

\paragraph{}
Table \ref{tab:E7_uWall_characters} summarizes the results in a character table for $\E{7}$ $u$-walls.

\begin{table}[h!]
    \centering
    {\tabulinesep=1.0mm
    \begin{tabu}{l|cc}
        & $[\identity]$ & $[\mathcal{T}]$ \\ \hline
        $u=1,17$ & 2 & 0 \\
        $u=2,16$ & 4 & 2 \\
        $u=3,15$ & 8 & 0 \\
        $u=4,14$ & 11 & 3 \\
        $u=5,13$ & 16 & 0 \\
        $u=6,12$ & 21 & 5 \\
        $u=7,11$ & 24 & 0 \\
        $u=8,10$ & 27 & 5 \\
        $u=9$ & 28 & 0
    \end{tabu}}
    \caption{Character table for $\E{7}$ $u$-walls.}
    \label{tab:E7_uWall_characters}
\end{table}

\subsubsection[$\E{8}$]{$\boldsymbol{\E{8}}$}
Both charge conjugation and center symmetry are trivial for $\E{8}$, so we just have to compute the character of the identity. The BPS condition for $\E{8}$ $u$-walls is
\begin{equation}
    u=q_0+2(q_1+q_7)+3(q_2+q_8)+4(q_3+q_6)+5q_4+6q_5.
\end{equation}
Solving by brute force the number of domain walls is
\begin{equation}
    \chi_u^{\E{8}}(\identity)=\begin{dcases}
        1 & u=1,29 \\ 2 & u=2,28 \\ 4 & u=3,27 \\ 5 & u=4,26 \\ 8 & u=5,25 \\ 11 & u=6,24 \\ 14 & u=7,23 \\ 17 & u=8,22 \\ 22 & u=9,21 \\ 25 & u=10,20 \\ 28 & u=11,19 \\ 32 & u=12,18 \\ 33 & u=13,17 \\ 35 & u=14,16 \\ 36 & u=15
    \end{dcases}.
\end{equation}

\subsubsection[$\F$]{$\boldsymbol{\F}$}
Both charge conjugation and center symmetry are trivial for $\F$, so we just have to compute the character of the identity. The BPS condition for $\F$ $u$-walls is
\begin{equation}
    u=q_0+q_4+2(q_1+q_3)+3q_2.
\end{equation}
Solving by brute force the number of domain walls is
\begin{equation}
    \chi_u^{\F}(\identity)=\begin{dcases}
        2 & u=1,8 \\ 3 & u=2,7 \\ 5 & u=3,6 \\ 5 & u=4,5
    \end{dcases}.
\end{equation}

\subsubsection[$\G$]{$\boldsymbol{\G}$}
Both charge conjugation and center symmetry are trivial for $\G$, so we just have to compute the character of the identity. The BPS condition for $\G$ $u$-walls is
\begin{equation}
    u=q_0+2q_1+q_2
\end{equation}
It is not too hard to see that for each $u=1,2,3$ there are two solutions,
\begin{equation}
    \chi_u^{\G}(\identity)=2.
\end{equation}

%% file: worldvolume.tex
For each $u$ there should be a worldvolume TQFT describing the BPS $u$-walls. The worldvolume TQFTs will be Chern-Simons theories defined on $\T^2\times\R$. Note that we interpret $\T^2$ as $\s^1_L\times \s^1_{L'}$ where we take $L'\rightarrow \infty$ so that the TQFT spacetime corresponds to the domain wall worldvolume of SYM on $\RS$. The states of the TQFT Hilbert space correspond to the $u$-walls of SYM, and are required to transform under center and charge conjugation symmetries in the same way as the domain walls they are associated with. 

\paragraph{}
In principle there is then a map between the $u$-walls of SYM and the Hilbert space of the corresponding worldvolume TQFT. Unfortunately, while such a map may exist it is usually not easy to find, and in general is not unique. To illustrate this point, consider $1$-walls of $\SU{N}$ which may be labelled by an integer $a=0,1,\dots,N-1$ corresponding to flux $\vec{\Phi}_a=\vec{w}_a-\frac{1}{N}\vec{\rho}$\footnote{Remember that we are using the convention $\vec{w}_0=0$.}. The $\Z_N^{(1),\s^1}$ center symmetry acts as $\mathcal{T}(\vec{\Phi}_a)=\vec{\Phi}_{a-1\bmod{N}}$, while the $\Z_2^{(0)}$ charge conjugation symmetry acts as $\mathcal{C}(\vec{\Phi}_a)=\vec{\Phi}_{N-a\bmod{N}}$. The corresponding worldvolume TQFT is $U(1)_N$, whose Hilbert space on $\T^2\times \R$ is spanned by states $\ket{a'}$ for $a'=0,1,\dots,N-1$. Under the $\Z_N^{(1)}$ symmetry in the first direction\footnote{The direction here is somewhat of an arbitrary choice.} the $U(1)_N$ states transform as $\operator{T}\ket{a'}=\ket{a'-1\bmod{N}}$, and under charge conjugation they transform as $\operator{C}\ket{a'}=\ket{N-a'\bmod{N}}$. It would then seem that the most natural choice for a map between the set of domain walls, $\left\{\vec{\Phi}_a\mid a=0,1,\dots,N-1\right\}$, to the set of $U(1)_N$ states, $\left\{\ket{a'}\mid a'=0,1,\dots,N-1\right\}$, would be the map taking $\vec{\Phi}_a$ to $\ket{a}$. Indeed such a map does preserve the symmetries of the theories. However, when $N$ is even we could also map $\vec{\Phi}_a$ to $\ket{\frac{N}{2}+a\bmod{N}}$:
\begin{align*}
    f&: \vec{\Phi}_a\mapsto\ket{\frac{N}{2}+a\bmod{N}}\\
    & f(\mathcal{T}(\vec{\Phi})_a)=f(\vec{\Phi}_{a-1})=\ket{\frac{N}{2}+a-1\bmod{N}}=\operator{T}\ket{\frac{N}{2}+a\bmod{N}}=\operator{T}f(\vec{\Phi}_a)\\
    & f(\mathcal{C}(\vec{\Phi})_a)=f(\vec{\Phi}_{N-a\bmod{N}})=\ket{\frac{N}{2}+N-a\bmod{N}}=\operator{C}\ket{\frac{N}{2}+a\bmod{N}}=\operator{C}f(\vec{\Phi}_a).
\end{align*}
We see that when $N$ is even there are two ways to map the $1$-walls of $\SU{N}$ SYM to the states of the worldvolume TQFT Hilbert space. In general there will not be a single unique way to map the domain walls of SYM to the TQFT Hilbert space in a way that preserves the symmetries. Instead, we must be content with simply computing and comparing the characters of the global symmetries between the domain walls of SYM and the TQFT Hilbert spaces. Here we will denote the character of a $G_k$ Chern-Simons theory by $\chi_{G_k}$.

\subsection{A quick review of Chern-Simons theory}
To begin, we will briefly review the construction of the Hilbert space of Chern-Simons theory on $\T^2\times \R$ and the action of the various symmetry operators for both a $U(1)$ gauge group and a simple non-Abelian gauge group.

\subsubsection*{Abelian Chern-Simons}
For $U(1)_k$ Chern-Simons theory the non-trivial operators are the Wilson lines in the two directions of $\T^2$ which commute up to a $\Z_{\abs{k}}$ phase,
\begin{equation*}
    \operator{W}_1\operator{W}_2=e^{2\pi i/k}\operator{W}_2\operator{W}_1.
\end{equation*}
The Hilbert space is then the $k$-dimensional space spanned by states $\ket{a}$ for $0\leq a\leq \abs{k}-1$ defined by
\begin{align}
    \operator{W}_1\ket{a}&=\ket{a-1\bmod{\abs{k}}}\label{eqn:U(1)_CS_1-form_1}\\
    \operator{W}_2\ket{a}&=e^{2\pi ia/k}\ket{a}\label{eqn:U(1)_CS_1-form_2}.
\end{align}
Notice that $\operator{W}_1$ and $\operator{W}_2$ each generate $\Z_{\abs{k}}$ symmetries, together forming a one-form $\Z_{\abs{k}}$ symmetry. Charge conjugation acts as in appendix \ref{appendix:charge_conjugation}, which in the Hilbert space is
\begin{equation*}
    \operator{C}\ket{a}=\ket{k-a}.
\end{equation*}

\subsubsection*{Non-Abelian Chern-Simons}
For $G_k$ Chern-Simons theory, where $G$ is a simple non-Abelian group, quantization is understood through the machinery of geometric quantization. Following Elitzur et al \cite{elitzur_remarks_1989}, the wave-functionals are given by Weyl-Kac characters at level $k$, which are in turn labelled by the so-called integrable representations at level $k$. The integrable representations are labelled by their highest weights which in addition to being dominant, as any highest weight must be, must satisfy $k+\vec{\alpha}_0^*\cdot\vec{\lambda}\geq 0$. In other words, the Hilbert space of $G_k$ Chern-Simons theory is spanned by
\begin{equation*}
    \left\{\ket{\vec{\lambda}}\mid \vec{\lambda}\in\Lambda_w,\  \vec{\alpha}_a^*\cdot\vec{\lambda}\geq -k\delta_{a,0},\ a=0,\dots,r\right\}.
\end{equation*}
Note that like domain walls, non-Abelian Chern-Simons states may be labelled by $r+1$ integers $(\lambda_0,\lambda_1,\dots,\lambda_r)$ with $\lambda_a=\vec{\alpha}_a^*\cdot\left(\vec{\lambda}-\frac{k}{c_2}\vec{\rho}\right)+\frac{k}{c_2}$ with $\sum_{a=0}^rk_a^*\lambda_a=k$, which must satisfy $\lambda_a\geq 0$\footnote{Unlike domain walls, which must have $q_a\in\{0,1\}$, the labels $\lambda_a$ of non-Abelian Chern-Simons states may be any non-negative integer so long as they satisfy the condition $\sum_{a=0}^rk_a^*\lambda_a=k$.}. The dimension of the Hilbert space is a function of the rank and the dual Kac labels of the gauge group. For $\SU{N+1}_k$ and $\Sp{N}_k$, the rank is $N$ and the dual Kac labels are all unity, so the dimension of the two Hilbert spaces, equal to the character of the identity, is the number of solutions $(\lambda_0,\lambda_1,\dots,\lambda_N)$ to $\sum_{a=0}^N\lambda_a=k$, which is solved using basic combinatorics to find
\begin{equation}
    \chi_{\SU{N+1}_k}(\identity)=\chi_{\Sp{N}_k}(\identity)=\binom{k+N}{k}\label{eqn:CS_dim_SU_Sp}
\end{equation}
The one-form center symmetry acts on the gauge fields by improper gauge transformations, which descend to the Hilbert space as phases and shifts,
\begin{align}
    \operator{T}_1\ket{\vec{\lambda}}&=\ket{\mathcal{T}_{c,k}(\vec{\lambda})}\label{eqn:NonAbelian_CS_1-form_1}\\
    \operator{T}_2\ket{\vec{\lambda}}&=e^{-2\pi i\vec{w}_c^*\cdot\vec{\lambda}}\ket{\vec{\lambda}}\label{eqn:NonAbelian_CS_1-form_2},
\end{align}
where $\mathcal{T}_{c,k}$ is the affine Weyl transformation defined in equation \eqref{eqn:center_Weyl_action} for an appropriate choice of $c$. Following a na\"{i}ve treatment\footnote{Charge conjugation may be complicated by spin structures. See the discussion in \cite{delmastro_domain_2021}, but for our purposes the na\"{i}ve picture is sufficient.}, charge conjugation acts on the gauge fields in the same was as described in \ref{appendix:charge_conjugation}. Acting on the Weyl-Kac characters, charge conjugation then acts as
\begin{equation}
    \operator{C}\ket{\vec{\lambda}}=\ket{\mathcal{C}(\vec{\lambda})},\label{eqn:NonAbelian_CS_CC}
\end{equation}
with the notable exception of $\Spin{4n}$, though that will not be relevant for this paper. See appendix \ref{appendix:charge_conjugation} for more details.

\subsection[$\SU{N}$]{$\boldsymbol{\SU{N}}$\label{sec:SU(N)_TQFT}}
The proposed worldvolume TQFT for $u$-walls in $\SU{N}$ is $U(u)_{N-u,N}$ given by
\begin{equation}
    U(u)_{N-u,N}=\frac{\SU{u}_{N-u}\times U(1)_{uN}}{\Z_u^{(1)}},
\end{equation}
where $\Z_u^{(1)}$ is the diagonal part of the $\Z_u^{(1)}$ one-form center of $\SU{u}_{N-u}$, and the $\Z_u^{(1)}$ subgroup of the $\Z_{uN}^{(1)}$ one-form symmetry of $U(1)_{uN}$. To construct the Hilbert space, we first construct the Hilbert space of $\SU{u}_{N-u}\times U(1)_{uN}$, which is simply the tensor product of the $\SU{u}_{N-u}$ and $U(1)_{uN}$ Hilbert spaces. We then gauge the diagonal $\Z_u^{(1)}$ by keeping only those states which are invariant under all possible insertions of $\Z_u^{(1)}$. We can accomplish this by starting with an arbitrary state $\ket{\vec{\lambda}}\otimes\ket{a}$ and acting with the projectors onto the $\Z_u^{(1)}$ invariant subspaces corresponding to the two directions along which we can insert $\Z_u^{(1)}$:
\begin{equation*}
    \operator{P}_i=\frac{1}{u}\sum_{m=0}^{u-1}\left(\operator{T}_i\otimes\operator{W}_i^N\right)^m,
\end{equation*}
where the action of $\operator{T}_i$ is given in equations \eqref{eqn:NonAbelian_CS_1-form_1} and \eqref{eqn:NonAbelian_CS_1-form_2} with $c=N-1$\footnote{Note that from here on we will drop the $c$ and $k$ subscripts on $\mathcal{T}$ in equation \eqref{eqn:NonAbelian_CS_1-form_2}.}, and the action of $\operator{W}_i$ is given in equations \eqref{eqn:U(1)_CS_1-form_1} and \eqref{eqn:U(1)_CS_1-form_2}. Note that $\operator{P}_1$ and $\operator{P}_2$ commute, otherwise taking the $\Z_u^{(1)}$ quotient would be ill-defined. Acting first with $\operator{P}_2$ on our test state $\ket{\vec{\lambda}}\otimes\ket{a}$ we see that only $a$ of the form $u(\xi+\vec{w}_{u-1}^*\cdot\vec{\lambda})$ for an integer $\xi$ survives,
\begin{equation*}
    \operator{P}_2\ket{\vec{\lambda}}\otimes\ket{a}=\frac{1}{u}\sum_{m=0}^{u-1}e^{2\pi im(a-u\vec{w}_{u-1}^*\cdot\vec{\lambda})/u}\ket{\vec{\lambda}}\otimes\ket{a}=\sum_{\xi\in\Z}\delta_{a,u(\xi+\vec{w}_{u-1}^*\cdot\vec{\lambda})}\ket{\vec{\lambda}}\otimes\ket{a}.
\end{equation*}
Acting with $\operator{P}_1$ imposes no further restrictions on $\vec{\lambda}$ or $a$, but simply averages over the $\Z_u^{(1)}$ orbit of $\ket{\vec{\lambda}}\otimes \ket{a}$ in the first direction. Thus we define the $U(u)_{N-u,N}$ state $\ket{[\vec{\lambda},\xi]}$ as the $\Z_u^{(1)}$ averaged version of $\ket{\vec{\lambda}}\otimes\ket{u(\xi+\vec{w}_{u-1}^*\cdot\vec{\lambda})}$,
\begin{align*}
    \ket{[\vec{\lambda},\xi]}\equiv \operator{P}_1\ket{\vec{\lambda}}\otimes\ket{u(\xi+\vec{w}_{u-1}^*\cdot\vec{\lambda})}&=\frac{1}{u}\sum_{m=0}^{u-1}\ket{\mathcal{T}^m(\vec{\lambda})}\otimes\ket{u(\xi+\vec{w}_{u-1}^*\cdot\vec{\lambda})-mN}\\
    &=\frac{1}{u}\sum_{m=0}^{u-1}\ket{\mathcal{T}^m(\vec{\lambda})}\otimes\ket{u\left(\xi-\sum_{a=0}^{m-1}\lambda_a-m+\vec{w}_{u-1}^*\cdot\mathcal{T}^m(\vec{\lambda})\right)},
\end{align*}
where we used the fact that $\vec{w}_{u-1}^*\cdot\mathcal{T}^m(\vec{\lambda})=\vec{w}_{u-1}^*\cdot\vec{\lambda}+\sum_{a=0}^{m-1}\lambda_a-m\frac{N-u}{u}$. The notation $[\vec{\lambda},\xi]$ is a shorthand used to indicate the $\Z_u^{(1)}$ orbit of the pair $(\vec{\lambda},\xi)$ in the first direction,
\begin{equation}
    [\vec{\lambda},\xi]=\left\{\left(\vec{\lambda},\xi\right),\ \left(\mathcal{T}(\vec{\lambda}),\xi-\lambda_0-1\right),\dots,\left(\mathcal{T}^{u-1}(\vec{\lambda}),\xi-\sum_{a=0}^{u-2}\lambda_a-(u-1)\right)\right\}.\label{eqn:U(u)_cs_label}
\end{equation}
Thus, each $U(u)_{N-u,N}$ state is labelled by a $\Z_u^{(1)}$ orbit and not by an individual representation.

\paragraph{}
The $\Z_N^{(1)}$ 1-form symmetry is the remainder of the $\Z_{uN}^{(1)}$ 1-form symmetry of $U(1)_{uN}$ after gauging $\Z_u^{(1)}$, and thus acts on the Hilbert space by $\operator{\mathsf{T}}_i=\identity\otimes\operator{W}_i^u$
\begin{align}
    \operator{\mathsf{T}}_1\ket{[\vec{\lambda},\xi]}&=\ket{[\vec{\lambda},\xi-1]}\label{eqn:U(u)_Center_1}\\
    \operator{\mathsf{T}}_2\ket{[\vec{\lambda},\xi]}&=e^{2\pi iu(\xi+\vec{w}_{u-1}^*\cdot\vec{\lambda})/N}\ket{[\vec{\lambda},\xi]}.\label{eqn:U(u)_Center_2}
\end{align}
Charge conjugation is unaffected by gauging $\Z_u^{(1)}$ and acts by
\begin{equation}
    \operator{C}\ket{[\vec{\lambda},\xi]}=\ket{[\mathcal{C}(\vec{\lambda}),-\xi-(N-u-\lambda_0)]}.\label{eqn:cc_unitary_cs}
\end{equation}
It is not too hard to verify using the data in appendix \ref{sec:all_groups_symmetries} that charge conjugation and the 1-form symmetry form a $D_{2N}$ dihedral algebra, which of course is required for the TQFT to match the $u$-walls of $\SU{N}$.

\paragraph{}
Let us for a moment comment on the sizes of the $\Z_u$ orbits of $\SU{u}_{N-u}$ representations. Suppose that $\vec{\lambda}$ has an orbit under $\mathcal{T}$ of size $l$\footnote{It is helpful to remember that $l$ must divide $u$ by Lagrange's theorem.}, so that $\mathcal{T}^l(\vec{\lambda})=\vec{\lambda}$, or in other words $\lambda_a=\lambda_{a+l}$. Then, knowing that $\vec{\lambda}$ is an integrable representation of $\SU{u}$ at level $N-u$ we have
\begin{equation}
    N-u=\sum_{a=0}^{u-1}\lambda_a=\frac{u}{l}\sum_{a=0}^{l-1}\lambda_a.\label{eqn:SU(u)_integrable_reps}
\end{equation}
Thus, we see that $u/l$ must divide $N-u$ and hence must divide $N$ as well. Since $u/l$ divides both $u$ and $N-u$, it must be a divisor of their $\gcd$, which allows us to write explicitly the allowed values of $l$ if necessary.

\paragraph{}
To determine the dimension of the Hilbert space we see that we must count each unique $\Z_u$ orbit $[\vec{\lambda},\xi]$. We start by considering an $\SU{N}$ integrable representation with highest weight $\vec{\lambda}$, and ask how many different orbits $[\vec{\lambda},\xi]$ we can construct. Note that since $U(1)_{uN}$ states are defined by integers $\bmod{uN}$, $\xi$ is equivalent to $\xi+N$, so there can be at most $N$ orbits $[\vec{\lambda},\xi]$. Now suppose that $\vec{\lambda}$ has a $\Z_u$ orbit under $\mathcal{T}$ of size $l$, so that $\operator{T}_1^l\ket{\vec{\lambda}}=\ket{\vec{\lambda}}$ and hence $\operator{T}_1^l\otimes\identity\ket{[\vec{\lambda},\xi]}=\ket{[\vec{\lambda},\xi]}$. We can use the fact that $\operator{T}_1\otimes\operator{W}_1^N$ acts trivially on the $U(N)_{N-u,N}$ Hilbert space to express $\operator{T}_1^l\otimes\identity$ as $\left(\operator{T}_1^l\otimes\identity\right)\left(\operator{T}_1\otimes \operator{W}_1^N\right)^{-l}=\identity\otimes \operator{W}_1^{-lN}$, giving us the following
\begin{equation*}
    \ket{[\vec{\lambda},\xi]}=\operator{T}_1^l\otimes\identity\ket{[\vec{\lambda},\xi]}=\identity\otimes\operator{W}_1^{-lN}\ket{[\vec{\lambda},\xi]}=\ket{\left[\vec{\lambda},\xi+\frac{N}{u/l}\right]},
\end{equation*}
where we used the fact that $\identity\otimes\operator{W}_1^{m}\ket{[\vec{\lambda},\xi]}=\ket{\left[\vec{\lambda},\xi-\frac{m}{u}\right]}$. We see that when $\vec{\lambda}$ has a $\Z_u$ orbit of size $l$, each orbit $[\vec{\lambda},\xi]$ is equal to $\left[\vec{\lambda},\xi+\frac{N}{u/l}\right]$, and hence there are $\frac{N}{u/l}$ $U(u)_{N-u,N}$ states for every orbit of size $l$ of $\SU{N}$ integrable representations under $\mathcal{T}$. If we let $\mathcal{O}^{\SU{N}}_{N-u}(l)$ be the number of $\SU{N}_{N-u}$ states with orbit size $l$, then the total number of $U(u)_{N-u,N}$ states $\ket{[\vec{\lambda},\xi]}$ is given by
\begin{equation}
    \chi_{U(u)_{N-u,N}}(\identity)=\sum_{l}\left(\frac{N}{u/l}\right)\left(\frac{1}{l}\mathcal{O}^{\SU{N}}_{N-u}(l)\right)=\frac{N}{u}\chi_{\SU{u}_{N-u}}(\identity)=\binom{N}{u},\label{eqn:U(u)_identity_character}
\end{equation}
where we used equation \eqref{eqn:CS_dim_SU_Sp}. This exactly matches the number of $\SU{N}$ $u$-walls given in table \ref{tab:SU_uWall_characters}.

\subsubsection*{1-form symmetry}
We will compute the character of $n$ applications of the 1-form symmetry generator in the first direction. Consider a state $\ket{[\vec{\lambda},\xi]}$, where $\vec{\lambda}$ has a $\Z_u$ orbit of size $l$ so that $\xi\equiv \xi+\frac{N}{u/l}$, and thus $\ket{[\vec{\lambda},\xi]}$ has a $\operator{\mathsf{T}}_1$ orbit size of $\frac{N}{u/l}$. Suppose that this state is invariant under $\operator{\mathsf{T}}_1^n$, so that $n$ is a multiple of $\frac{N}{u/l}$, say $n=\frac{N}{u/l}m$ for some $m\in\Z$. Rearranging we find that $\frac{nu}{N}=lm$, so that $\vec{\lambda}$ must be invariant under $\mathcal{T}^{nu/N}$. Before pressing on, notice that $N$ must divide $nu$, which means that $\frac{N}{\gcd(N,n)}$ must divide $\frac{n}{\gcd(N,n)}u$, implying that $\frac{N}{\gcd(N,n)}$ divides $u$ since $\frac{N}{\gcd(N,n)}$ and $\frac{n}{\gcd(N,n)}$ have no common factors. Recall that this is the same condition on $n$ that we derived for $u$-walls of $\SU{N}$ SYM, shown in table \ref{tab:SU_uWall_characters}. Thus for each $\Z_u$ orbit of size $l$, $[\vec{\lambda}]=\left\{\mathcal{T}^m(\vec{\lambda})\mid m=0,1,\dots,l-1\right\}$, (where $\vec{\lambda}$ is invariant under $\mathcal{T}^{nu/N}$) there are $\frac{N}{u/l}$ states $\ket{[\vec{\lambda},\xi]}$ invariant under $\operator{\mathsf{T}}_1^n$. Since each weight in $[\vec{\lambda}]$ will be invariant under $\mathcal{T}^{nu/N}$, we can count weights invariant under $\mathcal{T}^{nu/N}$ instead of orbits if for each weight we add a factor of $\frac{1}{l}$. The two factors of $l$ cancel out, and we find that the character of $\operator{\mathsf{T}}_1^n$ is $\frac{N}{u}\chi_{\SU{u}}\left(\operator{T}_1^{nu/N}\right)$. From our analysis of center symmetry characters in $\SU{N}$ SYM, we know that when $\vec{\lambda}$ is invariant under $\mathcal{T}^{nu/N}$ equation \eqref{eqn:SU(u)_integrable_reps} becomes
\begin{equation*}
    N-u=\frac{u}{\gcd\left(u,\frac{nu}{N}\right)}\sum_{a=0}^{\gcd\left(u,\frac{nu}{N}\right)-1}\lambda_a=\frac{N}{\gcd\left(N,n\right)}\sum_{a=0}^{\frac{u}{N}\gcd\left(N,n\right)-1}\lambda_a.
\end{equation*}
The character is $\frac{N}{u}$ times the number of solutions, which gives us
\begin{equation}
    \chi_{U(u)_{N-u,N}}\left(\operator{\mathsf{T}}_1^n\right)=\begin{dcases}
        \binom{\gcd(N,n)}{\frac{u}{N}\gcd(N,n)} & \frac{N}{\gcd(N,n)}\mid u \\ 0 & \text{otherwise}
    \end{dcases},\label{eqn:U(u)_center_character}
\end{equation}
exactly matching that of $\SU{N}$ $u$-walls from table \ref{tab:SU_uWall_characters}.

\subsubsection*{Charge conjugation}
We will compute the character of $\operator{C}$ in several steps. First, we will show that if $\ket{[\vec{\lambda},\xi]}$ is $\operator{C}$ invariant, then there is some $(\vec{\lambda}',\xi')\in[\vec{\lambda},\xi]$ such that $\mathcal{C}(\vec{\lambda}')=\vec{\lambda}'$. Notice that $\mathcal{C}$ and $\mathcal{T}$ obey a $D_{2u}$ dihedral algebra
\begin{equation}
    \mathcal{C}\circ\mathcal{T}\circ\mathcal{C}=\mathcal{T}^{-1},\label{eqn:cc_algebra_SU}
\end{equation}
which can be worked out explicitly using the action of $\mathcal{C}$ and $\mathcal{T}$ given in appendix \ref{sec:all_groups_symmetries}. If $\ket{[\vec{\lambda},\xi]}$ is $\operator{C}$-invariant, then we must have
\begin{equation*}
    (\mathcal{C}(\vec{\lambda}),-\xi-(N-u-\lambda_0))\in[\vec{\lambda},\xi]\iff \operator{C}\ket{\vec{\lambda}}\otimes\ket{u(\xi+\vec{w}_{u-1}^*\cdot\vec{\lambda})}=\operator{T}^m\ket{\vec{\lambda}}\otimes\ket{u(\xi+\vec{w}_{u-1}^*\cdot\vec{\lambda})}
\end{equation*}
where $0\leq m< u$ is an integer. If $m$ is even, we can apply $\mathcal{T}^{m/2}$ to $\vec{\lambda}$ to get the desired $\vec{\lambda}'$
\begin{equation*}
    \mathcal{C}\left(\mathcal{T}^{m/2}(\vec{\lambda})\right)=\mathcal{T}^{-m/2}\circ\mathcal{C}(\vec{\lambda})=\mathcal{T}^{m/2}(\vec{\lambda}).
\end{equation*}
Letting $l$ be the size of the $\mathcal{T}$ orbit of $\vec{\lambda}$, if both $m$ and $l$ are odd, we can instead apply $\mathcal{T}^{(m+l)/2}$ and get a similar result. In appendix \ref{appendix:U(N)_CS} we prove that the last case, when $l$ is even and $m$ is odd, is not possible. In short, we may assume without loss of generality that $\vec{\lambda}=\vec{\lambda}'$, ie we may assume that $\vec{\lambda}$ is $\mathcal{C}$ invariant. Then, the full action of charge conjugation on the $U(u)_{N-u,N}$ state is
\begin{equation*}
    \operator{C}\ket{[\vec{\lambda},\xi]}=\ket{[\vec{\lambda},-\xi-(N-u-\lambda_0)]},
\end{equation*}
so that our state $\ket{[\vec{\lambda},\xi]}$ is charge conjugation invariant when $\xi$ satisfies the following
\begin{equation}
    \xi\equiv -\xi-(N-u-\lambda_0)\bmod{\frac{N}{u/l}},\quad 0\leq \xi<\frac{N}{u/l}.\label{eqn:xi_charge_conjugation}
\end{equation}
Generically, solutions will be of the form $2\xi=\frac{N}{u/l}k-(N-u-\lambda_0)$ for some integer $k$ such that $0\leq \xi<\frac{N}{u/l}$. When $\frac{N}{u/l}$ is even, it is clear that $N-u-\lambda_0$ must also be even to get solutions. In particular, there are two solutions which are related to each other by addition/subtraction of $\frac{1}{2}\frac{N}{u/l}$. When $\frac{N}{u/l}$ is odd there is only one solution.

\paragraph{}
To count the number of independent charge conjugation invariant $U(u)_{N-u,N}$ states we will first count the number of charge conjugation invariant pairs $(\vec{\lambda},\xi)$ for a given $l$, then count the number of orbits $[\vec{\lambda},\xi]$. Finding the number of charge conjugation invariant $\SU{u}_{N-u}$ weights is done by solving equation \eqref{eqn:SU(u)_integrable_reps} subject to $\mathcal{C}(\vec{\lambda})=\vec{\lambda}$,
\begin{equation}
    \frac{N-u}{u/l}=\begin{dcases}\lambda_0+2\sum_{a=1}^{l/2-1}\lambda_a+\lambda_{l/2}&l\text{ even} \\
    \lambda_0+2\sum_{a=1}^{(l-1)/2}\lambda_a&l\text{ odd}\end{dcases}.\label{eqn:charge_invariant_weights}
\end{equation}
Notice that the above equation applies to weights with orbit sizes of $l$ \textit{and} divisors of $l$. Namely, if we set $l$ to it's maximum value, $u$, of which all orbits sizes are divisors, we obtain an equation for all $\mathcal{C}$ invariant weights. We split our analysis into cases when $N$ and $u$ are even/odd.

\paragraph{$\boldsymbol{N}$ and $\boldsymbol{u}$ odd}
In this case, $\frac{N}{u/l}$ is always odd, so for each $\mathcal{C}$ invariant weight there is one solution to \eqref{eqn:xi_charge_conjugation}. Furthermore, since $u$ is odd, $l$ too must be odd, and thus we do not have to worry about over-counting. The number of charge-conjugation invariant pairs $(\vec{\lambda},\xi)$ is equal to the number of charge-conjugation invariant orbits, and hence the corresponding number of $U(u)_{N-u,N}$ states. Thus, we just have to find the number of independent solutions to $\mathcal{C}(\vec{\lambda})=\vec{\lambda}$, which amounts to solving equation \eqref{eqn:charge_invariant_weights} with $l=u$,
\begin{equation*}
    N-u=\lambda_0+2\sum_{a=1}^{(u-1)/2}\lambda_a.
\end{equation*}
The number of solutions is the number of ways to choose $\frac{u+1}{2}$ non-negative integers $(\lambda_0,\dots,\lambda_{(u-1)/2})$ which sum to $\frac{N-u}{2}$,
\begin{equation*}
    \binom{\frac{N-1}{2}}{\frac{u-1}{2}}.
\end{equation*}

\paragraph{$\boldsymbol{N}$ odd and $\boldsymbol{u}$ even}
Like the last case, $\frac{N}{u/l}$ is always odd, so for each $\mathcal{C}$ invariant weight there is one solution to \eqref{eqn:xi_charge_conjugation}. Since $u/l$ must divide $N$, $l$ must be even in order to cancel out the ``evenness" from $u$. The number of charge-conjugation invariant pairs $(\vec{\lambda},\xi)$ is then twice the number of charge-conjugation invariant orbits, and hence the corresponding number of $U(u)_{N-u,N}$ states. Thus, we first have to find the number of independent solutions to $\mathcal{C}(\vec{\lambda})=\vec{\lambda}$, which corresponds to solving equation \eqref{eqn:charge_invariant_weights} with $l=u$,
\begin{equation*}
    N-u=\lambda_0+2\sum_{a=1}^{u/2-1}\lambda_a+\lambda_{u/2},
\end{equation*}
then we must divide by two to account for over-counting. With $N-u$ odd, we see that $\lambda_0+\lambda_{u/2}$ must also be odd, so we can redefine $\lambda_0+\lambda_{u/2}$ to be $2\lambda_0+2\lambda_{u/2}+1$, multiplying our final answer by two to account for the two possible parities of $\lambda_0$ and $\lambda_{u/2}$ (either even/odd or odd/even). The number of weights is then two times the number of ways to choose $\frac{u}{2}+1$ non-negative integers which sum to $\frac{N-u-1}{2}$, and the number of independent charge-conjugation invariant states is half of that,
\begin{equation*}
    \binom{\frac{N-1}{2}}{\frac{u}{2}}.
\end{equation*}

\paragraph{$\boldsymbol{N}$ even and $\boldsymbol{u}$ odd}
Now $\frac{N}{u/l}$ is always even, since $u/l$ must always be odd. Since $N-u$ is odd, we count only the weights with $\lambda_0$ odd, which have two solutions each to equation \eqref{eqn:xi_charge_conjugation}. Referring to equation \eqref{eqn:charge_invariant_weights} we see that indeed all the $\mathcal{C}$ invariant weights will have $\lambda_0$ odd. Thus, the number of charge conjugation invariant orbits $[\vec{\lambda},\xi]$ is twice the number of $\mathcal{C}$ invariant weights,
\begin{equation*}
    N-u=\lambda_0+2\sum_{a=1}^{(u-1)/2}\lambda_a.
\end{equation*}
Knowing that $\lambda_0$ is odd we redefine $\lambda_0\rightarrow 2\lambda_0+1$ and see that the number of solutions is the number of ways to choose $\frac{u+1}{2}$ non-negative integers which sum to $\frac{N-u-1}{2}$. Thus, the number of orbits is
\begin{equation*}
    2\binom{\frac{N-2}{2}}{\frac{u-1}{2}}.
\end{equation*}

\paragraph{$\boldsymbol{N}$ and $\boldsymbol{u}$ even}
When $N$ and $u$ are even, things can become a bit more complicated. First we set $u=2^\alpha \tilde{u}$ and $N-u=k=2^\beta \tilde{k}$, for positive integers $\alpha$ and $\beta$, and odd integers $\tilde{u}$ and $\tilde{k}$. Then, we know that $u/l$ must be a divisor of $\gcd(u,k)$, so we set $u/l=n=2^\gamma \tilde{n}$ where $0\leq \gamma\leq \min(\alpha,\beta)$ and $\tilde{n}$ divides $\gcd(\tilde{u},\tilde{k})$. We split our analysis into four cases:
\begin{enumerate}
    \item $0\leq \gamma <\min(\alpha,\beta)$: $\frac{N}{u/l}$, $\frac{N-u}{u/l}$, and $l$ are all even, so $\lambda_0$ must be even and there are two charge conjugation invariant pairs, $(\vec{\lambda},\xi)$ and $(\mathcal{T}^{l/2}(\vec{\lambda}),\xi')$, per $\mathcal{C}$ invariant weight\footnote{Note that equation \eqref{eqn:charge_invariant_weights} says that if $\lambda_0$ is even then $\lambda_{l/2}$ is also even, so that $\mathcal{T}^{l/2}(\vec{\lambda})$ also has two solutions to equation \eqref{eqn:xi_charge_conjugation}.}. Then, each $\mathcal{C}$ invariant weight with $\lambda_0$ even contributes $\frac{1}{2}\times 2$ to the character. The two factors of two cancel out and the contribution to the character is the number of $\mathcal{C}$ invariant weights with $\lambda_0$ even.

    \item $\gamma=\alpha=\beta$: $\frac{N}{u/l}$ is even, so $\lambda_0$ must be even, while $l$ and $\frac{N-u}{u/l}$ are odd. From equation \eqref{eqn:charge_invariant_weights}, we see that $\lambda_0$ cannot be even, and hence there are no charge conjugation invariant states.

    \item $\gamma=\alpha < \beta$: $\frac{N-u}{u/l}$ is even while $\frac{N}{u/l}$ and $l$ are both odd, so the number of invariant orbits is equal to the number of $\mathcal{C}$ invariant weights. From equation \eqref{eqn:charge_invariant_weights} we see that $\lambda_0$ must be even, so the counting matches that of the first case.

    \item $\gamma=\beta < \alpha$: $\frac{N-u}{u/l}$ and $\frac{N}{u/l}$ are both odd, while $l$ is even so the number of invariant orbits is half the number of $\mathcal{C}$ invariant weights. From equation \eqref{eqn:charge_invariant_weights}, $\lambda_0$ and $\lambda_{l/2}$ must have opposite parity $\bmod{2}$. Further, it is clear that the number of solutions is the same for the two cases, ie the number of $\mathcal{C}$ invariant weights with $\lambda_0$ even is the same as that with $\lambda_0$ odd. Thus we may assume that $\lambda_0$ is even and multiply our counting by two, cancelling out the factor of $\frac{1}{2}$ from the evenness of $l$. In the end, we again find the number of charge conjugation invariant orbits equal to the number of $\mathcal{C}$ invariant weights with $\lambda_0$ even.
\end{enumerate}
We see that we simply need to count all the solutions to equation \eqref{eqn:charge_invariant_weights} with $\lambda_0$ even, which is accomplished by simply setting $l=u$,
\begin{equation*}
    N-u=\eval{\lambda_0+2\sum_{a=1}^{u/2-1}\lambda_a+\lambda_{u/2}}_{\lambda_0\text{ even}}.
\end{equation*}
The number of solutions is the number of ways to choose $\frac{u}{2}+1$ positive integers which sum to $\frac{N-u}{2}$,
\begin{equation*}
    \binom{\frac{N}{2}}{\frac{u}{2}}.
\end{equation*}

\paragraph{}
We thus get our final result for the number of independent charge conjugation invariant $U(u)_{N-u,N}$ states,
\begin{equation}
    \chi_{U(u)_{N-u,N}}(\operator{C})=\begin{dcases}
        2\binom{\frac{N-2}{2}}{\frac{u-1}{2}} & N\text{ even, }u\text{ odd}\\
        \binom{\left\lfloor\frac{N}{2}\right\rfloor}{\left\lfloor\frac{u}{2}\right\rfloor} & \text{otherwise}
    \end{dcases},\label{eqn:U(u)_CC_character}
\end{equation}
which exactly matches the number of charge conjugation invariant $u$-walls of $\SU{N}$ from table \ref{tab:SU_uWall_characters}.

\subsubsection*{Combined charge conjugation and 1-form symmetry}
Finally, we want to compute the character for the combined action of charge conjugation and 1-form symmetry when $N$ is even. Here it will be most convenient to use the action of the 1-form symmetry generator in the second direction\footnote{One can do a change-of-basis to a basis where $\operator{\mathsf{T}}_1$ is diagonal and acts exactly as $\operator{\mathsf{T}}_2$ does on the basis we use here, so it is clear that the character of the two operators should be the same. Further, there really is no reason why the two $\T^2$ directions should be different, so based on these symmetry principles alone we expect that the characters will be equal.}, $\operator{\mathsf{T}}_2$, so that the combined action of charge conjugation and the 1-form symmetry is
\begin{equation*}
    \operator{C}\operator{\mathsf{T}}_2\ket{[\vec{\lambda},\xi]}=e^{2\pi iu(\xi+\vec{w}_{u-1}^*\cdot\vec{\lambda})/N}\operator{C}\ket{[\vec{\lambda},\xi]}.
\end{equation*}
It is then clear that the only states that will contribute to the character are those which are invariant under charge conjugation, which, conveniently, we have already found. First consider $u$ odd, where we will see that the character vanishes. From our computation of the character of charge conjugation, we know that when $N$ is even and $u$ is odd, $\frac{N}{u/l}$ is always even and thus there are two charge conjugation invariant states for every $\mathcal{C}$ invariant weight with $\lambda_0$ odd, which are related by adding/subtracting $\frac{1}{2}\frac{N}{u/l}$ from $\xi$. Given such a weight, letting the two charge conjugation invariant orbits be $[\vec{\lambda},\xi]$ and $\left[\vec{\lambda},\xi+\frac{1}{2}\frac{N}{u/l}\right]$, the contribution to the character is
\begin{equation*}
    e^{2\pi iu\left(\xi+\vec{w}_{u-1}^*\cdot\vec{\lambda}\right)/N}\left(1+e^{i\pi l}\right).
\end{equation*}
Since $u$ is odd, $l$ must be odd and hence the contribution to the character is always zero, leading to a vanishing character.

\paragraph{}
When $u$ is even, we have to apply the same analysis that we did for charge conjugation, but can recycle many of our results. Again we set $u=2^\alpha\tilde{u}$, $N-u=k=2^\beta\tilde{k}$, and $u/l=n=2^\gamma\tilde{n}$, where the numbers with $\tilde{}$ are odd, and $n$ must divide $\gcd(u,k)$. We have
\begin{enumerate}
    \item $0\leq \gamma<\min(\alpha,\beta)$: For every $\mathcal{C}$-invariant weight with $\lambda_0$ even there are two $\xi$ which give charge conjugation invariant states, but both $\vec{\lambda}$ and $\mathcal{T}^{l/2}(\vec{\lambda})$ produce the same states\footnote{It can be verified that $\vec{\lambda}$ and $\mathcal{T}^{l/2}$ both produce charge conjugation invariant states, which have the same phase under $\operator{\mathsf{T}}_2$.}, so we must divide their contributions to the character by two. Then, each such weight contributes the following to the character
    \begin{equation*}
        \frac{1}{2}e^{2\pi iu\left(\xi+\vec{w}_{u-1}^*\cdot\vec{\lambda}\right)/N}\left(1+e^{i\pi l}\right)=e^{2\pi iu\left(\xi+\vec{w}_{u-1}^*\cdot\vec{\lambda}\right)/N},
    \end{equation*}
    where we used the fact that $l$ is even, and took $\xi$ to be the lesser of the two solutions to equation \eqref{eqn:xi_charge_conjugation}, though it really doesn't matter which of the two solutions is taken.

    \item $\gamma=\alpha=\beta$: No charge conjugation invariant states.

    \item $\gamma=\alpha<\beta$: Each $\mathcal{C}$-invariant weight has $\lambda_0$ even and produces one charge conjugation invariant state, so the contribution to the character is the same as the first case.\label{case:l_odd}

    \item $\gamma=\beta<\alpha$: $l$ is even so for each charge conjugation invariant state there are two $\mathcal{C}$-invariant weights, related by $\mathcal{T}^{l/2}$. The action of $\mathcal{T}^{l/2}$ swaps $\lambda_0$ and $\lambda_{l/2}$, whose sum must be odd. Thus for each charge conjugation invariant state there is just one $\mathcal{C}$-invariant weight with $\lambda_0$ even. The contribution to the character is then the same as the first case.
\end{enumerate}
In summary, the character is computed by summing $e^{2\pi iu\left(\xi+\vec{w}_{u-1}^*\cdot\vec{\lambda}\right)/N}$ over each $\mathcal{C}$-invariant weight with $\lambda_0$ even, where $\xi$ is any solution to equation \eqref{eqn:xi_charge_conjugation}. Further, we will show that the phase is always trivial and thus the character is equal to character of charge conjugation. First, using the data in \ref{sec:all_groups_symmetries}, and the fact that $\vec{\lambda}$ is $\mathcal{C}$-invariant, along with the fact that $\vec{\lambda}$ is a weight of $\SU{u}_{N-u}$ we find
\begin{equation*}
    u\vec{w}_{u-1}^*\cdot\vec{\lambda}=\frac{u}{2}\left(N-u-\lambda_0\right).
\end{equation*}
From equation \eqref{eqn:xi_charge_conjugation}, we let $2\xi=k\frac{N}{u/l}-(N-u-\lambda_0)$, where $k$ is an integer with the requirement that $k\frac{N}{u/l}$ must be even. Then, we find that the phase is
\begin{equation*}
    \frac{u}{N}\left(\xi+\vec{w}_{u-1}^*\cdot\vec{\lambda}\right)=\frac{1}{2}kl,
\end{equation*}
which we claim is always an integer. When $l$ is even, it is obvious that we get an integer. When $l$ is odd, we must be in case \ref{case:l_odd} where $\frac{N}{u/l}$ is also odd. From the requirement that $k\frac{N}{u/l}$ is even, we see that $k$ is even, and again we find that $\frac{1}{2}kl$ is an integer. Thus we see that the character is in fact the same as the character of charge conjugation when $u$ is even. In summary we find
\begin{equation}
    \chi_{U(u)_{N-u,N}}(\operator{C}\operator{\mathsf{T}})=\begin{dcases}
        0 & u\text{ odd} \\ \binom{N/2}{u/2} & u\text{ even},
    \end{dcases}\label{eqn:U(u)_center_CC_character}
\end{equation}
exactly matching that of $\SU{N}$ $u$-walls given in table \ref{tab:SU_uWall_characters}.

\subsection[$\Sp{N}$]{$\boldsymbol{\Sp{N}}$}
The proposed worldvolume theory for $\Sp{N}$ $u$-walls is $\Sp{u}_{N+1-u}$. As noted earlier in equation \eqref{eqn:CS_dim_SU_Sp}, $\Sp{N}_k$ has a Hilbert space of dimension $\binom{N+k}{k}$, so we see that $\Sp{u}_{N+1-u}$ has dimension 
\begin{equation}
    \chi_{\Sp{u}_{N+1-u}}(\identity)=\binom{N+1}{N+1-u}=\binom{N+1}{u},\label{eqn:Sp_CS_identity_character}
\end{equation}
which is exactly equal to the number of $\Sp{N}$ $u$-walls from table \ref{tab:Sp_uWall_characters}.

\subsubsection*{1-form center symmetry}
The $\Z_2^{(1)}$ center symmetry in $\Sp{u}_{N+1-u}$ acts on states $\ket{\vec{\lambda}}=\ket{(\lambda_0,\lambda_1,\dots,\lambda_u)}$ as
\begin{align*}
    \operator{T}_1\ket{\vec{\lambda}=\sum_{a=1}^{u}\lambda_a\vec{w}_a}&=\ket{\sum_{a=1}^{u}\lambda_{u-a}\vec{w}_a}\\
    \operator{T}_2\ket{\vec{\lambda}=\sum_{a=1}^{u}\lambda_a\vec{w}_a}&=(-1)^{\sum_{a=0}^{\left\lfloor (u-1)/2\right\rfloor}\lambda_{2a+1}}\ket{\vec{\lambda}}.
\end{align*}
The character of $\operator{T}_1$ is the number of representations $(\lambda_0,\lambda_1,\dots,\lambda_u)$ with $\lambda_a=\lambda_{u-a}$ and $\sum_{a=0}^u\lambda_a=N+1-u$, giving us
\begin{equation}
    N+1-u=\begin{dcases}
        2\sum_{a=0}^{u/2-1}\lambda_a+\lambda_{u/2} & u\text{ even}\\
        2\sum_{a=0}^{(u-1)/2}\lambda_a & u\text{ odd}
    \end{dcases}.\label{eqn:Sp_center_invariant_weights}
\end{equation}

\paragraph{$\boldsymbol{u}$ even}
From \eqref{eqn:Sp_center_invariant_weights}, $N+1-\lambda_{u/2}$ must be even, in other words $\lambda_{u/2}$ must have the opposite parity $\bmod{2}$ of $N$. We can redefine $\lambda_{u/2}$ to be $2\lambda_{u/2}+\delta$ with $\delta=2\left(\frac{N+1}{2}-\left\lfloor\frac{N+1}{2}\right\rfloor\right)$\footnote{In other words, $\delta$ is the remainder of $N+1$ when dividing by two.}, so that equation \eqref{eqn:Sp_center_invariant_weights} becomes
\begin{equation*}
    \left\lfloor\frac{N+1}{2}\right\rfloor-\frac{u}{2}=\sum_{a=0}^{u/2}\lambda_a.
\end{equation*}
The number of solutions is simply the number of ways to choose $u/2+1$ non-negative integers which sum to $\left\lfloor\frac{N+1}{2}\right\rfloor-\frac{u}{2}$,
\begin{equation*}
    \binom{\left\lfloor(N+1)/2\right\rfloor}{u/2}.
\end{equation*}

\paragraph{$\boldsymbol{u}$ odd}
From \eqref{eqn:Sp_center_invariant_weights}, when $N$ is odd there are no $\operator{T}_1$ invariant states. For $N$ even, the number of states is the number of ways to choose $(u+1)/2$ non-negative integers which sum to $\frac{N}{2}-\frac{u-1}{2}$,
\begin{equation*}
    \binom{N/2}{(u-1)/2}.
\end{equation*}

\paragraph{}
Combining the above results, we find that the number of $\operator{T}_1$ invariant states is
\begin{equation}
    \chi_{\Sp{u}_{N+1-u}}(\operator{T}_1)=\begin{dcases}
        0 & N,\ u\text{ odd}\\
        \binom{\left\lfloor(N+1)/2\right\rfloor}{\left\lfloor u/2\right\rfloor}&\text{ otherwise}
    \end{dcases},\label{eqn:Sp_CS_center_character}
\end{equation}
exactly matching that of $\Sp{N}$ $u$-walls from table \ref{tab:Sp_uWall_characters}.

\subsection[$\E{6}$]{$\boldsymbol{\E{6}}$}
Consider $(\E{6})_{3}$ Chern-Simons theory, which has a $\Z_3^{(1)}$ 1-form center symmetry and a $\Z_2^{(0)}$ charge conjugation symmetry, which form a $D_6$ dihedral global symmetry whose conjugacy classes are represented by $\identity$, $\operator{T}_i$, and $\operator{C}$. The states satisfy $\lambda_0+\lambda_1+\lambda_5+2(\lambda_2+\lambda_4+\lambda_6)+3\lambda_3=3$. One sees that there are 10 solutions with $\lambda_0+\lambda_1+\lambda_5=3$, 9 solutions with $\lambda_0+\lambda_1+\lambda_5=1$ and $\lambda_2+\lambda_4+\lambda_6=1$, and one solution with $\lambda_3=1$. In total we find the character of the identity to be 20,
\begin{equation}
    \chi_{(\E{6})_3}\left(\identity\right)=20,\label{eqn:E6_CS_identity_character}
\end{equation}
exactly matching the number of $u=6$ walls of $\E{6}$ from table \ref{tab:E6_uWall_characters}.

\subsubsection*{1-form center symmetry}
The $\Z_3^{(1)}$ 1-form center symmetry acts in the first direction as $\lambda_0\rightarrow \lambda_5\rightarrow\lambda_1\rightarrow\lambda_0$ and $\lambda_2\rightarrow\lambda_6\rightarrow \lambda_4\rightarrow\lambda_2$, so the states invariant under $\operator{T}_1$ have $\lambda_0=\lambda_1=\lambda_5$ and $\lambda_2=\lambda_4=\lambda_6$. We are then left with $3=3\lambda_0+6\lambda_2+3\lambda_3$, meaning that either $\lambda_0=1$ or $\lambda_3=1$, and hence the only $\operator{T}_1$ invariant states are $\ket{\vec{w}_1+\vec{w}_5}$ and $\ket{\vec{w}_3}$, giving us the character of $\operator{T}_1$
\begin{equation}
    \chi_{(\E{6})_3}\left(\operator{T}_1\right)=2,\label{eqn:E6_CS_center_character}
\end{equation}
consistent with the observation that there are two $6$-walls invariant under $\Z_3^{(1),\s^1}$ in SYM from table \ref{tab:E6_uWall_characters}.

\subsubsection*{0-form charge conjugation symmetry}
The $\Z_2^{(0)}$ charge conjugation symmetry acts as $\lambda_1\leftrightarrow \lambda_5$ and $\lambda_2\leftrightarrow \lambda_4$, so the only charge conjugation invariant states have $\lambda_1=\lambda_5$ and $\lambda_2=\lambda_4$, giving us $3=\lambda_0+2\lambda_1+4\lambda_2+3\lambda_3+2\lambda_6$. We see that there are then four charge conjugation invariant states, $\ket{0}$, $\ket{\vec{w}_1+\vec{w}_5}$, $\ket{\vec{w}_3}$, and $\ket{\vec{w}_6}$,
\begin{equation}
    \chi_{(\E{6})_3}\left(\operator{C}\right)=4,\label{eqn:E6_CS_CC_character}
\end{equation}
exactly matching the number of charge conjugation invariant $6$-walls in SYM from table \ref{tab:E6_uWall_characters}.

%% file: deconfinement.tex
\begin{figure}
    \centering
    \includegraphics{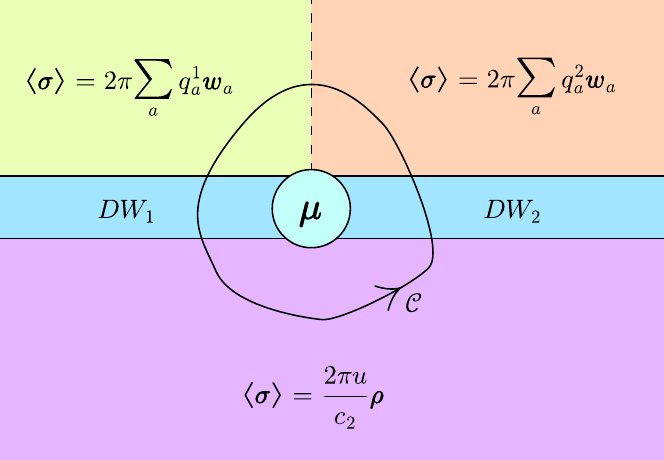}
    \caption{A quark supported by two domain walls with fluxes $\vec{\Phi}_i=\sum_{a=1}^rq_a^i\vec{w}_a-\frac{u}{c_2}\vec{\rho}$, so that $\vec{\Phi}_2-\vec{\Phi}_1=\vec{\mu}$. The two upper vacua (yellow and orange) are $u=0$ vacua, while the lower vacuum (purple) is the $u^{th}$ vacuum. The insertion of the static quark is equivalent to demanding $\oint_{\mathcal{C}}\dd{\vec{\sigma}}=2\pi\vec{\mu}$ for any $\mathcal{C}$ which encloses the charge.}
    \label{fig:quark_DW}
\end{figure}
\begin{figure}[h]
    \centering
    \begin{subfigure}{0.45\textwidth}
        \centering
        \includegraphics[width=\linewidth]{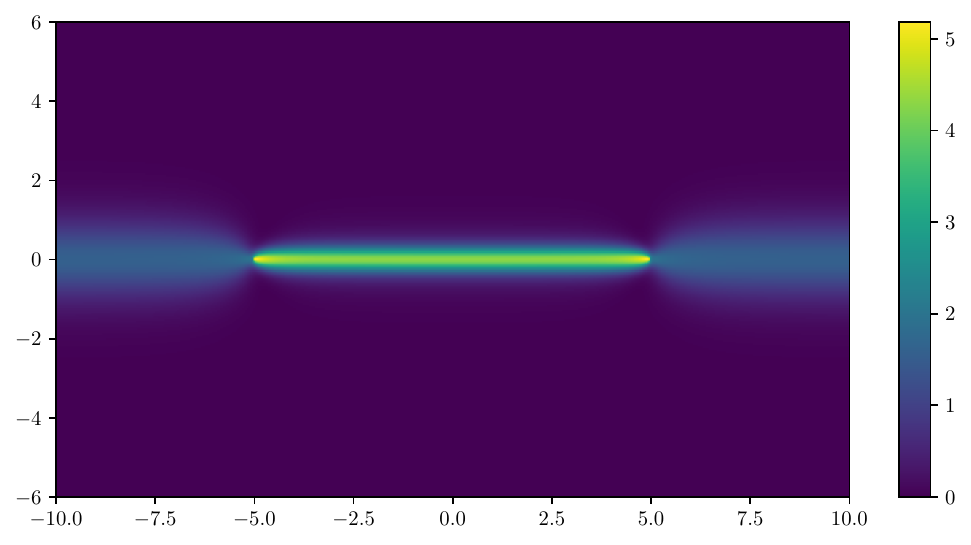}
        \caption{}
        \label{fig:G2_Deconfinement}
    \end{subfigure}
    \hspace{0.05\textwidth}
    \begin{subfigure}{.45\textwidth}
        \centering
        \includegraphics[width=\linewidth]{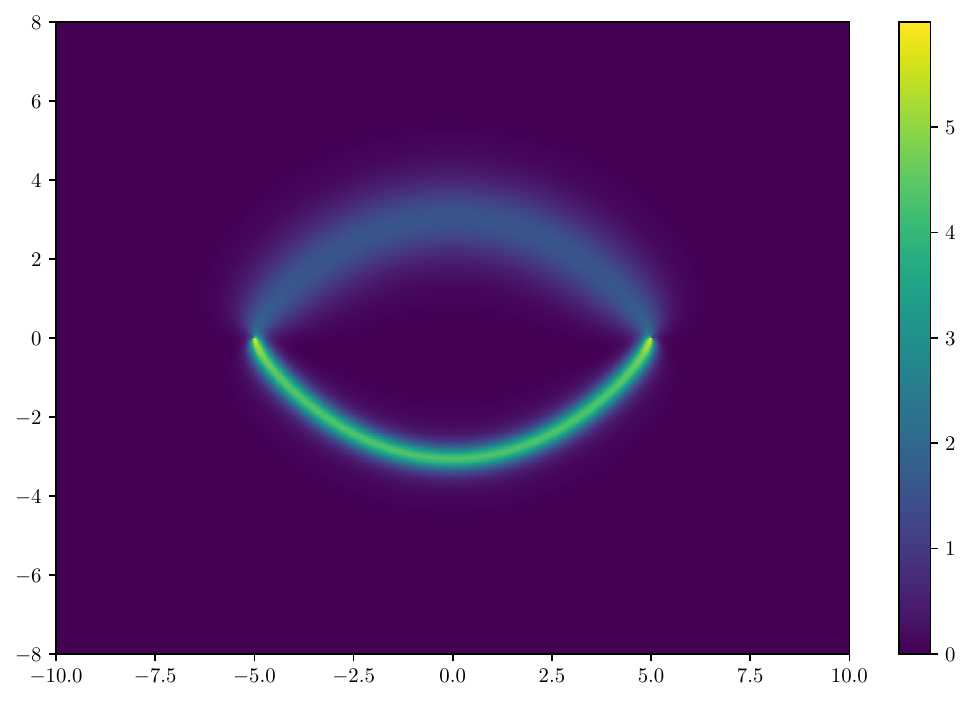}
        \caption{}
        \label{fig:G2_Confinement}
    \end{subfigure}
    \caption{(a): Potential energy density of two heavy probe quarks of charge $\pm\vec{w}_2$ for $\G$ showing deconfinement of quarks on domain walls for $\G$. (b): The same quarks are now confined by ``wrapping" the two domain walls in (a) around to form confining strings.}
    \label{fig:G2_2D}
\end{figure}

We now turn to studying deconfinement of static (heavy) quarks in the presence of a $u$-wall. The insertion of a static quark with charge $\vec{\mu}$ at position $\svec{r}_0=(x_0,y_0)\in\R^2$ corresponds to inserting a static Wilson line, which modifies the classical equations of motion for $\vec{\sigma}$ by adding a term which forces $\vec{\sigma}$ to ``jump" by $2\pi\vec{\mu}$ as it crosses $x=x_0$ in the upper half-plane\footnote{This is just one way to insert the jump in $\vec{\sigma}$ which is useful for illustrating the salient points, but other choices can be made. See \cite{Cox2019} for more details.} \cite{Cox2019}
\begin{equation*}
    \nabla^2{\vec{\sigma}}\supset 2\pi\vec{\mu}\partial_x\delta(x-x_0)\int_{y_0}^\infty\dd{y'}\delta(y-y')=\begin{dcases}
        2\pi\vec{\mu}\partial_x\delta(x-x_0) & y \geq y_0 \\ 0 & y<y_0
    \end{dcases}.
\end{equation*}
Then, taking a contour $\mathcal{C}$ which winds positively around $\svec{r}_0$, we find that inserting the static charge amounts to imposing $\oint_c\dd{\vec{\sigma}}=2\pi\vec{\mu}$. Such a configuration can be achieved by suspending the quark between $u$-walls whose fluxes differ by $2\pi\vec{\mu}$, as in figure \ref{fig:quark_DW}. If the two $u$-walls are both BPS, and hence have the same tension, the quark will be free to move along the domain walls: moving to one side lengthens one of the domain walls by the exact same amount that the other is shortened. We may then consider adding a quark with weight $-\vec{\mu}$ as in figure \ref{fig:G2_Deconfinement}. We see that the second quark is also free to move, and hence the two quarks can move independently and are deconfined, so long as they do not cross each other.

\paragraph{}
Mathematically, a weight $\vec{\mu}$ may be supported by two $u$-walls with fluxes $\vec{\Phi}_1$ and $\vec{\Phi}_2$ if $\vec{\mu}=\vec{\Phi}_1-\vec{\Phi}_2$. In appendix \ref{appendix:deconProofs} we prove that
\begin{equation}
    \vec{\mu}\text{ deconfined on domain walls}\iff \mu_a=\vec{\alpha}_a^*\cdot\vec{\mu}\in\{-1,0,+1\}\ \forall \ a=0,1,\dots,r\label{eqn:deconfined_weights}.
\end{equation}
Further, we show that the $u$-walls which deconfine a given $\vec{\mu}$ are given by
\begin{equation}
    u(\vec{\mu})=\left\{\sum_{\vec{\alpha}_a^*\cdot\vec{\mu}=+1}k_a^*+\sum_{\vec{\alpha}_a^*\cdot\vec{\mu}=0}k_a^*q_a\mid q_a\in\{0,1\}\right\},\label{eqn:u_deconfined}
\end{equation}
where $\sum_{\vec{\alpha}_a^*\cdot\vec{\mu}=x}$ means to sum over the co-roots $\vec{\alpha}_a^*$ such that $\vec{\alpha}_a^*\cdot\vec{\mu}=x$. Moreover, if two quarks can be deconfined on domain walls, then they can also be confined by ``wrapping" the domain walls around to form a double string as in figure \ref{fig:G2_Confinement}. When the probe quarks have $N$-ality 0, which always is the case when the gauge group has a trivial center, the double string picture only holds for sufficiently small quark separations; there is a point where it is energetically favourable to pair-produce $W$-bosons (of mass $\sim \frac{1}{L}$) screening the quarks and breaking the double string.

\paragraph{}
In \cite{Cox2019} it was argued that to determine if a representation will be deconfined it is enough to find a weight of that representation which is deconfined. Here we will briefly summarize the argument for completeness. Consider a Wilson loop in a representation $R_{\vec{\lambda}}$, which in the Abelian limit studied here is given by
\begin{equation*}
    \expval{W_{R_{\vec{\lambda}}}[\mathcal{C}]}=\sum_{\vec{\mu}\in R_{\vec{\lambda}}}\expval{e^{i\vec{\mu}\cdot\oint_{\mathcal{C}}\vec{A}}}.
\end{equation*}
We are interested in the behaviour of $\expval{W_{R_{\vec{\lambda}}}[\mathcal{C}]}$ as $\mathcal{C}$ becomes large. In the limit of large $\mathcal{C}$, perimeter law terms $e^{-P(\mathcal{C})}$ due to deconfined weights dominate over area law terms $e^{-A(\mathcal{C})}$. Thus, if $R_{\vec{\lambda}}$ has any weights which are deconfined on $u$-walls, then $\expval{W_{R_{\vec{\lambda}}}[\mathcal{C}]}$ will exhibit perimeter law in the background of a $u$-wall, and quarks will be deconfined.

\paragraph{}
We can classify representations then by whether or not they deconfine in the background of $u$-walls, determining which (if any) $u$ do the job. To do so, we will organize representations by their $N$-ality, defined as the charge under the center of the representation\footnote{For a $\Z_n$ center, with generator $g=e^{2\pi i/n}$, a representation $R$ has $N$-ality $k$ if $R(g)=e^{2\pi ik/n}$. See appendix \ref{appendix:center} for more details.}, where for $\Spin{4n}$ representations will have two $N$-alities, corresponding to the two copies of $\Z_2$ in the center. Then, for each $N$-ality, $k$, we will find a weight of all representations of that $N$-ality, which we will call a \textit{universal weight} of $N$-ality $k$. Finally, using equation \eqref{eqn:u_deconfined} we will find the $u$-walls which deconfine the universal weights for each $N$-ality. For all non-zero $N$-alities, we will find universal weights which are deconfined on all $u$-walls for all $u$. For $N$-ality zero, in some groups we will find universal weights which are again deconfined on $u$-walls for all $u$, and in other groups we will find universal weights which are deconfined on $u$-walls for all $u$ \textit{except} $u=1$ and $u=c_2-1$. In the latter case, we will show that there are no $N$-ality 0 weights deconfined on $u=1,c_2-1$ walls. We want to stress that our treatment here is valid in the abelianized regime. Notably, the statement that $N$-ality 0 quarks are not deconfined on $u=1,c_2-1$ walls is not equivalent to the statement that $N$-ality 0 quarks are confined outside the abelianized regime. As discussed above, at energies of order $\frac{1}{L}$ $W$-bosons can be pair-produced and screen $N$-ality 0 quarks, leading to deconfinement as expected.

\subsection[$\SU{N}$]{$\boldsymbol{\SU{N}}$}
The $N$-ality of the irreducible representation of $\SU{N}$ with highest weight $\vec{\lambda}=\sum_{a=1}^r\lambda_a\vec{w}_a$ is
\begin{equation*}
    N\text{-ality}(\vec{\lambda})= N\vec{w}_{N-1}^*\cdot\vec{\lambda}\equiv \sum_{a=1}^ra\lambda_a\bmod{N}.
\end{equation*}
Then $-\vec{\alpha}_0=\vec{w}_1+\vec{w}_{N-1}$ is a universal weight of $N$-ality 0, while $\vec{w}_q$ is a universal weight of $N$-ality $q$ for $q=1,\dots,N-1$ \cite{Poppitz2018}. We see that $\vec{w}_q$ is of the form of equation \eqref{eqn:deconfined_weights}, and may be deconfined by all $u$-walls, as noted in \cite{Cox2019}. While $-\vec{\alpha}_0$ is not of the form of equation \eqref{eqn:deconfined_weights}, performing a Weyl reflection with respect to $\vec{\alpha}_{N-1}$ gives us $\vec{w}_1+\vec{w}_{N-2}-\vec{w}_{N-1}$ which is deconfined on all $u$-walls \textit{except} for $u=1,N-1$. Moreover, there are no weights of \textit{any} $N$-ality 0 representations deconfined on $u=1$ or $u=N-1$ domain walls. We illustrate this last point for $u=1$, but the process for $u=N-1$ is more or less the same. First, notice that $u=1$ domain walls are of the form $\vec{\Phi}=\vec{w}_a-\frac{1}{N}\vec{\rho}$, where $a=0$ corresponds to $\vec{w}_0=0$. Then, a weight deconfined on $1$-walls is of the form $\vec{\mu}=\vec{w}_a-\vec{w}_b$, where $a\neq b$ and $0\leq a,b<N$. The $N$-ality of $\vec{\mu}$ is then $a-b\bmod{N}$, which works even when $a$ or $b$ is zero. The only way to get $N$-ality 0, is to have $a-b=kN$ for some integer $k$, and the only $k$ which is compatible with $0\leq a,b<N$ is $k=0$. In other words, the only way to get $N$-ality 0 is if $\vec{\mu}$ is zero, and hence there are no non-trivial weights of $N$-ality 0 deconfined on $1$-walls. When $N=2,3$, there are no $N$-ality 0 weights deconfined on any domain walls.

\subsection[$\Sp{N}$]{$\boldsymbol{\Sp{N}}$}
The $N$-ality of $\vec{\lambda}$ is
\begin{equation*}
    N\text{-ality}(\vec{\lambda})= 2\vec{w}_N^*\cdot\vec{\lambda}\bmod{2}\equiv \sum_{a=0}^{\left\lfloor(N-1)/2\right\rfloor}\lambda_{2a+1}\bmod{2}.
\end{equation*}
Then $\vec{w}_2$ is a universal weight of $N$-ality 0, while $\vec{w}_1$ is a universal weight of $N$-ality 1. Further, both of these weights are of the form of equation \eqref{eqn:deconfined_weights}, and from equation \eqref{eqn:u_deconfined} are deconfined on $u$-walls for all $u$.

\subsection[$\Spin{4n+2}$]{$\boldsymbol{\Spin{4n+2}}$}
For $\Spin{4n+2}$, in this section only we will take the $\Z_4$ center to be generated by $e^{2\pi i\vec{w}_-^*\cdot\vec{H}}$ for $n$ even and $e^{2\pi i\vec{w}_+^*\cdot\vec{H}}$ for $n$ odd, so that the $N$-ality of $\vec{\lambda}$ is
\begin{equation*}
    N\text{-ality}(\vec{\lambda})=\begin{dcases}
        4\vec{w}_-^*\cdot\vec{\lambda}\bmod{4} & n\text{ even} \\ 4\vec{w}_+^*\cdot\vec{\lambda}\bmod{4} & n\text{ odd}
    \end{dcases}\equiv2\sum_{a=0}^{n-1}\lambda_{2b+1}+\lambda_-+3\lambda_+\bmod{4}.
\end{equation*}
Note that the choice of center generator is in some sense arbitrary, and here we just choose a convenient generator for calculating $N$-ality. Then $\vec{w}_2$ is a universal weight of $N$-ality 0, $\vec{w}_1$ of $N$-ality 2, $\vec{w}_-$ of $N$-ality 1, and $\vec{w}_+$ of $N$-ality 3. All of $\vec{w}_1$, $\vec{w}_-$, and $\vec{w}_+$ are of the form of equation \eqref{eqn:deconfined_weights}, and from equation \eqref{eqn:u_deconfined} are deconfined on $u$-walls for all $u$. While $\vec{w}_2$ is not deconfined, $\vec{w}_3-\vec{w}_1$ is in its Weyl group orbit and is deconfined on $u$-walls for all $u$ except $u=1$ and $u=4n-1$. Indeed there are no deconfined weights of $N$-ality 0 deconfined on 1- or $4n-1$-walls, where the proof is much the same as that done for $\SU{N}$.

\subsection[$\Spin{4n}$]{$\boldsymbol{\Spin{4n}}$}
Like $\Spin{4n+2}$, we will take the generators of each $\Z_2^\pm$ copy of the $\Z_2^+\times\Z_2^-$ center symmetry to be generated by $e^{2\pi i\vec{w}_\mp^*\cdot\vec{H}}$ for $n$ even and $e^{2\pi i\vec{w}_\pm^*\cdot\vec{H}}$ for $n$ odd, so that the $N$-ality of $\vec{\lambda}$ with respect to $\Z_2^\pm$ is
\begin{equation*}
    N\text{-ality}_\pm(\vec{\lambda})=\begin{dcases}
        2\vec{w}_\mp^*\cdot\vec{\lambda}\bmod{2} & n\text{ even} \\ 2\vec{w}_\pm^*\cdot\vec{\lambda}\bmod{2} & n\text{ odd}
    \end{dcases}\equiv \sum_{a=0}^{n-2}\lambda_{2a+1}+\lambda_\pm\bmod{2}.
\end{equation*}
Then, labelling the $N$-ality of a representation by a tuple of $N$-alities with respect to $\Z_2^\pm$, we find that $\vec{w}_2$ is a universal weight of $N$-ality $(0,0)$, $\vec{w}_1$ of $N$-ality $(1,1)$, $\vec{w}_+$ of $N$-ality $(1,0)$, and $\vec{w}_-$ of $N$-ality $(0,1)$. All of $\vec{w}_1$, $\vec{w}_-$, and $\vec{w}_+$ are of the form of equation \eqref{eqn:deconfined_weights}, and from equation \eqref{eqn:u_deconfined} are deconfined on $u$-walls for all $u$. Meanwhile, $\vec{w}_2$ has in its Weyl group orbit $\vec{w}_1+\vec{w}_--\vec{w}_+$ which is deconfined on $u$-walls for all $u$ except $u=1$ and $u=4n-3$, and indeed there are no $N$-ality $(0,0)$ weights deconfined on $1$- and $4n-3$-walls, where the proof is much the same as that done for $\SU{N}$.

\subsection[$\Spin{2N+1}$]{$\boldsymbol{\Spin{2N+1}}$}
The $N$-ality of $\vec{\lambda}$ is
\begin{equation*}
    N\text{-ality}(\vec{\lambda})= 2\vec{w}_1^*\cdot\vec{\lambda}\bmod{2}\equiv \lambda_N\bmod{2}.
\end{equation*}
Then $\vec{w}_1$ is a universal weight of $N$-ality 0, and $\vec{w}_N$ of $N$-ality 1. Further, both $\vec{w}_1$ and $\vec{w}_N$ are of the form of equation \eqref{eqn:deconfined_weights}, and from equation \eqref{eqn:u_deconfined} are deconfined on $u$-walls for all $u$.

\subsection[$\E{6}$]{$\boldsymbol{\E{6}}$}
The $N$-ality of $\vec{\lambda}$ is
\begin{equation*}
    N\text{-ality}(\vec{\lambda})=\vec{w}_1^*\cdot\vec{\lambda}\bmod{3}\equiv \lambda_1+\lambda_4+2(\lambda_2+\lambda_5)\bmod{3}.
\end{equation*}
Then $\vec{w}_6$ is a universal weight of $N$-ality 0, $\vec{w}_1$ of $N$-ality 1, and $\vec{w}_5$ of $N$-ality 2. Further, both $\vec{w}_1$ and $\vec{w}_5$ are of the form of equation \eqref{eqn:deconfined_weights}, and from equation \eqref{eqn:u_deconfined} are deconfined on all $u$-walls. While $\vec{w}_6$ is not of the form of equation \eqref{eqn:deconfined_weights}, $\vec{w}_1+\vec{w}_4-\vec{w}_2$ is in its Weyl group orbit and is a deconfined weight, being deconfined on $u$-walls for all $u$.

\subsection[$\E{7}$]{$\boldsymbol{\E{7}}$}
The $N$-ality of $\vec{\lambda}$ is
\begin{equation*}
    N\text{-ality}(\vec{\lambda})= 2\vec{w}_6^*\cdot\vec{\lambda}\bmod{2}\equiv \lambda_4+\lambda_6+\lambda_7\bmod{2}.
\end{equation*}
Then $\vec{w}_1$ is a universal weight of $N$-ality 0, and $\vec{w}_6$ of $N$-ality 1. We see that $\vec{w}_6$ is of the form of equation \eqref{eqn:deconfined_weights}, and from equation \eqref{eqn:u_deconfined} is deconfined on $u$-walls for all $u$. While $\vec{w}_1$ is not of the form of equation \eqref{eqn:deconfined_weights}, $\vec{w}_6+\vec{w}_7-\vec{w}_5$ is in its Weyl group orbit and is deconfined on $u$-walls for all $u$.

\subsection[$\E{8}$]{$\boldsymbol{\E{8}}$}
Since $\E{8}$ has trivial center, all representations have $N$-ality 0, and $\vec{w}_1$ is a universal weight of $N$-ality 0. While $\vec{w}_1$ is not of the form of equation \eqref{eqn:deconfined_weights}, there are weights in its Weyl group orbit which are, namely $\vec{w}_7-\vec{w}_1$ and $\vec{w}_6-\vec{w}_3$. These weights are deconfined on $u$-walls for all $u$ except $u=1$ and $u=29$. Further, there are no weights which are deconfined on $1$-walls, which is seen from the fact that there is only one $u=1$ (or $u=29$) domain wall.

\subsection[$\F$]{$\boldsymbol{\F}$}
Since $\F$ has trivial center, all representations have $N$-ality 0, and $\vec{w}_4$ is a universal weight of $N$-ality 0. We observe that $\vec{w}_4$ is of the form of equation \eqref{eqn:deconfined_weights}, and from equation \eqref{eqn:u_deconfined} is deconfined on $u$-walls for all $u$ except $u=2$ and $u=7$. For $u=2$ and $u=7$, we find $\vec{w}_1-\vec{w}_4$ in the Weyl group orbit of $\vec{w}_4$ which is deconfined on $2$- and $7$-walls.

\subsection[$\G$]{$\boldsymbol{\G}$}
Since $\G$ has trivial center, all representations have $N$-ality 0, and $\vec{w}_2$ is a universal weight of $N$-ality 0. We observe that $\vec{w}_2$ is of the form of equation \eqref{eqn:deconfined_weights}, and from equation \eqref{eqn:u_deconfined} is deconfined on $u$-walls for all $u$ except $u=2$. For $u=2$, we find $\vec{w}_1-\vec{w}_2$ in the Weyl group orbit of $\vec{w}_2$ which is deconfined on $2$-walls.

%% file: group.tex
For a Lie algebra $\lie{g}$ of a Lie group $G$, a Cartan subalgebra $\lie{h}\subseteq \lie{g}$ is a maximally commuting subalgebra of dimension $r$, where $r$ is the rank of $G$. In the defining representation we can select $r$ real-symmetric generators as a basis of $\lie{h}$, $\{H^a\mid a=1,2,\dots,r\}$ satisfying $\Tr(H^aH^b)=\delta^{ab}$. The rest of $\lie{g}$ is spanned by root vectors $E_\alpha$ defined as the eigenvectors of the Cartan generators in the adjoint representation,
\begin{equation*}
    \operatorname{ad}_{\vec{H}}(E_{\vec{\alpha}})=\comm{\vec{H}}{E_{\vec{\alpha}}}=\vec{\alpha}E_{\vec{\alpha}},
\end{equation*}
where the eigenvalues, $\vec{\alpha}\in\R^r$ are the roots which form a root system which we denote $\Delta$. We select a basis of $r$ roots for $\Delta$, called the simple roots which we denote $\Pi=\left\{\vec{\alpha}_a\mid a=1,\dots,r\right\}$, so that each root in $\Delta$ is written as a linear combination of simple roots with integer coefficients that are either all non-negative (the positive roots, $\Delta^+$), or all non-positive (the negative roots, $\Delta^-$). Note that the positive and negative roots are in one-to-one correspondence, ie $\vec{\alpha}\in\Delta\iff -\vec{\alpha}\in\Delta$ or equivalently $\Delta^-=-\Delta^+$. In the defining representation we take the $E_{\vec{\alpha}}$ to be real, so that $E_{\vec{\alpha}}^T=E_{-\vec{\alpha}}$. We normalize so that $\comm{E_{\vec{\alpha}}}{E_{-\vec{\alpha}}}=\vec{\alpha}^*\cdot\vec{H}$, and otherwise take $\comm{E_{\vec{\alpha}}}{E_{\vec{\beta}}}=C_{\vec\alpha,\vec\beta}E_{\vec{\alpha}+\vec{\beta}}$, where $C_{\vec\alpha,\vec\beta}$ is a group dependent factor which is only non-zero when $\vec{\alpha}+\vec{\beta}$ is a root.

\paragraph{}
The roots generate a root lattice, $\Lambda_r=\{\sum_{a=1}^r c_a\vec{\alpha}_a\mid c_a\in\Z\}$. Dual to the root lattice is the co-root lattice $\Lambda_r^*$, spanned by the simple co-roots $\vec{\alpha}_a^*=\frac{2}{\abs{\vec{\alpha}_a}^2}\vec{\alpha}_a$. Fundamental weights $\vec{w}_a$ are defined by requiring $\vec{w}_a\cdot\vec{\alpha}_b^*=\delta_{a,b}$, and span the weight lattice $\Lambda_w$. Finally, fundamental co-weights are defined by $\vec{w}_a^*=\frac{2}{\abs{\vec{\alpha}_a}^2}\vec{w}_a$ so that $\vec{w}_a^*\cdot\vec{\alpha}_b=\delta_{a,b}$, and span the co-weight lattice $\Lambda_w^*$.

\paragraph{}
In addition to the simple roots, there is one more important root that we will worry about: the affine, or lowest, root $\vec{\alpha}_0$ which is the root which is lower than all other roots, ie $\vec{\alpha}_0\preceq\vec{\beta}$ for all $\vec{\beta}\in\Delta$, where $\preceq$ is the partial ordering on $\R^r$ defined by the simple roots\footnote{The partial ordering $\preceq$ on $\R^r$ is defined so that $\vec{\lambda}\preceq\vec{\mu}$ (or $\vec{\mu}\succeq\vec{\lambda}$) if and only if $\vec{\mu}-\vec{\lambda}$ when written as a linear combination of simple roots has only non-negative coefficients. Equivalently, we say that $\vec{\lambda}\preceq\vec{\mu}$ if and only if $\vec{w}_a^*\cdot(\vec{\mu}-\vec{\lambda})\geq 0$ for all $a=1,\dots,r$.}. Note that here we will take the affine root, and hence all the long roots, to have length 2, so that they are identified with their co-roots.

\paragraph{}
It is often useful to represent a root system by its Dynkin diagram, which assigns a node to each simple root. Two nodes, say $a$ and $b$, are connected if $\vec{\alpha}_a^*\cdot\vec{\alpha}_b$ is non-zero. Further, the nodes are connected by one line if $\vec{\alpha}_a^*\cdot\vec{\alpha}_b=-1$, two lines if $\vec{\alpha}_a^*\cdot\vec{\alpha}_b=-2$, and three lines if $\vec{\alpha}_a^*\cdot\vec{\alpha}_b=-3$, where in the latter two cases an arrow is also drawn pointing towards the smaller (with respect to the Euclidean length) root. Note that, along with $\vec{\alpha}_a^*\cdot\vec{\alpha}_a$, it is a well established fact that these are the only possible inner products between simple co-roots and simple roots. We can extend the Dynkin diagram to include the affine node using the same rules. Appendix \ref{sec:all_groups_symmetries} gives the Dynkin diagrams and our labelling conventions for all simple Lie groups.

\subsection{The Weyl group and gauge transformations\label{appendix:gauge_Weyl}}
It is often stated but rarely shown in the physics literature that Weyl group elements are constant gauge transformations, here we will break down how that works. To begin, recall that the Weyl reflection with respect to $\vec{\alpha}\in\Delta$, which we denote $s_{\vec{\alpha}}$, acts on $\vec{v}\in\R^r$ as
\begin{equation*}
    s_{\vec{\alpha}}(\vec{v})=\vec{v}-\left(\vec{v}\cdot\vec{\alpha}^*\right)\vec{\alpha}.
\end{equation*}
The Weyl group, which we denote $W$, is then the group generated by all such Weyl reflections. In fact, one can show that the Weyl group is generated by Weyl reflections only with respect to the simple roots \cite{Humphreys_1990}. Thus each $\weyl{w}\in W$ can be written as a product of some number of simple Weyl reflections, say $\weyl{w}=\prod_{i=1}^n s_{\vec{\alpha}_{a_i}}$ for indices $a_i$. It is clear that there is no upper bound on $n$ since we can always multiply $\weyl{w}$ by $\identity=s_{\vec{\alpha}_b}^2$ for some $b$ to effectively extend $n$ by two. There is however a lower bound on $n$, which we call the length of $\weyl{w}$, where we define the length of the identity to be zero. Practically, one may compute the length of a given Weyl group element by counting how many positive roots it maps to negative roots \cite{Humphreys_1990}.

\paragraph{}
All we really need to do now is to show how to obtain a simple Weyl reflection as a gauge transformation. We will see that there are in fact multiple ways to obtain a Weyl reflection which act differently on non-Cartan degrees of freedom. To begin, we recall that for every positive root $\vec{\alpha}\in\Delta^+$ we may define two self-adjoint generators $T_1^{\vec{\alpha}}$ and $T_2^{\vec{\alpha}}$,
\begin{equation}
    T_1^{\vec{\alpha}}=\frac{1}{2}\left(E_{\vec{\alpha}}+E_{-\vec{\alpha}}\right),\qquad T_2^{\vec{\alpha}}=\frac{1}{2i}\left(E_{\vec{\alpha}}-E_{-\vec{\alpha}}\right),
\end{equation}
where $\{T_i^{\vec{\alpha}}\mid i=1,2,\ \vec{\alpha}\in\Delta^+\}$ spans all of $\lie{g}\setminus\lie{h}$. Note that $T_1^{\vec{\alpha}}$, $T_2^{\vec{\alpha}}$, and $\frac{1}{2}\vec{\alpha}^*\cdot\vec{H}$ form an $\su{2}$ algebra, taking for example $\tau_1^{\vec{\alpha}}=T_1^{\vec{\alpha}}$, $\tau_2^{\vec{\alpha}}=T_2^{\vec{\alpha}}$, and $\tau_3^{\vec{\alpha}}=\vec{\alpha}^*\cdot\vec{H}$ we get $\comm{\tau_i^{\vec{\alpha}}}{\tau_j^{\vec{\alpha}}}=i\varepsilon_{ijk}\tau_k^{\vec{\alpha}}$. Further, it is not too hard to show that $\left(\operatorname{ad}_{\tau_i^{\vec{\alpha}}}\right)^{2n}(\vec{H})=\frac{1}{2}\vec{\alpha}^*\left(\vec{\alpha}\cdot\vec{H}\right)$ for $n>0$, and $\left(\operatorname{ad}_{\tau_i^{\vec{\alpha}}}\right)^{2n+1}(\vec{H})=-i\vec{\alpha}\varepsilon_{ij}\tau_j^{\vec{\alpha}}$ for $n\geq 0$ with $i=1,2$ taking $\varepsilon_{ij}$ to be the two-dimensional Levi-Cevita symbol with $\varepsilon_{12}=+1$. Then, using the BCH formula we can evaluate the action of the 1-parameter family of constant gauge transformations $g_i(\xi)=e^{i\pi \xi \tau_i^{\vec{\alpha}}}$ on a Cartan configuration $\vec{a}\cdot\vec{H}$,
\begin{equation*}
    e^{i\pi\xi \tau_i^{\vec{\alpha}}}\vec{a}\cdot\vec{H}e^{-i\pi\xi \tau_i^{\vec{\alpha}}}=\vec{a}\cdot\vec{H}+\vec{a}\cdot\vec{\alpha}\tau_3^{\vec{\alpha}}\left(\cos(\pi \xi)-1\right)+\vec{a}\cdot\vec{\alpha}\varepsilon_{ij}\tau_j^{\vec{\alpha}}\sin(\pi \xi).
\end{equation*}
Taking $\xi=1$ (or $\xi=-1$), we see that $g_i=g_i(1)$ gives us the simple Weyl reflection acting on $\vec{a}$,
\begin{equation*}
    e^{i\pi\tau_i^{\vec{\alpha}}}\vec{a}\cdot\vec{H}e^{-i\pi\tau_i^{\vec{\alpha}}}=\left(\vec{a}-(\vec{a}\cdot\vec{\alpha}^*)\vec{\alpha}\right)\cdot\vec{H}=s_{\vec{\alpha}}(\vec{a})\cdot\vec{H}, \quad i=1,2.
\end{equation*}
We notice that there are at least two ways to do a simple Weyl reflection on the Cartan degrees of freedom, meaning that the map from the Weyl group to the gauge group is many-to-one. In fact, we could conjugate $g_i$ by any $U(1)^r$ element, say $e^{i\vec{v}\cdot\vec{H}}$, to get a new constant gauge transformation which acts in the same way on the Cartan elements. In this way, for each simple Weyl reflection there is a whole family of corresponding gauge transformations given by
\begin{equation*}
    g_{\vec{\alpha}}(\vec{v})=e^{i\vec{v}\cdot\vec{H}}g_1e^{-i\vec{v}\cdot\vec{H}}=e^{i\pi \cos\left(\vec{v}\cdot\vec{\alpha}\right)\tau_1^{\vec{\alpha}} - i\pi\sin\left(\vec{v}\cdot\vec{\alpha}\right)\tau_2^{\vec{\alpha}}},
\end{equation*}
where $g_1=g_{\vec{\alpha}}(0)$ and $g_2=g_{\vec{\alpha}}\left(\frac{\pi}{4}\vec{\alpha}^*\right)$. Note that we have essentially demonstrated the well-known result that the Weyl group is isomorphic to the quotient group $N\left(U(1)^r\right)/U(1)^r$, where $N\left(U(1)^r\right)=\left\{g\in G\mid gU(1)^rg^{-1}=U(1)^r\right\}$ is the normalizer of $U(1)^r$ \cite{Hall_Lie}.

\subsection{Two important Weyl group elements\label{appendix:Special_Weyl}}
There are two Weyl group elements that are especially important for our work, $\weyl{w}_\Pi$ and $\weyl{w}_{\Pi_c}$, so named because they are the Weyl group elements which map setwise $\Pi\rightarrow -\Pi$ and $\Pi\setminus\{\vec{\alpha}_c\}\rightarrow-\left(\Pi\setminus\{\vec{\alpha}_c\}\right)$, where $1\leq c\leq r$ is an index such that $\vec{\alpha}_c$ is a long root with $k_c^*=1$. The specific actions of $\weyl{w}_\Pi$ and $\weyl{w}_{\Pi_c}$ depend on the group in question, but in general they both permute the simple and affine roots, they both (almost) permute the fundamental weights, and the permutations preserve dual Kac labels and of course root lengths,
\begin{align}
    &\begin{dcases}
    \weyl{w}_\Pi(\vec{\alpha}_a)=-\vec{\alpha}_{\varpi(a)}\\
    \weyl{w}_\Pi(\vec{w}_a)=-\vec{w}_{\varpi(a)}\\
    \weyl{w}_\Pi(\vec{\rho})=-\vec{\rho}\\
    k_a=k_{\varpi(a)}^*\\
    \varpi\in S_{r+1},\ \varpi(0)=0, \varpi^2=\identity
    \end{dcases}\label{eqn:w_Pi}\\
    &\begin{dcases}
    \weyl{w}_{\Pi_c}(\vec{\alpha}_a)=-\vec{\alpha}_{\gamma_c(a)}\\
    \weyl{w}_{\Pi_c}(\vec{w}_a)=k_a^*\vec{w}_c-\vec{w}_{\gamma_c(a)}\\
    \weyl{w}_{\Pi_c}(\vec{\rho})=c_2\vec{w}_c-\vec{\rho}\\
    k_a=k_{\gamma_c(a)}^*\\
    \gamma_c\in S_{r+1},\ \gamma_c(0)=c,\ \gamma_c^2=\identity
    \end{dcases}\label{eqn:w_Pic},
\end{align}
where we use the convention that $\vec{w}_0=0$.

\paragraph{}
Practically we can determine $\weyl{w}_\Pi$ and $\weyl{w}_{\Pi_c}$ by using the fact that they are the longest Weyl group elements generated by all simple Weyl reflections, and all simple Weyl reflections except the one with respect to $\vec{\alpha}_c$ respectively. Then, given a set of simple Weyl reflections we want to use, all of the simple Weyl reflections for $\weyl{w}_\Pi$ and all of them except the one with respect to $\vec{\alpha}_c$ for $\weyl{w}_{\Pi_c}$, we can construct our element of interest by starting with a single Weyl reflection and iteratively adding more, making sure that each reflection we add increases the length of the element, until we can no longer make the element any longer. The specific permutations $\varpi$ and $\gamma_c$ are listed for each group in table \ref{tab:Weyl_longest}.

\begin{table}[h]
    \centering
    \begin{tabular}{lll}\hline
        Group & $\varpi$ & $\gamma_c$ \\ \hline
        $\SU{N}$ & $\varpi(a)=N-a$ & $\gamma_c(a)=c-a\bmod{N}$, $c=1,\dots,N-1$ \\
        $\Sp{N}$ & $\identity$ & $\gamma_{N}(a)=N-a$ \\
        \multirow{2}{*}{$\Spin{4n}$} & \multirow{2}{*}{$\identity$} & $\gamma_{2n-1}=(0,2n-1)(1,2n)\prod_{a=2}^{n-1}(a,2n-a)$\\
        & & $\gamma_{2n}=(0,2n)(1,2n-1)\prod_{a=2}^{n-1}(a,2n-a)$ \\
        \multirow{2}{*}{$\Spin{4n+2}$} & \multirow{2}{*}{$(2n,2n+1)$} & $\gamma_{2n}=(0,2n)(1,2n+1)\prod_{a=2}^{n}(a,2n+1-a)$\\
        & & $\gamma_{2n+1}=(0,2n+1)(1,2n)\prod_{a=2}^{n}(a,2n+1-a)$ \\
        $\Spin{2N+1}$ & $\identity$ & $\gamma_1=(0,1)$ \\
        \multirow{2}{*}{$\E{6}$} & \multirow{2}{*}{$(1,5)(2,4)$} & $\gamma_1=(0,1)(2,6)$ \\
        & & $\gamma_5=(0,5)(4,6)$ \\
        $\E{7}$ & $\identity$ & $\gamma_6=(0,6)(1,5)(2,4)$ \\
        $\E{8}$ & $\identity$ & - \\
        $\F$ & $\identity$ & - \\
        $\G$ & $\identity$ & - \\ \hline
    \end{tabular}
    \caption{Permutations corresponding to the special Weyl group elements $\weyl{w}_\Pi$ and $\weyl{w}_{\Pi_c}$ (where the $c^{th}$ (dual) Kac label is one), so that $\weyl{w}_\Pi(\vec{\alpha}_a)=-\vec{\alpha}_{\varpi(a)}$ and $\weyl{w}_{\Pi_c}(\vec{\alpha}_a)=-\vec{\alpha}_{\gamma_c(a)}$. Note that we use a standard notation for permutations, where $(a,b,c)(d)=(a,b,c)$ is the permutation where $a\rightarrow b\rightarrow c\rightarrow a$ and $d\rightarrow d$. See section \ref{sec:all_groups_symmetries} for labelling conventions for the roots.}
    \label{tab:Weyl_longest}
\end{table}

%% file: symmetries.tex
In this appendix we review the group-theoretic origins of the center and charge conjugation symmetries used in the text acting on Cartan degrees of freedom which are identified under $W\ltimes m\Lambda_r^*$, for some integer $m$. In addition, we will show how center symmetry acts in arbitrary irreducible representations of a group. This will prove useful for both SYM and non-abelian Chern-Simons theory, both of which have Cartan fields with a moduli space of the form above.

\subsection{Center symmetry\label{appendix:center}}
We will consider three different realizations of the center of a simply connected Lie group $G$: as matrices in irreducible representations, as a subgroup of the extended affine Weyl group, and as a subgroup of the Weyl group itself.

\subsubsection*{The center in an irreducible representation}
Let $R_{\vec{\lambda}}$ be an irreducible representation of $G$ with highest weight $\vec{\lambda}$, letting $u(\vec{\nu})$ be a weight vector associated to the weight $\vec{\nu}$. Recall that the elements of $Z(G)$ are of the form $e^{2\pi i\vec{\mu}^*\cdot\vec{H}}$ for $\vec{\mu}^*\in\Lambda_w^*$, and act on weights of $R_{\vec{\lambda}}$ as
\begin{equation*}
    R_{\vec{\lambda}}\left(e^{2\pi i\vec{\mu}^*\cdot\vec{H}}\right)u(\vec{\nu})=e^{2\pi i\vec{\mu}^*\cdot R_{\vec{\lambda}}(\vec{H})}u(\vec{\nu})=e^{2\pi i\vec{\mu}^*\cdot\vec{\nu}}u(\vec{\nu})=e^{2\pi i\vec{\mu}^*\cdot\vec{\lambda}}u(\vec{\nu}),
\end{equation*}
where in the last step we used the fact that weights of $R_{\vec{\lambda}}$ can differ from $\vec{\lambda}$ by at most roots, which would not contribute to the phase. It should also be clear here that $\vec{\mu}^*$ and $\vec{\mu}^*+\vec{\alpha}^*$ for any $\vec{\alpha}^*\in\Lambda_r^*$ produce the same center element, which leads to the well-known result that $Z(G)\cong \Lambda_w^*/\Lambda_r^*$ when $G$ is simply connected. 

\paragraph{}
When $Z(G)$ is cyclic, say $Z(G)=\Z_n$, there will be a co-weight $\vec{w}_c^*$\footnote{The specific value of $c$ depends on the group, see \ref{sec:all_groups_symmetries} for the details for each group.} such that $e^{2\pi i\vec{w}_c^*\cdot\vec{H}}$ generates $\Z_n$. We can then characterize $R_{\vec{\lambda}}$ by its $N$-\textit{ality}, defined as the ``charge" of $R_{\vec{\lambda}}$ under the center: if $R_{\vec{\lambda}}$ has $N$-ality $k$, then $R_{\vec{\lambda}}\left(e^{2\pi i\vec{w}_c^*\cdot\vec{H}}\right)=e^{2\pi ik/n}$. Starting with the action of the center on $u(\vec{\nu})$ given above, it is not too hard to show that the $N$-ality of $R_{\vec{\lambda}}$, which we write as a function of $\vec{\lambda}$, is given by
\begin{equation}
    N\text{-ality}(\vec{\lambda})=n\vec{w}_c^*\cdot\vec{\lambda}\bmod{n}.\label{eqn:N-ality}
\end{equation}
Note that all weights of $R_{\vec{\lambda}}$ will have the same $N$-ality, calculated with respect to equation \eqref{eqn:N-ality}, as $\vec{\lambda}$.

\subsubsection*{The center as a subgroup of the (extended affine) Weyl group}
We consider a generalized extended affine Weyl group $W\ltimes m\Lambda_w^*$, for some positive integer $m$ acting on a Cartan field $\vec{\eta}\in\R^r/(W\ltimes m\Lambda_r^*)$, defined within the fundamental domain
\begin{equation}
    \hat{T}=\left\{\vec{v}\in\R^r\mid \vec{\alpha}_a^*\cdot\vec{v}\geq 0,\ a=1,\dots,r,\ -\vec{\alpha}_0^*\cdot\vec{v}<m\right\}.\label{eqn:fundamental_affineWeyl}
\end{equation}
Since $\Lambda_r^*\subseteq \Lambda_w^*$, we have $W\ltimes m\Lambda_r^*$ as a subgroup of $W\ltimes m\Lambda_w^*$. One can show that the subgroup of $W\ltimes m\Lambda_w^*$ which preserves $\hat{T}$ is isomorphic to the center of $G$ \cite{Iwahori1965},
\begin{equation*}
    Z(G)\cong \left\{\identity\right\}\cup\left\{\left(\weyl{w}_{\Pi_c}\circ \weyl{w}_\Pi,m\vec{w}_c^*\right)\mid -\vec{w}_c^*\cdot\vec{\alpha}_0=1\right\}\equiv \mathcal{Z}(G),
\end{equation*}
which defines an action on $\vec{\eta}$, which we denote $\mathcal{T}_{c,m}$,
\begin{equation}
    Z(G): \vec{\eta}\rightarrow \mathcal{T}_{c,m}(\vec{\eta})=\weyl{w}_{\Pi_c}\circ\weyl{w}_\Pi(\vec{\eta}) + m\vec{w}_c^*.\label{eqn:center_Weyl_action}
\end{equation}
Thus, associated to each non-trivial $g\in Z(G)$ is an integer $1\leq c\leq r$ with $-\vec{w}_c^*\cdot\vec{\alpha}_0=1$, or equivalently with $k_c=1$. In almost all cases, except $\Spin{4n}$ which is easily accommodated, $Z(G)$ will be a cyclic group, say $\Z_q$, and thus it is enough to find an element of order $q$ in $\mathcal{Z}(G)$ to act as a generator of $\Z_q$, remembering that the group operation in $W\ltimes m\Lambda_w^*$ is given by
\begin{equation*}
    \left(\weyl{w}_1,\vec{\mu}_1\right)\cdot \left(\weyl{w}_2,\vec{\mu}_2\right) = \left(\weyl{w}_1\circ\weyl{w}_2,\ \vec{\mu}_1+\weyl{w}_1(\vec{\mu}_2)\right).
\end{equation*}
Alternatively, we can use the fact that $Z(G)$ is isomorphic to it's image in the defining representation $R_{\vec{\lambda}_D}(Z(G))$, where $\vec{\lambda}_D$ is the highest weight of the defining representation, to construct an isomorphism mapping $(\weyl{w}_{\Pi_c}\circ\weyl{w}_\Pi,m\vec{w}_c^*)$ to $e^{2\pi i\vec{w}_c^*\cdot\vec{\lambda}_D}$. This isomorphism easily enables us to find the order of $(\weyl{w}_{\Pi_c}\circ\weyl{w}_\Pi,m\vec{w}_c^*)$ simply by computing the inner product $\vec{w}_c^*\cdot\vec{\lambda}_D$.

\paragraph{}
Now, let $\vec{\eta}=\sum_{a=1}^r\eta_a\vec{w}_a$ and define $\eta_0$ such that $m=\sum_{a=0}^rk_a^*\eta_a$. Following from equations \eqref{eqn:w_Pi} and \eqref{eqn:w_Pic}, the action of $Z(G)$ on $\vec{\eta}$ is
\begin{align*}
    \mathcal{T}_{c,m}(\vec{\eta})&=\sum_{a=1}^r\eta_a\left(\vec{w}_{\gamma_c\circ\varpi(a)}-k_a^*\vec{w}_c\right)+m\vec{w}_c=\sum_{a=1}^r\eta_a\vec{w}_{\gamma_c\circ\varpi(a)}+\left(m-\sum_{a=1}^rk_a^*\eta_a\right)\vec{w}_c\\
    &=\sum_{a=1}^r\eta_a\vec{w}_{\gamma_c\circ\varpi(a)}+\eta_0\vec{w}_c.
\end{align*}
Notice that not all of $\vec{w}_{\gamma_c\circ\varpi(a)}$ are necessarily non-trivial, in particular there could be an $a$ for which $\gamma_c\circ\varpi(a)=0$. Letting $\pi_c=\gamma_c\circ\varpi$, using the fact that $\vec{w}_0=0$ we find $\sum_{a=0}^r\eta_a\vec{w}_{\pi_c(a)}=\sum_{a=0}^r\eta_{\pi_c^{-1}(a)}\vec{w}_a=\sum_{a=1}^r\eta_{\pi_c^{-1}(a)}\vec{w}_a$, or in other words $\sum_{a=1}^r\eta_a\vec{w}_{\pi_c(a)}=\sum_{a=1}^r\eta_{\pi_c^{-1}(a)}\vec{w}_a-\eta_0\vec{w}_{\pi_c(0)}$. Using $\pi_c(0)=c$ we find
\begin{equation*}
    \mathcal{T}_{c,m}(\vec{\eta})=\sum_{a=1}^r\eta_{\pi_c^{-1}(a)}\vec{w}_a.
\end{equation*}
In other words, the action of $\mathcal{T}_{c,m}$ on the labels $\eta_a$ is
\begin{equation*}
    \mathcal{T}_{c,m}: \eta_a\rightarrow \eta_{\varpi\circ\gamma_c(a)},
\end{equation*}
where we substituted $\pi_c^{-1}$ in terms of $\varpi$ and $\gamma_c$. Notice that $m$ does not appear explicitly in the action on the labels, but implicitly through the constraint $\sum_{a=0}^rk_a^*\eta_a=m$. Thus, we can drop the subscript $m$ on $\mathcal{T}_{c,m}$ so long as we remember that the constraint must be enforced. In practice, we will often drop the $c$ subscript as well after fixing a $c$ such that $\mathcal{T}_{c,m}$ generates $\mathcal{Z}(G)$.

\paragraph{}
We can then consider fluctuations about center symmetric points, letting $\vec{\eta}=\vec{\eta}_0+\vec{\delta}$, where $\vec{\eta}_0$ is invariant under the action of the center defined above\footnote{Note that we can always find such an invariant point, one of which is the vacuum of the scalar holonomy in SYM. See \cite{Argyres2012} for explicit expressions for the center invariant points for all the relevant groups.}. We see that the resulting action of the center on $\vec{\delta}$ is just the Weyl group element $\weyl{w}_{\Pi_c}\circ\weyl{w}_\Pi$,
\begin{equation*}
    \weyl{w}_{\Pi_c}\circ\weyl{w}_\Pi(\vec{\eta}_0+\vec{\delta})+m\vec{w}_c^*=\left[\weyl{w}_{\Pi_c}\circ\weyl{w}_\Pi(\vec{\eta}_0)+m\vec{w}_c^*\right]+\weyl{w}_{\Pi_c}\circ\weyl{w}_\Pi(\vec{\delta})=\vec{\eta}_0+\weyl{w}_{\Pi_c}\circ\weyl{w}_\Pi(\vec{\delta}).
\end{equation*}
Thus, the center of $G$ is also isomorphic to just the Weyl group part of $\mathcal{Z}(G)$,
\begin{equation*}
    Z(G)\cong \left\{\identity\right\}\cup\left\{\weyl{w}_{\Pi_c}\circ \weyl{w}_\Pi\mid -\vec{w}_c^*\cdot\vec{\alpha}_0=1\right\}.
\end{equation*}
Note that the above action of the center could be obtained from $\mathcal{Z}(G)$ by simply setting $m=0$.

\subsubsection*{An example: $\boldsymbol{\SU{N}}$}
The center of $\SU{N}$ is $\Z_N$, and each Kac label is one, so every $c$ from $1$ to $N-1$ appears in $\mathcal{Z}(\SU{N})$. By either brute force, or using the fact that $\weyl{w}_{\Pi_c}$ must permute the nodes of the extended Dynkin diagram in a way which preserves the diagram and swaps the $0^{th}$ and $c^{th}$ nodes, we find that $\weyl{w}_{\Pi_c}$ acts as
\begin{equation*}
    \weyl{w}_{\Pi_c}(\vec{\alpha}_a)=-\vec{\alpha}_{c-a\bmod{N}}.
\end{equation*}
Similarly, we find that $\weyl{w}_\Pi$ acts as
\begin{equation*}
    \weyl{w}_\Pi(\vec{\alpha}_a)=-\vec{\alpha}_{N-a\bmod{N}}.
\end{equation*}
Thus, the combined action $\weyl{w}_{\Pi_c}\circ\weyl{w}_\Pi$ is given by
\begin{equation*}
    \weyl{w}_{\Pi_c}\circ\weyl{w}_\Pi(\vec{\alpha}_a)=\vec{\alpha}_{a+c\bmod{N}}
\end{equation*}
Thus, we see that for any $N$ we can either take $\weyl{w}_{\Pi_{N-1}}\circ\weyl{w}_\Pi$ or $\weyl{w}_{\Pi_{1}}\circ\weyl{w}_\Pi$ to generate $\Z_N$.

\paragraph{}
The defining representation of $\SU{N}$, often called the fundamental representation, has highest weight $\vec{w}_1$. To determine the $\Z_N$ generator we just need to find the center element $e^{2\pi i\vec{w}_c^*\cdot\vec{H}}$ which has order $N$ in the fundamental representation. In other words we need to find $c$ such that $N\vec{w}_c^*\cdot\vec{w}_1\equiv 1\bmod{N}$. For $\SU{N}$ the inner product between co-weights and weights is given by
\begin{equation*}
    \vec{w}_a^*\cdot\vec{w}_b=\frac{\min(a,b)(N-\max(a,b))}{N},
\end{equation*}
so we see that taking $c=1$ or $c=N$ gives $N\vec{w}_c^*\cdot\vec{w}_1\equiv 1\bmod{N}$, and thus we can say that $e^{2\pi i\vec{w}_c^*\cdot\vec{H}}$ for both $c=1,N-1$ generate the $\Z_N$ center of $\SU{N}$.

\subsection{Charge conjugation\label{appendix:charge_conjugation}}
Typically, charge conjugation is taken to act on gauge fields by complex conjugation and reversing the sign: $A\rightarrow -A^*$. For a basis of self-adjoint gauge group generators, $T^a=(T^a)^\dagger$, we have $A=A^aT^a$, where $A^a$ are real 1-forms. Then, we can take charge conjugation to act completely on the generators as $T^a\rightarrow -(T^a)^T$. We write $A$ in terms of Cartan generators and root vectors,
\begin{equation*}
    A=\vec{A}\cdot\vec{H}+\sum_{\vec{\alpha}\in\Delta}A^{\vec{\alpha}} E_{\vec{\alpha}},
\end{equation*}
where in the defining representation the Cartan generators are taken to be real and symmetric, and the root vectors are taken to be real so that $(E_{\vec{\alpha}})^T=E_{-\vec{\alpha}}$, requiring $(A^{\vec{\alpha}})^*=A^{-\vec{\alpha}}$. Acting with charge conjugation we see that $\vec{H}\rightarrow -\vec{H}$ and $E_{\vec{\alpha}}\rightarrow -E_{-\vec{\alpha}}$, or equivalently $\vec{A}\rightarrow -\vec{A}$ and $A^{\vec{\alpha}}\rightarrow -A^{-\vec{\alpha}}$. The field strength then transforms in the same way. We can then consider a Cartan field like we did for the center, $\vec{\eta}\in\R^r/(W\ltimes m\Lambda_r^*)$, defined within $\hat{T}$ \eqref{eqn:fundamental_affineWeyl}. We can think of this Cartan field as coming from the gauge field, like the holonomy or dual photon in SYM, so we take it to transform under charge conjugation in the same way, $\vec{\eta}\rightarrow-\vec{\eta}$. We see however that such a transformation does not preserve $\hat{T}$, but if supplemented by the Weyl transformation $\weyl{w}_\Pi$, which in the gauge field picture is just a constant gauge transformation, the action of charge conjugation does preserve $\hat{T}$,
\begin{equation*}
    \mathcal{C}: \vec{\eta}\rightarrow-\weyl{w}_\Pi(\vec{\eta}).
\end{equation*}
We note that this definition of charge conjugation also corresponds to the notion of charge conjugation in representation theory. If we have an irrep $R_{\vec{\lambda}}$, with highest weight $\vec{\lambda}$, then the conjugate representation, obtained by complex conjugation from $R_{\vec{\lambda}}$, has highest weight $-\weyl{w}_\Pi(\vec{\lambda})$.

\subsubsection{The special case of $\Spin{4n}$\label{sec:Spin4n_CC}}
For $\Spin{4n}$ we find that $-\weyl{w}_\Pi$ is trivial, yet we can still define a non-trivial ``charge conjugation" symmetry as the group of symmetries of the Dynkin diagram which preserves the affine node\footnote{This is also true for the other groups with charge conjugation, except in those cases $-\weyl{w}_\Pi$ is non-trivial and indeed does generate the charge conjugation symmetry.}. For $n>2$, the charge conjugation symmetry is $\Z_2^{(0)}$ and acts on the Cartan degrees of freedom by swapping $\vec{\alpha}_+\leftrightarrow \vec{\alpha}_-$, and on the non-Cartan by swapping $E_{\vec{\alpha}_+}$ and $E_{\vec{\alpha}_-}$. In the $\SO{4n}$ representation, this is done by a diagonal matrix with determinant -1. For $n=2$, the charge conjugation symmetry is $S_3$, the group of permutations on $\{\vec{\alpha}_1,\vec{\alpha}_-,\vec{\alpha}_+\}$.

\subsection{All groups\label{sec:all_groups_symmetries}}
\subsubsection*{$\boldsymbol{\SU{N}}$}
The extended Dynkin diagram for $\SU{N}$ with our labelling conventions is
\begin{equation*}
    \dynkin[label macro/.code={\vec{\alpha}_{\drlap{#1}}},edge length=1cm,labels={0,1,2,N-2,N-1},labels*={1,1,1,1,1}]{A}[1]{},
\end{equation*}
where all the dual Kac labels are one as indicated above the nodes. The inner product between weights and co-weights is
\begin{equation*}
    \vec{w}_a^*\cdot\vec{w}_b=\frac{\min(a,b)\left(N-\max(a,b)\right)}{N}.
\end{equation*}
From table \ref{tab:Weyl_longest}, and discussed above, we can take the $\Z_N$ center to be generated by either $c=N-1$ or $c=1$, but here we will take $c=N-1$, where $\gamma_{N-1}(a)=N-1-a$. The $N$-ality of the irreducible representation $R_{\vec{\lambda}}$ with $\vec{\lambda}=\sum_{a=1}^{N-1}\lambda_a\vec{w}_a$ is
\begin{equation*}
    N\text{-ality}(\vec{\lambda})=N\vec{w}_{N-1}^*\cdot\vec{\lambda}\bmod{N}\equiv \sum_{a=1}^{N-1}a\lambda_a\bmod{N}.
\end{equation*}
Consider now an arbitrary $\vec{\mu}=\sum_{a=1}^{N-1}\mu_a\vec{w}_a$, with $\sum_{a=0}^{N-1}\mu_a=m$. The action of $\mathcal{T}=\mathcal{T}_{N-1,m}$ on $\vec{\mu}$ is
\begin{equation*}
    \mathcal{T}(\vec{\mu})=\weyl{w}_{\Pi_{N-1}}\circ\weyl{w}_\Pi(\vec{\mu})+m\vec{w}_{N-1}^*=\sum_{a=1}^{N-2}\mu_{a+1}\vec{w}_a+\mu_0\vec{w}_{N-1},
\end{equation*}
or in other words,
\begin{equation*}
    \mathcal{T}: \mu_a\rightarrow \mu_{a+1\bmod{N}}.
\end{equation*}
From table \ref{tab:Weyl_longest}, we see that $\mathcal{C}$ acts on $\vec{\mu}$ as
\begin{equation*}
    \mathcal{C}(\vec{\mu})=\sum_{a=1}^{N-1}\mu_a\vec{w}_{N-a}=\sum_{a=1}^{N-1}\mu_{N-a}\vec{w}_a,
\end{equation*}
or equivalently,
\begin{equation*}
    \mathcal{C}: \mu_a\rightarrow \mu_{N-a\bmod{N}}.
\end{equation*}

\subsubsection*{$\boldsymbol{\Sp{N}}$}
The extended Dynkin diagram for $\Sp{N}$ with our labelling conventions is
\begin{equation*}
    \dynkin[label macro/.code={\vec{\alpha}_{\drlap{#1}}},edge length=1cm,labels={0,1,2,N-2,N-1,N},labels*={1,1,1,1,1,1}]{C}[1]{},
\end{equation*}
where all the dual Kac labels are one as indicated above the nodes. The inner product between weights and co-weights is
\begin{equation*}
    \vec{w}_a^*\cdot\vec{w}_b=\begin{dcases}
        \min(a,b) & a < N \\ \frac{b}{2} & a=N
    \end{dcases}.
\end{equation*}
The $\Z_2$ center is generated with $c=N$. The $N$-ality of the irreducible representation $R_{\vec{\lambda}}$ with $\vec{\lambda}=\sum_{a=1}^{N}\lambda_a\vec{w}_a$ is
\begin{equation*}
    N\text{-ality}(\vec{\lambda})=2\vec{w}_{N}^*\cdot\vec{\lambda}\bmod{2}\equiv \sum_{a=0}^{\left\lfloor(N-1)/2\right\rfloor}\lambda_{2a+1}\bmod{2}.
\end{equation*}
Referring to table \ref{tab:Weyl_longest}, the action of $\mathcal{T}=\mathcal{T}_{N,m}$ on $\vec{\mu}=\sum_{a=1}^{N}\mu_a\vec{w}_a$ is
\begin{equation*}
    \mathcal{T}(\vec{\mu})=\weyl{w}_{\Pi_N}\circ\weyl{w}_\Pi(\vec{\mu})+m\vec{w}_N^*=\sum_{a=1}^{N}\mu_{a}\vec{w}_{N-a},
\end{equation*}
or
\begin{equation*}
    \mathcal{T}: \mu_a\rightarrow \mu_{N-a}.
\end{equation*}
From table \ref{tab:Weyl_longest}, the action of $\mathcal{C}$ is trivial.

\subsubsection*{$\boldsymbol{\Spin{2N}}$}
The extended Dynkin diagram for $\Spin{2N}$ with our labelling conventions and corresponding dual Kac lables is
\begin{equation*}
    \dynkin[label macro/.code={\vec{\alpha}_{\drlap{#1}}},edge length=1cm,labels={0,1,2,3,N-3,N-2,-,+},labels*={1,1,2,2,2,2,1,1},label* directions={right,right,above right,above,above,above left,left,left},label directions={left,left,below right,below,below,right,right,right}]{D}[1]{}.
\end{equation*}
The inner product between weights and co-weights is
\begin{equation*}
    \vec{w}_a^*\cdot\vec{w}_b=\begin{dcases}
        \min(a,b) & a,b\leq N-2 \\
        \frac{a}{2} & a\leq N-2 < b\\
        \frac{b}{2} & b\leq N-2 < a\\
        \frac{N}{4}-\frac{\abs{b-a}}{2} & a,b\geq N-1
    \end{dcases}.
\end{equation*}
For convenience, we take $\vec{w}_-=\vec{w}_{N-1}$ and $\vec{w}_+=\vec{w}_N$.

\paragraph{$\boldsymbol{N}$ odd}
The $\Z_4$ center symmetry is generated by both $c=\pm$. The $N$-ality of the irreducible representation $R_{\vec{\lambda}}$ with $\vec{\lambda}=\sum_{a=1}^{N}\lambda_a\vec{w}_a$ generated by $c=\pm$ is
\begin{equation*}
    N\text{-ality}_{c=\pm}(\vec{\lambda})=4\vec{w}_{\pm}^*\cdot\vec{\lambda}\bmod{4}\equiv 2\sum_{a=0}^{(N-3)/2}\lambda_{2a+1}+N\lambda_\pm+(N-2)\lambda_\mp\bmod{4}.
\end{equation*}
From table \ref{tab:Weyl_longest}, the action of $\mathcal{T}_\pm=\mathcal{T}_{\pm,m}$ on $\vec{\mu}=\sum_{a=1}^N\mu_a\vec{w}_a$ is
\begin{equation*}
    \mathcal{T}_\pm(\vec{\mu})=\weyl{w}_{\Pi_\pm}\circ\weyl{w}_\Pi(\vec{\mu})+m\vec{w}_\pm^*=\mu_1\vec{w}_\pm+\mu_\mp\vec{w}_1+\mu_0\vec{w}_\pm+\sum_{a=2}^{N-2}\mu_{a}\vec{w}_{N-a},
\end{equation*}
or
\begin{equation*}
    \mathcal{T}_\pm: \begin{dcases}
        \mu_0\rightarrow \mu_\pm\rightarrow \mu_1\rightarrow \mu_\mp & \\
        \mu_a \rightarrow \mu_{N-a} & 2\leq a\leq N-2
    \end{dcases}.
\end{equation*}
From table \ref{tab:Weyl_longest}, we see that $\mathcal{C}$ acts on $\vec{\mu}$ as
\begin{equation*}
    \mathcal{C}(\vec{\mu})=\sum_{a=1}^{N-2}\mu_a\vec{w}_{a}+\mu_-\vec{w}_++\mu_+\vec{w}_-,
\end{equation*}
or equivalently,
\begin{equation*}
    \mathcal{C}: \mu_+\leftrightarrow \mu_-.
\end{equation*}

\paragraph{$\boldsymbol{N}$ even}
The $\Z_2^\pm$ factor of the $\Z_2^+\times\Z_2^-$ center symmetry is generated by $c=\pm$. The $N$-ality with respect to $\Z_2^\pm$ of the irreducible representation $R_{\vec{\lambda}}$ with $\vec{\lambda}=\sum_{a=1}^{N}\lambda_a\vec{w}_a$ is
\begin{equation*}
    N\text{-ality}_\pm(\vec{\lambda})=2\vec{w}_{\pm}^*\cdot\vec{\lambda}\bmod{2}\equiv \sum_{a=0}^{(N-4)/2}\lambda_{2a+1}+\frac{N}{2}\lambda_\pm+\frac{N-2}{2}\lambda_\mp\bmod{2}.
\end{equation*}
From table \ref{tab:Weyl_longest}, the action of $\mathcal{T}_\pm=\mathcal{T}_{\pm,m}$ on $\vec{\mu}=\sum_{a=1}^N\mu_a\vec{w}_a$ is
\begin{equation*}
    \mathcal{T}_\pm(\vec{\mu})=\weyl{w}_{\Pi_\pm}\circ\weyl{w}_\Pi(\vec{\mu})+m\vec{w}_\pm^*=\mu_1\vec{w}_\mp + \mu_\mp\vec{w}_1+\mu_0\vec{w}_\pm + \sum_{a=2}^{N-2}\mu_a\vec{w}_{N-a},
\end{equation*}
or
\begin{equation*}
    \mathcal{T}_\pm:\begin{dcases}
        \mu_0\leftrightarrow \mu_\pm & \\
        \mu_1\leftrightarrow \mu_\mp & \\
        \mu_a\leftrightarrow \mu_{N-a} & 2\leq a\leq N-2
    \end{dcases}.
\end{equation*}
As discussed above, $\mathcal{C}$ is trivial, but a charge conjugation operation may still be defined. See section \ref{sec:Spin4n_CC} for specific details.

\subsubsection*{$\boldsymbol{\Spin{2N+1}}$}
The extended Dynkin diagram for $\Spin{2N+1}$ with our labelling conventions and corresponding dual Kac labels is
\begin{equation*}
    \dynkin[label macro/.code={\vec{\alpha}_{\drlap{#1}}},edge length=1cm,labels={0,1,2,3,N-2,N-1,N},labels*={1,1,2,2,2,2,1},label* directions={right,right,,,,,}]{B}[1]{}.
\end{equation*}
The inner product between weights and co-weights is
\begin{equation*}
    \vec{w}_a^*\cdot\vec{w}_b=\begin{dcases}
        \min(a,b) & b<N \\
        \frac{a}{2} & b=N
    \end{dcases}.
\end{equation*}
The $\Z_2$ center symmetry is generated by $c=1$. The $N$-ality of the irreducible representation $R_{\vec{\lambda}}$ with $\vec{\lambda}=\sum_{a=1}^{N}\lambda_a\vec{w}_a$ is
\begin{equation*}
    N\text{-ality}(\vec{\lambda})=2\vec{w}_{1}^*\cdot\vec{\lambda}\bmod{2}\equiv \lambda_N\bmod{2}.
\end{equation*}
Referring to table \ref{tab:Weyl_longest}, the action of $\mathcal{T}=\mathcal{T}_{1,m}$ on $\vec{\mu}=\sum_{a=1}^N\mu_a\vec{w}_a$ is
\begin{equation*}
    \mathcal{T}(\vec{\mu})=\weyl{w}_{\Pi_1}\circ\weyl{w}_\Pi(\vec{\mu})+m\vec{w}_1^*=\mu_0\vec{w}_1+\sum_{a=2}^N\mu_a\vec{w}_a,
\end{equation*}
or
\begin{equation*}
    \mathcal{T}:\mu_0\leftrightarrow \mu_1.
\end{equation*}
From table \ref{tab:Weyl_longest}, the action of $\mathcal{C}$ is trivial.

\subsubsection*{$\boldsymbol{\E{6}}$}
The extended Dynkin diagram for $\E{6}$ with our labelling conventions and corresponding dual Kac labels is
\begin{equation*}
    \dynkin[label macro/.code={\vec{\alpha}_{\drlap{#1}}},extended,label,ordering=Dynkin,labels*={1,1,2,3,2,1,2},label* directions={left,,,,,,left}] E6.
\end{equation*}
The inner product between weights and co-weights, written as a matrix, is
\begin{equation*}
    [\vec{w}_a^*\cdot\vec{w}_b]=\begin{pmatrix}4/3 & 5/3 & 2 & 4/3 & 2/3 & 1 \\
 5/3 & 10/3 & 4 & 8/3 & 4/3 & 2 \\
 2 & 4 & 6 & 4 & 2 & 3 \\
 4/3 & 8/3 & 4 & 10/3 & 5/3 & 2 \\
 2/3 & 4/3 & 2 & 5/3 & 4/3 & 1 \\
 1 & 2 & 3 & 2 & 1 & 2 \end{pmatrix}_{a,b}.
\end{equation*}
The $\Z_3$ center symmetry is generated by $c=1$ or $c=5$, but here we take $c=1$. The $N$-ality of the irreducible representation $R_{\vec{\lambda}}$ with $\vec{\lambda}=\sum_{a=1}^{6}\lambda_a\vec{w}_a$ is
\begin{equation*}
    N\text{-ality}(\vec{\lambda})=3\vec{w}_{1}^*\cdot\vec{\lambda}\bmod{3}\equiv \lambda_1+\lambda_4+2(\lambda_2+\lambda_5)\bmod{3}.
\end{equation*}
Referring to table \ref{tab:Weyl_longest}, the action of $\mathcal{T}=\mathcal{T}_{1,m}$ on $\vec{\mu}=\sum_{a=1}^6\mu_a\vec{w}_a$ is
\begin{equation*}
    \mathcal{T}(\vec{\mu})=\weyl{w}_{\Pi_1}\circ\weyl{w}_\Pi(\vec{\mu})+m\vec{w}_1^*=\mu_0\vec{w}_1+\mu_6\vec{w}_2+\mu_3\vec{w}_3 + \mu_2\vec{w}_4 + \mu_1\vec{w}_5 + \mu_4\vec{w}_6
\end{equation*}
or
\begin{equation*}
    \mathcal{T}:\begin{dcases}
        \mu_0\rightarrow \mu_5\rightarrow \mu_1\rightarrow \mu_0 \\
        \mu_2\rightarrow \mu_6\rightarrow \mu_4\rightarrow \mu_2
    \end{dcases}.
\end{equation*}
From table \ref{tab:Weyl_longest}, the action of $\mathcal{C}$ is
\begin{equation*}
    \mathcal{C}(\vec{\mu})=\mu_1\vec{w}_5+\mu_2\vec{w}_4+\mu_3\vec{w}_3+\mu_4\vec{w}_2+\mu_5\vec{w}_1+\mu_6\vec{w}_6,
\end{equation*}
or
\begin{equation*}
    \mathcal{C}: \begin{dcases}
        \mu_1\leftrightarrow \mu_5 \\
        \mu_2\leftrightarrow \mu_4
    \end{dcases}.
\end{equation*}

\subsubsection*{$\boldsymbol{\E{7}}$}
The extended Dynkin diagram for $\E{7}$ with our labelling conventions and corresponding dual Kac labels is
\begin{equation*}
    \dynkin[label macro/.code={\vec{\alpha}_{\drlap{#1}}},extended,label,ordering=Kac,labels*={1,2,3,4,3,2,1,2},label* directions={left,,,,,,,left}] E7.
\end{equation*}
The inner product between weights and co-weights, written as a matrix, is
\begin{equation*}
    [\vec{w}_a^*\cdot\vec{w}_b]=\frac{1}{2}\begin{pmatrix}
 4 & 6 & 8 & 6 & 4 & 2 & 4 \\
 6 & 12 & 16 & 12 & 8 & 4 & 8 \\
 8 & 16 & 24 & 18 & 12 & 6 & 12 \\
 6 & 12 & 18 & 15 & 10 & 5 & 9 \\
 4 & 8 & 12 & 10 & 8 & 4 & 6 \\
 2 & 4 & 6 & 5 & 4 & 3 & 3 \\
 4 & 8 & 12 & 9 & 6 & 3 & 7 \end{pmatrix}_{a,b}.
\end{equation*}
The $\Z_2$ center symmetry is generated by $c=6$. The $N$-ality of the irreducible representation $R_{\vec{\lambda}}$ with $\vec{\lambda}=\sum_{a=1}^{7}\lambda_a\vec{w}_a$ is
\begin{equation*}
    N\text{-ality}(\vec{\lambda})=2\vec{w}_{6}^*\cdot\vec{\lambda}\bmod{2}\equiv \lambda_4+\lambda_6+\lambda_7\bmod{2}.
\end{equation*}
Referring to table \ref{tab:Weyl_longest}, the action of $\mathcal{T}=\mathcal{T}_{6,m}$ on $\vec{\mu}=\sum_{a=1}^7\mu_a\vec{w}_a$ is
\begin{equation*}
    \mathcal{T}(\vec{\mu})=\weyl{w}_{\Pi_6}\circ\weyl{w}_\Pi(\vec{\mu})+m\vec{w}_6^*=\sum_{a=1}^6\mu_{6-a}\vec{w}_a+\mu_7\vec{w}_7
\end{equation*}
or
\begin{equation*}
    \mathcal{T}:\mu_a\leftrightarrow \mu_{6-a},\quad 0\leq a\leq 6.
\end{equation*}
From table \ref{tab:Weyl_longest}, the action of $\mathcal{C}$ is trivial.

\subsubsection*{$\boldsymbol{\E{8}}$}
The extended Dynkin diagram for $\E{8}$ with our labelling conventions and corresponding dual Kac labels is
\begin{equation*}
    \dynkin[label macro/.code={\vec{\alpha}_{\drlap{#1}}},extended,label,ordering=Kac,labels*={1,2,3,4,5,6,4,2,3},label* directions={left,,,,,,,left}] E8.
\end{equation*}
The inner product between weights and co-weights, written as a matrix, is
\begin{equation*}
    [\vec{w}_a^*\cdot\vec{w}_b]=\begin{pmatrix}
    2 & 3 & 4 & 5 & 6 & 4 & 2 & 3 \\
 3 & 6 & 8 & 10 & 12 & 8 & 4 & 6 \\
 4 & 8 & 12 & 15 & 18 & 12 & 6 & 9 \\
 5 & 10 & 15 & 20 & 24 & 16 & 8 & 12 \\
 6 & 12 & 18 & 24 & 30 & 20 & 10 & 15 \\
 4 & 8 & 12 & 16 & 20 & 14 & 7 & 10 \\
 2 & 4 & 6 & 8 & 10 & 7 & 4 & 5 \\
 3 & 6 & 9 & 12 & 15 & 10 & 5 & 8\end{pmatrix}_{a,b}.
\end{equation*}
There is no center symmetry, and from table \ref{tab:Weyl_longest} $\mathcal{C}$ is trivial.

\subsubsection*{$\boldsymbol{\F}$}
The extended Dynkin diagram for $\F$ with our labelling conventions and corresponding dual Kac labels is
\begin{equation*}
    \dynkin[label macro/.code={\vec{\alpha}_{\drlap{#1}}},extended,label,labels*={1,2,3,2,1},label* directions={left,,,,,,left}] F4.
\end{equation*}
The inner product between weights and co-weights, written as a matrix, is
\begin{equation*}
    [\vec{w}_a^*\cdot\vec{w}_b]=\begin{pmatrix}
    2 & 3 & 2 & 1 \\
    3 & 6 & 4 & 2 \\
    4 & 8 & 6 & 3 \\
    2 & 4 & 3 & 2
    \end{pmatrix}_{a,b}.
\end{equation*}
There is no center symmetry, and from table \ref{tab:Weyl_longest} $\mathcal{C}$ is trivial.

\subsubsection*{$\boldsymbol{\G}$}
The extended Dynkin diagram for $\G$ with our labelling conventions and corresponding dual Kac labels is
\begin{equation*}
    \dynkin[label macro/.code={\vec{\alpha}_{\drlap{#1}}},extended,label,labels*={1,2,1},label* directions={left,,,,,,left}] G2.
\end{equation*}
The inner product between weights and co-weights, written as a matrix, is
\begin{equation*}
    [\vec{w}_a^*\cdot\vec{w}_b]=\begin{pmatrix}2 & 1 \\ 3 & 2\end{pmatrix}_{a,b}.
\end{equation*}
There is no center symmetry, and from table \ref{tab:Weyl_longest} $\mathcal{C}$ is trivial.

%% file: unitaryCS.tex
Recall from equation \eqref{eqn:cc_unitary_cs} how charge conjugation acts on $U(u)_{N-u,N}$ states, and that if a state $\ket{[\vec{\lambda},\xi]}$ is charge conjugation invariant then $\mathcal{C}(\vec{\lambda})$ must be in the $\Z_u$ orbit of $\vec{\lambda}$ under $\mathcal{T}$. In other words, there is an integer $0\leq m<l$ such that $\mathcal{C}(\vec{\lambda})=\mathcal{T}^m(\vec{\lambda})$, where $l$ is the size of the $\mathcal{T}$ orbit of $\vec{\lambda}$. Here we will prove the statement made in the text that if $\ket{[\vec{\lambda},\xi]}$ is invariant under charge conjugation, then $m$ cannot be odd when $l$ is even. Taking $l$, and hence $u$, to be even, assume towards contradiction that $m$ is odd. We then apply $\mathcal{T}^{(m-1)/2}$ to $\vec{\lambda}$ to get a weight which is invariant under $\mathcal{C}$ up to one application of $\mathcal{T}$,
\begin{equation*}
    \mathcal{C}\left(\mathcal{T}^{(m-1)/2}(\vec{\lambda})\right)=\mathcal{T}^{-(m-1)/2}\circ\mathcal{C}(\vec{\lambda})=\mathcal{T}^{m/2+1}(\vec{\lambda}).
\end{equation*}
We may then assume without loss of generality that $m=1$, so that $\mathcal{C}(\vec{\lambda})=\mathcal{T}(\vec{\lambda})$, or in other words $\lambda_a=\lambda_{l-a+1}$. We then use equation \eqref{eqn:SU(u)_integrable_reps} to get $N-u=2\frac{u}{l}\sum_{a=1}^{l/2}\lambda_a$, so that $N-u$ must be an even multiple of $\frac{u}{l}$. The action of charge conjugation on our state is then
\begin{equation*}
    \operator{C}\ket{[\vec{\lambda},\xi]}=\ket{[\mathcal{T}(\vec{\lambda}),-\xi-(N-u-\lambda_0)]}.
\end{equation*}
If our state is $\operator{C}$ invariant, then referring to equation \eqref{eqn:U(u)_cs_label} we see that $\xi$ must satisfy $\xi-\lambda_0-1\equiv -\xi-(N-u-\lambda_0)\bmod{\frac{N}{u/l}}$, or equivalently $2(\xi-\lambda_0)\equiv 1-(N-u)\bmod{\frac{N}{u/l}}$. Since $N-u$ is an even multiple of $\frac{u}{l}$, and both $u$ and $l$ are even, we know that $\frac{N}{u/l}$ must be even, and hence there are no integers $\xi$ which make $\ket{[\vec{\lambda},\xi]}$ charge conjugation invariant.

%% file: deconfinementProofs.tex
\subsection{Proofs of equations \eqref{eqn:deconfined_weights} and \eqref{eqn:u_deconfined}}
We want to prove that $\vec{\mu}$ can be written as the difference of two BPS $u$-wall fluxes if and only if $\mu_a=\vec{\alpha}_a^*\cdot\vec{\mu}$ is one of $\{-1,0,+1\}$ for each $a=0,1,\dots,r$.

\paragraph{}
For the forward direction suppose that $\vec{\mu}$ is deconfined so that $\vec{\mu}=\vec{\Phi}_1-\vec{\Phi}_2$ for two BPS $u$-walls $\vec{\Phi}_1$ and $\vec{\Phi}_2$. Let the domain walls be such that $\vec{\Phi}_i=\sum_{a=1}^rq_a^i\vec{w}_i-\frac{u}{c_2}\vec{\rho}$, with $q_a^i=\vec{\alpha}_a^*\cdot\vec{\Phi}_i+\frac{u}{c_2}\in\{0,1\}$ for $a=0,1,\dots,r$ by the BPS condition \eqref{eqn:BPS_condition}. The condition $\vec{\mu}=\vec{\Phi}_1-\vec{\Phi}_2$ tells us that $\mu_a\equiv\vec{\alpha}_a^*\cdot\vec{\mu}=q_a^1-q_a^2$ must be in the set $\{-1,0,+1\}$.

\paragraph{}
For the other direction suppose that $\mu_a\in\{-1,0,+1\}$. To show that $\vec{\mu}$ is deconfined we just have to show that there are two BPS $u$-walls, $\vec{\Phi}_1$ and $\vec{\Phi}_2$, such that $\vec{\mu}=\vec{\Phi}_1-\vec{\Phi}_2$. To that end, let $\vec{\Phi}_i=\sum_{a=1}^rq_a^i\vec{w}_i-\frac{u}{c_2}\vec{\rho}$ as before, and define $q_a^i$ in the following way,
\begin{align*}
    q_a^1&=\begin{dcases} 1 & \mu_a=1 \\ 0 & \mu_a=-1 \\ q_a & \mu_a=0 \end{dcases}\\
    q_a^2&=\begin{dcases} 0 & \mu_a=1 \\ 1 & \mu_a=-1 \\ q_a & \mu_a=0 \end{dcases},
\end{align*}
which automatically guarantees that $q_a^1-q_a^2=\mu_a$. Note that at this point, $q_a\in\{0,1\}$ but is not otherwise constrained. Now we just have to show that we may choose $u$ and $q_a$ such that both the $\vec{\Phi}_i$ are BPS. To proceed, let $\zeta\in S_{r+1}$ be a permutation on the labels $\{0,1,\dots,r\}$ so that $\mu_{\zeta(a)}$ is
\begin{equation*}
    \mu_{\zeta(a)}=\begin{dcases}
        +1 & a=0,1,\dots,m-1\\
        -1 & a=m,\dots,l\\
        0 & a = l+1,\dots,r
    \end{dcases}.
\end{equation*}
Because $\sum_{a=0}^rk_a^*\vec{\alpha}_a^*=0$, we know that $0=\sum_{a=0}^rk_a^*\mu_a^*=\sum_{a=0}^{m-1}k_{\zeta(a)}^*-\sum_{a=m}^{l}k_{\zeta(a)}^*$, so we automatically get $\sum_{a=0}^rk_a^*q_a^1=\sum_{a=0}^rk_a^*q_a^2$. We then simply take $u=\sum_{a=0}^rk_a^*q_a^i$ and we've proved equation \eqref{eqn:deconfined_weights}. To get the full set of allowed values of $u$, we simply consider all possible values of $q_{\zeta(a>l)}$, giving us equation \eqref{eqn:u_deconfined}.

\subsection[Finding weights of all $N$-alities]{Finding weights of all $\boldsymbol{N}$-alities}
To find weights of all $N$-alities we use the fact that every irreducible representation, $R_{\vec{\lambda}}$, is labelled by its highest weight, $\vec{\lambda}$ which must be dominant\footnote{A weight is dominant when it has only non-negative coefficients when written as a sum of fundamental weights. Equivalently, $\vec{\lambda}$ is dominant if and only if $\vec{\alpha}_a^*\cdot\vec{\lambda}\geq 0$ for all $a=1,\dots,r$.}. Further, we use the well known result that a dominant weight $\vec{\mu}$ is a weight of $R_{\vec{\lambda}}$ if and only if $\vec{\lambda}\succeq\vec{\mu}$ and $\vec{\lambda}-\vec{\mu}$ is in the root lattice, or in other words if $\vec{w}_a^*\cdot(\vec{\lambda}-\vec{\mu})$ is a non-negative integer for each $a=1,\dots,r$. We may apply this result to an arbitrary weight $\vec{\nu}$ by first applying a Weyl group element which makes $\vec{\nu}$ dominant, $\weyl{w}_\nu$, and then checking if $\weyl{w}_\nu(\vec{\nu})$ is a weight of $R_{\vec{\lambda}}$, remembering that the set of weights of a representation is invariant under the Weyl group.

\paragraph{}
Note that the fact that weights of an irrep have the same $N$-ality as the highest weight, as noted in appendix \ref{appendix:center}, allows us to check if a dominant weight $\vec{\mu}$ is a weight of $R_{\vec{\lambda}}$ by verifying that $N$-ality of $R_{\vec{\mu}}$ is the same as that of $R_{\vec{\lambda}}$, along with checking that $\vec{\lambda}\succeq\vec{\mu}$.

\paragraph{An example: $\boldsymbol{\Sp{N}}$}
Recall from appendix \ref{sec:all_groups_symmetries} that the $N$-ality of $R_{\vec{\lambda}}$ is
\begin{equation*}
    N\text{-ality}(R_{\vec{\lambda}})=2\vec{w}_N^*\cdot\vec{\lambda}\bmod{2}=\sum_{j=0}^{\left\lfloor(N-1)/2\right\rfloor}\lambda_{2j+1}\bmod{2}.
\end{equation*}
We claim that $\vec{w}_2$ is a weight of all $N$-ality 0 representations and $\vec{w}_1$ is a weight of all $N$-ality 1 representations. We will just prove the first claim, as the proof of the second is essentially identical.

\paragraph{}
We start with an $N$-ality 0 representation $R_{\vec{\lambda}}$. To see that $\vec{w}_2$ is a weight of $R_{\vec{\lambda}}$ we have to show that $\vec{w}_2$ has $N$-ality 0, which is clear from the inner product above, and that $\vec{\lambda}\succeq\vec{w}_2$. Using the inner product above, we calculate $\vec{w}_a^*\cdot(\vec{\lambda}-\vec{w}_2)$,
\begin{equation*}
    \vec{w}_a^*\cdot(\vec{\lambda}-\vec{w}_2)=\begin{dcases}
        \sum_{b=1}^{N}\lambda_b-1 & a=1\\
        \sum_{b=1}^{N}\min(a,b)\lambda_b-2 & 2\leq a<N \\
        \sum_{b=1}^{\left\lfloor(N-1)/2\right\rfloor}\left(j+\frac{1}{2}\right)\lambda_{2j+1}+\sum_{j=1}^{\left\lfloor N/2\right\rfloor}j\lambda_{2j}-1 & a=N
    \end{dcases}.
\end{equation*}
For $a=1$, we see that the only way to get a negative result is if $\vec{\lambda}$ vanishes, which by assumption does not happen. For $2\leq a<N$ and $a=N$, we would need $\vec{\lambda}=\vec{w}_1$ in order to get a negative result, which does not have $N$-ality 0 so can be ruled out, thus showing that $\vec{w}_2$ must be a weight of $R_{\vec{\lambda}}$.